\documentclass[twocolumn]{aastex63}

\usepackage{color}
\usepackage{graphicx, epstopdf} 
\usepackage{gensymb}
\usepackage{natbib}
\usepackage{amsmath}
\usepackage{amsfonts}
\usepackage{amsthm}
\usepackage{xspace}

\newcommand{\dmunits}{\ensuremath{{\,{\rm pc}\,{\rm cm}^{-3}}}\xspace}
\newcommand{\src}{FRB~121102}
\newcommand{\frb}{FRB~121102}

\received{-}
\revised{-}
\accepted{-}
\submitjournal{ApJ}

\shorttitle{FRB~121102 Dense bursts Analysis}
\shortauthors{Aggarwal et al.}

\graphicspath{{./}{figures/}}

\begin{document}

\title{Comprehensive analysis of a dense sample of FRB~121102 bursts}

\correspondingauthor{Kshitij Aggarwal}
\email{ka0064@mix.wvu.edu}

\author[0000-0002-2059-0525]{Kshitij Aggarwal}
\affil{West Virginia University, Department of Physics and Astronomy, P. O. Box 6315, Morgantown, WV, USA}
\affil{Center for Gravitational Waves and Cosmology, West Virginia University, Chestnut Ridge Research Building, Morgantown, WV, USA}

\author[0000-0003-0385-491X]{Devansh Agarwal}
\affil{West Virginia University, Department of Physics and Astronomy, P. O. Box 6315, Morgantown, WV, USA}
\affil{Center for Gravitational Waves and Cosmology, West Virginia University, Chestnut Ridge Research Building, Morgantown, WV, USA}

\author[0000-0002-2972-522X]{Evan F. Lewis}
\affil{West Virginia University, Department of Physics and Astronomy, P. O. Box 6315, Morgantown, WV, USA}
\affil{Center for Gravitational Waves and Cosmology, West Virginia University, Chestnut Ridge Research Building, Morgantown, WV, USA}

\author[0000-0001-8057-0633]{Reshma Anna-Thomas}
\affil{West Virginia University, Department of Physics and Astronomy, P. O. Box 6315, Morgantown, WV, USA}
\affil{Center for Gravitational Waves and Cosmology, West Virginia University, Chestnut Ridge Research Building, Morgantown, WV, USA}

\author[0000-0001-9852-6825]{Jacob Cardinal Tremblay}
\affil{West Virginia University, Department of Physics and Astronomy, P. O. Box 6315, Morgantown, WV, USA}
\affil{Center for Gravitational Waves and Cosmology, West Virginia University, Chestnut Ridge Research Building, Morgantown, WV, USA}

\author[0000-0003-4052-7838]{Sarah Burke-Spolaor}
\affil{West Virginia University, Department of Physics and Astronomy, P. O. Box 6315, Morgantown, WV, USA}
\affil{Center for Gravitational Waves and Cosmology, West Virginia University, Chestnut Ridge Research Building, Morgantown, WV, USA}
\affiliation{CIFAR Azrieli Global Scholars program, CIFAR, Toronto, Canada}

\author[0000-0001-7697-7422]{Maura A. McLaughlin}
\affil{West Virginia University, Department of Physics and Astronomy, P. O. Box 6315, Morgantown, WV, USA}
\affil{Center for Gravitational Waves and Cosmology, West Virginia University, Chestnut Ridge Research Building, Morgantown, WV, USA}

\author[0000-0003-1301-966X]{Duncan R. Lorimer}
\affil{West Virginia University, Department of Physics and Astronomy, P. O. Box 6315, Morgantown, WV, USA}
\affil{Center for Gravitational Waves and Cosmology, West Virginia University, Chestnut Ridge Research Building, Morgantown, WV, USA}

\begin{abstract}
We present an analysis of a densely repeating sample of bursts from the first repeating fast radio burst, \src.
%detection of 93 additional bursts from \src, the first repeating FRB, 
We reanalysed the data used by \citet{Gourdji2019} and detected 93 additional bursts using our single-pulse search pipeline. In total, we detected 133 bursts in three hours of data at a center frequency of 1.4~GHz using the Arecibo telescope, and
%We report the detection of 93 additional bursts from \src, the first repeating FRB, at a center frequency of 1.4~GHz using Arecibo Telescope. We 
develop robust modeling strategies to constrain the spectro-temporal properties of all the bursts in the sample. Most of the burst profiles show a scattering tail, and burst spectra are well modeled by a Gaussian with a median width of 230~MHz. We find a lack of emission below 1300~MHz, consistent with previous studies of \src. We also find that the peak of the log-normal distribution of wait times decreases from 207~s to 75~s using our larger sample of bursts, as compared to that of \citet{Gourdji2019}. Our observations do not favor either Poissonian or Weibull distributions for the burst rate distribution. We searched for periodicity in the bursts %(on timescales less than one minute)} periodicities 
using multiple techniques, but did not detect any significant period. The cumulative burst energy distribution exhibits a broken power-law shape, with the lower and higher-energy slopes of $-0.4\pm0.1$ and $-1.8\pm0.2$, with the break at $(2.3\pm0.2)\times 10^{37}$~ergs.
%MAM: What about the lower-energy slope? Also could just say 2.3\pm0.2.
%KA: We didn't mention the lower energy slope here as we suspect that it could be due to incompleteness.  
%We compare the pipelines used for the previous result published by \citep{Gourdji2019} and our search pipeline to explain the extra burst detections. 
We provide our burst fitting routines as a python package \textsc{burstfit}\footnote{\url{https://github.com/thepetabyteproject/burstfit}} that can be used to model the spectrogram of any complex FRB or pulsar pulse using robust fitting techniques.
All the other analysis scripts and results are publicly available\footnote{\url{https://github.com/thepetabyteproject/FRB121102}}.    
\end{abstract}


\section{Introduction}
\label{sec:intro}
Repeating sources of fast radio bursts (FRBs) have helped broaden the horizons of FRB astronomy since their discovery \citep{Spitler2014, Spitler2016, r2}. \src\ was first detected in the PALFA survey using the Arecibo telescope, and ten repeat bursts were later detected during targeted observations of the source. While \src\ remains the most extensively observed and studied repeater, to date, 24 repeaters have now been reported \citep{frb_letter, Spitler2016, luo2020, Kumar2019, Kumar2021, chime_cat, fonseca2020}. Moreover, two of these repeating FRBs (\src\ and FRB~180916) show long-term (i.e., days to months) periodicity in their activity \citep{Aggarwal2020, Rajwade121102, Marazuela2020}. Some FRBs have also been localized to a variety of host galaxies, ranging from dwarf galaxies to massive elliptical galaxies to luminous spiral galaxies \citep{Heintz2020}.

Both repeating and non-repeating FRBs  show a variety of spectro-temporal features and polarization properties: frequency modulation, sub-millisecond structure, drifting sub-pulses, varying polarization position angle,
%MAM: What does diverse polarization angle mean?
% KA: It was supposed to be polarization position angle, from Luo et al: https://www.nature.com/articles/s41586-020-2827-2 
etc \citep{Shannon2018, farah2019, luo2020}. Detection of multiple bursts from repeaters (like \src) have facilitated detailed studies of their properties and their environment. However, even after extensive follow-up and detection of hundreds of bursts from \src, the intrinsic emission mechanism remains uncertain, and many progenitor models have been proposed to explain the observational results. As theoretical models lack robust predictions for the observed properties of bursts, several empirical techniques have been employed to model the observed properties of the bursts. Some of those are: (1) using Weibull and Poisson distributions to model the clustering of repeater bursts; (2) using a truncated and broken power-law to model the flux-density distribution; (3) using 2D Gaussians to model the burst spectro-temporal properties; (4) using Gaussians convolved with an exponential tail to model scattering; (5) using a statistical spectral index to compare burst rates at multiple frequencies; (6) using signal-to-noise and structure to maximize DM, etc. \citep[for further details, see][]{li2021, Cruces2020, Marazuela2020, Gourdji2019, Hessels2019, Houben2019, Gajjar2018}.

In view of all these considerations, it is necessary to detect and carefully investigate a large number of bursts from the repeaters to improve the understanding of their emission mechanism. In this paper, we reanalyze the observations for \src\ previously reported by \citet{Gourdji2019} and present the detection of an additional 93 bursts, for a full sample of 133 bursts in these observations. We detail a thorough burst modeling procedure and report extreme frequency modulation in burst spectra and a dearth of burst emission below 1300~MHz. We present and compare the updated burst energy and wait-time distributions and demonstrate how these estimates change dramatically for an incomplete search. We also perform exhaustive short-period periodicity tests to detect any possible rotational period as predicted by neutron star-based progenitor models. We also discuss various differences between our single-pulse search pipeline and the one used by \citet{Gourdji2019} to explain the extra bursts detected in our search (see Section~\ref{sec:comparison}).

% \fixme{With the detection of these additional bursts, these two observations constitute the highest \src\ detection rate reported so far for any FRB at 1.4~GHz.}
%MAM: Is this true? And the previous result w/o your work wasn't already the highest detection rate?
% KA: Maybe we should remove this statement. Zhang et al found most of their bursts (out of 92) within 30 min. And the FAST 121102 paper will anyway have a much higher burst rate. 
This paper is laid out as follows: In \S2 we briefly discuss the data used in this work and discuss the search and spectro-temporal modeling procedure in \S3. We then present our modeling results in \S4, followed by a discussion of those results and conclusions  in \S5 and \S6. 

% do we need this paragraph?
% \citet{Chatterjee2017} used the Very Large Array to localize the FRB to $\sim100$~mas position and reported coincident persistent radio and optical sources. The coincide between the persistent source, and R1 was further improved to $\sim12$~mas by \citet{Marcote2017} from detection using the European VLBI Network. Using optical observations \citet{Tendulkar2017} reported the host galaxy of R1 as a low mass and metallicity dwarf galaxy at a redshift of $z=0.19$. \citet{Bassa2017} used the Hubble Space Telescope to show that the position of R1 is within the half-light radius of a star-forming region.

% Multi-wavelength observations of the source have spanned 

\section{Data}
\label{subsec:data}
The data reported here were originally collected, searched for FRBs, and reported by \citet{Gourdji2019}. Here we provide only a brief summary of the data used in this analysis, and refer the reader to \citet{Gourdji2019} for further details. The observations were carried out with the 305-m William E. Gordon Telescope at the Arecibo Observatory with the L-Wide receiver and recorded using the Puerto Rican Ultimate Pulsar Processing Instrument (PUPPI). \src\ was observed  with 800~MHz bandwidth at a center frequency of 1375~MHz on MJDs 57644 and 57645 for 5967~s and 5545~s respectively. The data were coherently dedispersed at a dispersion measure (DM) value of 557~\dmunits during the observations and were recorded with 1.56-MHz channel bandwidth and 10.25-$\mu$s sampling resolution. The data used for this study were further decimated to 12.5-MHz channel bandwidth with 64 total channels and 81.92-$\mu$s sampling interval.

\section{Methods}
\subsection{The Petabyte Project}
Characterizing the diversity and event rates of FRBs as a function of observing frequency is critical for understanding their nature,
the extreme emission physics responsible for FRB and pulsar emission, and the relationship between these two classes of objects. Many surveys have sought comprehensive estimates of these values, all using different observing frequencies, telescopes, and search algorithms but without characterizing the completeness of their search. The Petabyte Project\footnote{\url{https://thepetabyteproject.github.io}} (TPP) aims to address these issues to provide robust event rate estimates and discoveries in several petabytes of new and archival radio data. TPP will perform a uniform search for FRBs in an unprecedented amount of archival data and better probe transients closer, farther, and at higher radio frequencies than previous searches. Our search will have a robust internal assessment of completeness, allowing us to confidently project the frequency-dependent rates of FRBs and other transients.

TPP will use \textsc{Your} \citep[a recursive acronym for ``your unified reader'';][]{your} to ingest the data and {\sc heimdall} \citep{barsdell2012heimdall}\footnote{\url{https://sourceforge.net/projects/heimdall-astro}} to search it for single pulses. The deep learning-based classifier {\sc fetch} \citep{fetch2020}\footnote{\url{https://github.com/devanshkv/fetch}} is then used to classify the candidates identified by  {\sc heimdall}. The data will be searched up to a DM of 5000~\dmunits (or more, if possible) and a pulse width of 32~ms. This pipeline can easily be modified to search for higher DMs and pulse widths, in specific cases. All the candidates above a signal-to-noise ratio (S/N) of 6 classified as astrophysical by {\sc fetch} will be manually verified. The maximum DM and the pulse width to be searched is governed by the observing frequency, data resolution, and GPU memory and hence would be dealt with on a case-by-case basis. The data and results presented in this paper were analyzed under TPP, using an early version of the TPP pipeline. In the following subsection, we discuss the details of the single-pulse search pipeline used in this analysis.

\subsection{Single-Pulse Search}
\begin{figure*}
    \centering
    \includegraphics{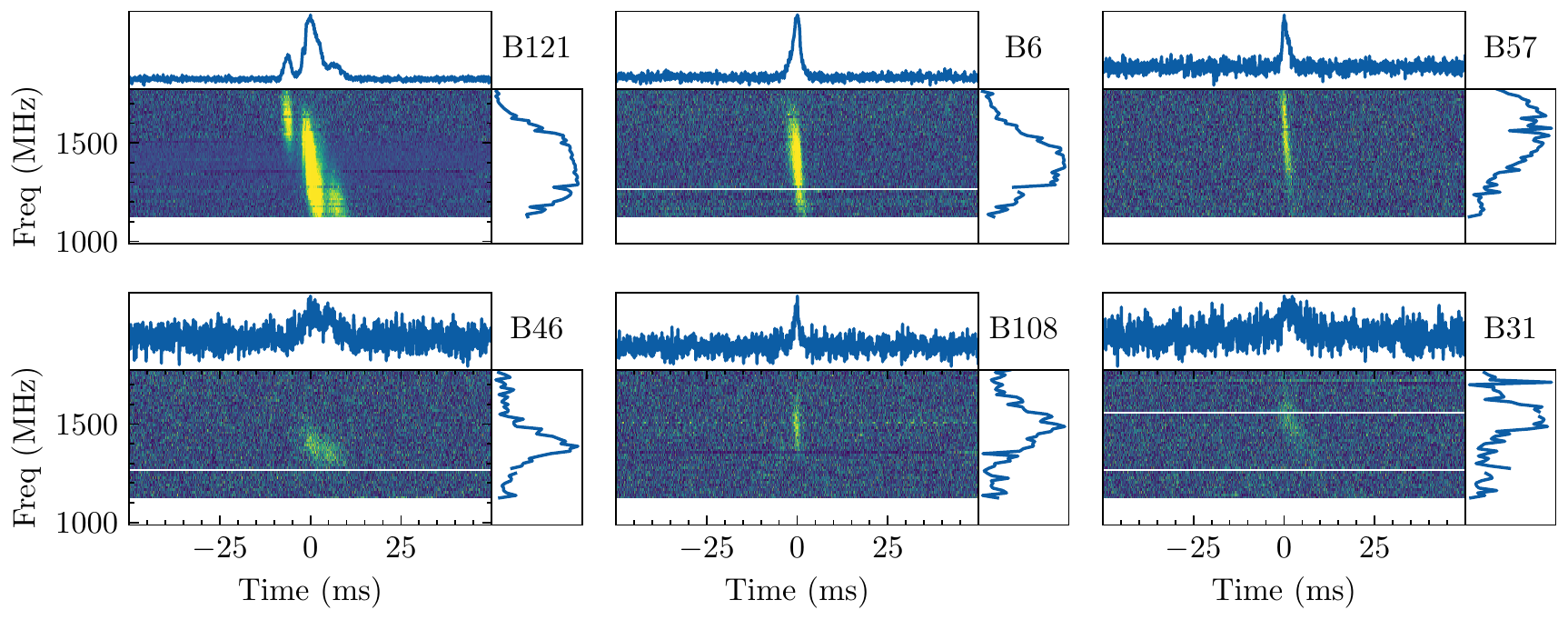}
    \caption{Dynamic spectra of six bursts, dedispersed at DM = 560.5\dmunits. For each burst, the top panel shows the burst profile obtained by averaging along the frequency axis, and the right panel shows the burst spectra obtained by averaging the burst data along the time axis. The white horizontal lines show the channels masked due to RFI. The color scale of each spectrogram has been set from mean to 3 times the standard deviation of the off-pulse region. The ranges of 1D plots are different for individual plots. Burst numbers are mentioned on the top right of each subplot.}
    \label{fig:bursts}
\end{figure*}

\begin{figure}
    \centering
    \includegraphics{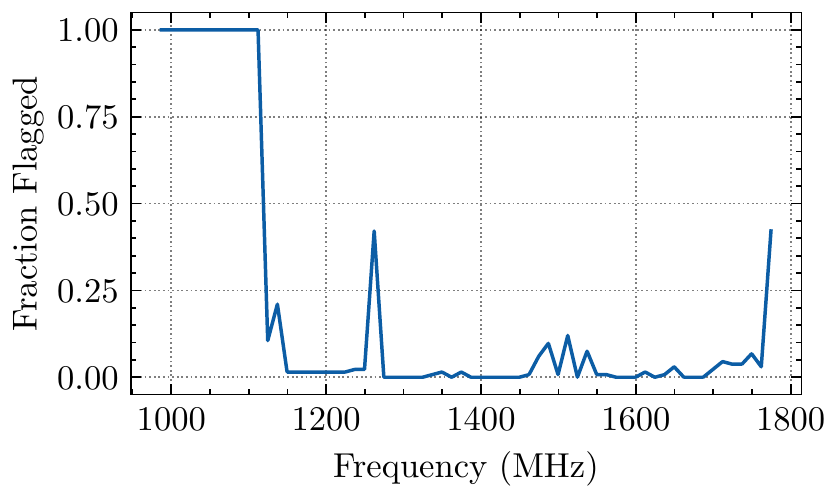}
    \caption{The fraction of bursts for which a frequency channel was flagged due to RFI.}
    \label{fig:rfifrac}
\end{figure}

Within \textsc{Your}, we used \texttt{your\_heimdall.py} which runs \textsc{heimdall} on the \textsc{psrfits} data files collected by PUPPI for the single-pulse search. We used two search strategies: (1) DM range between 450--650~\dmunits with a DM tolerance\footnote{DM tolerance is the acceptable sensitivity loss between DM trials for a single-pulse search \citep[for further details, see][]{rfclustering, levin2012}} of 1\% and (2) DM range between 10--5000~\dmunits with a tolerance of 25\%, with a maximum pulse width of 84~ms in both cases \citep[note that the widest \src\ pulse reported at this frequency had a width of $\sim$35~ms, see][]{Cruces2020}. 
%MAM: DM tolerance needs to be defined.
% KA: Done. 
The searches resulted in 1,428 and 11,276 candidates, respectively. 
%We used \textsc{fetch} to label radio frequency interference (RFI) and FRB bursts. 
%MAM: I commented this out, as I found it confusing to say this first, and then again later in the paragraph.
% KA: Looks better now. Thanks!  
For each candidate, we extracted a segment of the data (which we hereafter refer to as a ``cutout'') centered at the arrival time (referenced to  the top of the observing band) as reported by \textsc{heimdall} with a time window equal to twice the dispersion delay using \texttt{your\_candmaker.py}. We then used spectral kurtosis RFI mitigation with a 3$\sigma$ threshold \citep{nita2010} to identify and excise frequency channels corrupted by RFI (see Figure~\ref{fig:rfifrac} for fraction of bursts for which a frequency channel was flagged due to RFI) and used this cleaned data to create dedispersed frequency-time images where a factor of width/2 was used to decimate the time axis. We created the DM--time image by dedispersing the data from zero to twice the reported DM and simultaneously decimating the time axis as above \citep[for more details on the candidate pre-processing, see][]{fetch2020}. These cutouts are then used by \textsc{fetch} to label FRBs and RFI, and were also manually verified. In total, we found 133 bursts with DMs consistent with that of \src\, (i.e., between 550-580~\dmunits) with 93 new bursts as compared to the previously published results. We highlight some important differences between our single-pulse search pipeline and the one used by \citet{Gourdji2019} in Section~\ref{sec:comparison} to explain the new burst detections. We did not detect any bursts at other DMs. Our search missed one burst, B33, reported by \citet{Gourdji2019}, probably because it was weak and narrowband (see Section~\ref{sec:caveats} for caveats of our search).  Figure~\ref{fig:bursts} shows the dynamic spectra of some of the bursts. Candidate cutouts for all the bursts are available on Github\footnote{\url{https://github.com/thepetabyteproject/FRB121102}}.  

\subsection{Completeness Limit}
\label{sec:completeness}
We define the completeness limit as the pulse energy (also known as fluence) value above which any burst emitted during the observation would be detected. Determining the completeness limit of any single-pulse search is, therefore, essential to defining the sample of bursts to be used for statistical analyses. The most robust method of determining the completeness limit involves an exhaustive injection analysis. In such analyses, simulated transients (with varying properties) are injected on background data, and by analyzing the transients that were recovered (or missed), one can determine the completeness limit of a search \citep{li2021, gupta2021, chime_cat, gb, farah2019}. Such an analysis requires access to a large amount of native-resolution data observed with the same telescope and observing configuration as the search data. 

As we had access to decimated data for just two observations, we could not do such an injection analysis. We, therefore, estimate the completeness limit from the radiometer equation. We use a conservative approach by including the effect of RFI mitigation, which reduces search sensitivity. We flagged more than 35\% of data for many candidates due to RFI, leading to a usable bandwidth of 500~MHz. Using this smaller bandwidth and nominal pulse width of 1~ms, the fluence limit above an S/N of 8 is 0.0216~Jy~ms. We use bursts with fitted fluence above this limit in the burst sample analysis.

\subsection{Spectro-temporal Burst Modelling}
\label{sec:stmodel}
To measure the properties of the bursts in our sample, we perform spectro-temporal modeling of all the detections. We model each component of the bursts' spectra with a Gaussian function and the profile using a Gaussian convolved with a one-sided exponential function to represent the Gaussian pulse and the scattering tail. Therefore, we model each component of the burst (at an observing frequency $f$ and time $t$), using the following function,
\begin{eqnarray}
    \mathcal{F}(f, t; {S}, \mu_f, \sigma_f, \mu_{\mathrm{DM}}, \sigma_t, \tau_\mathrm{sc} ) =&&  S \times 2.355 \sigma_f \nonumber \\
    && \times \mathcal{P}_f(t;\mu_{\mathrm{DM}},\sigma_t,\tau_\mathrm{sc}) \nonumber \\
    && \times \mathcal{G}(f; \mu_f, \sigma_f).
\end{eqnarray}
Here, $S$ is the fluence of the component, $\mathcal{G}$ is the Gaussian function to model the spectra
\begin{eqnarray}
    \mathcal{G}(x;\mu_x, \sigma_x) = \frac{1}{\sigma_x \sqrt{2 \pi}} \exp{\bigg(-\frac{1}{2} \frac{(x-\mu_x)^2}{\sigma_x^2}\bigg)}
\end{eqnarray}
and $\mathcal{P}$ is used to model the profile. For bursts with scattering, we use the Gaussian convolved with a one-sided exponential function \citep{McKinnon2014}, 
\begin{eqnarray}
    \label{eq:pscat}
    \mathcal{P}^\mathrm{scat}_f(t;\mu_{\mathrm{DM}},\sigma_t,\tau_\mathrm{sc}) =&& \frac{1}{2 \tau_\mathrm{sc}} \left\{ 1 + \mathrm{erf}\bigg[ \frac{t - (\mu_{\mathrm{DM}} + \sigma_t^2/\tau_\mathrm{sc})}{\sigma_t \sqrt{2}}\bigg] \right\} \nonumber \\ 
    && \times \exp{\bigg( \frac{\sigma_t^2}{2 \tau_\mathrm{sc}^2}\bigg)} \exp{\bigg( \frac{t-\mu_{\mathrm{DM}}}{\tau_\mathrm{sc}}\bigg)}.
    % \mathcal{P}^\mathrm{scat}_f(t;\mu_{\mathrm{DM}},\sigma_t,\tau_\mathrm{sc}) =&& \frac{1}{2 \tau_\mathrm{sc}} \exp{\bigg( \frac{\sigma_t^2}{2 \tau_\mathrm{sc}^2}\bigg)} \exp{\bigg( \frac{t-\mu_{\mathrm{DM}}}{\tau_\mathrm{sc}}\bigg)} \nonumber \\
    % && \times \left\{ 1 + \mathrm{erf}\bigg[ \frac{t - (\mu_{\mathrm{DM}} + \sigma_t^2/\tau_\mathrm{sc})}{\sigma_t \sqrt{2}}\bigg] \right\}.
\end{eqnarray}
where the mean of the Gaussian pulse after accounting for the dispersion delay at that respective frequency,  
\begin{eqnarray}
    \mu_{\mathrm{DM}} = \mu_t - 4.148808 \times 10^3 ~ \mathrm{DM}\, \Big( \frac{1}{f^2} - \frac{1}{f_{\rm top}^2} \Big) \,\mathrm{s}.
\end{eqnarray}
Here $f_{\rm top}$ is the highest frequency in the band (in MHz), $f$ is the frequency of a channel (in MHz) and $\mu_t$ is the mean of the Gaussian pulse at $f_{\rm top}$. At each frequency channel, the scattering time scale 
\begin{equation}
    \tau_\mathrm{sc}(f) = \tau_\mathrm{sc} \bigg( \frac{f}{f_\mathrm{ref}} \bigg)^{-4}.
\end{equation}
In this expression, $f_\mathrm{ref}$ is the reference frequency and is set to be 1~GHz. The exponent (--4) is assuming a normal distribution of plasma-density inhomogeneities. % {\bf DL:SAY WHY THE --4 EXPONENT HAS BEEN CHOSEN}.
Finally, we defined $\mathcal{P}_f(t;\mu_{\mathrm{DM}},\sigma_t,\tau_\mathrm{sc})$ to be a Gaussian $\mathcal{G}(t;\mu_\mathrm{DM}, \sigma_t)$ for $\tau_\mathrm{sc}/\sigma_t < 6$, and $\mathcal{P}^\mathrm{scat}_f(t;\mu_{\mathrm{DM}},\sigma_t,\tau_\mathrm{sc})$ for $\tau_\mathrm{sc}/\sigma_t > 6$. We used this value for the cutoff in order to maintain numerical stability while calculating Equation~\ref{eq:pscat}.
% Using this, we were able to use a single pulse function to model the bursts that do not show scattering with the same function, 
% We used this value for the cutoff because for $\tau_\mathrm{sc}/\sigma_t < 6$ we 

It follows from the above discussion that our model $\mathcal{F}$ is generated using  seven parameters: $S, \mu_f, \sigma_f$, $\mu_t, \sigma_t, \tau_\mathrm{sc}$ and DM. Using this model we fit the burst spectrograms, as described in the next section.

\subsection{\textsc{burstfit}}
\label{sec:burstfit}
While fitting for complex FRB bursts is an arduous task, we scrupulously automate the entire procedure and present it as python package \textsc{burstfit}\footnote{\url{https://github.com/thepetabyteproject/burstfit}}. \textsc{burstfit} provides a framework to model any spectrogram consisting of any complex FRB or pulsar pulse using robust methods. It can easily incorporate any user-defined python function(s) to model the profile, spectra, and spectrogram and is not limited to the functions we have implemented for this current analysis. \textsc{burstfit} primarily consists of the following five steps. 

% Here we outline the modelling procedure implemented in \textsc{burstfit}. 

\subsubsection{Data Preparation}
First, we dedispersed the burst spectrogram at the DM obtained from the single-pulse search. This DM is usually accurate enough to correct for most of the dispersion. We cut out a time window of 200~ms encompassing the burst from this dedispersed spectrogram and normalized this data to zero mean and unit standard deviation using the off-pulse region. Both the cutout and normalized data were then used for fitting. We also masked all the channels flagged as RFI during the search and candidate pre-processing so as not to influence the fitting procedure. 

\subsubsection{Stage 1: Single-component Fitting}
In this first stage of fitting, we used \texttt{scipy.curve\_fit}\footnote{This routine is a part of the python-based \texttt{scipy} package. We used version 1.5.2 of this package in our analysis.} to perform the fits and got an initial estimate of the fitted parameters. We created the time-series profile by summing along the frequency axis, and modelled it using $S\times\mathcal{P}_f(t;\mu_t,\sigma_t,\tau_\mathrm{sc})$. We then used the fitted values of $\mu_t$ and $\sigma_t$ to identify the time samples with the burst signal and average them to produce the burst spectra. We normalised this spectra to unit area, and modelled it using $\mathcal{G}(f;\mu_f, \sigma_f)$. Following this, we modelled the complete spectrogram. We first generated the model spectrogram by stacking N$_f$ model pulse profiles ($S\times\mathcal{P}_f(t;\mu_\mathrm{DM},\sigma_t,\tau_\mathrm{sc})$) together, where N$_f$ is the number of frequency channels. Note that the mean of each profile was already corrected for the dispersion delay at that respective frequency. This gave us a scattered and dispersed spectrogram at a given set of profile parameters and a DM. 

% \fixme{We correct for dispersion delay within the time samples using a $1\times2$ convolution kernel}. 
We then multiplied this spectrogram with the model spectra to obtain the model spectrogram. Additionally, we clipped the model spectrogram at the estimated saturation level (see Section~\ref{sec:saturation}) and masked the RFI channels. Using this model we fit for all the seven parameters in $\mathcal{F}(f, t; \mathrm{S}, \mu_f, \sigma_f, \mu_t, \sigma_t, \tau_\mathrm{sc})$ along with DM by comparing the model with the dedispersed cutout spectrogram obtained in the previous step. Again, we used \texttt{scipy.curve\_fit} for fitting, and used the estimates from individual profile and spectra fits as initial guesses for the parameters. 

\subsubsection{Stage 2: Statistical Tests}
Following Stage 1, we obtained the residual spectrogram by subtracting the fitted model from the original spectrogram. Then we performed several statistical tests \citep[see, e.g.,][]{kramer1994} to compare the properties of on-pulse residual with respect to the off-pulse regions in the original spectrogram. We performed the following three comparisons: left off-pulse vs. right off-pulse, on-pulse vs. left off-pulse, and on-pulse vs. right off-pulse. 
%MAM: Do ON and OFF needs to be capitalized? I find it distracting.
% KA: I also found it distracting, and I somehow thought it was necessary to capitalize them. Fixed it now. 
We used the following four tests (all implemented within \texttt{scipy}):  Kolmogorov–Smirnov test (for distribution comparison), F-test (for variance comparison), T-test (for mean comparison), and Kruskal test (for median comparison). 

We considered the two distributions similar if at least two of the four tests had a $p$-value above 0.05 (i.e. we did not have significant support for the non-similarity of the distributions). Comparing left off-pulse with right off-pulse region gave us confidence in our choice of off-pulse region. We terminated the single component fitting procedure if either of the off-pulse vs. on-pulse comparisons demonstrates that the distributions are similar. 
If the distributions were different, we used the residual spectrogram and repeated Stage 1 and 2 to fit another component and compared the on-pulse residual with off-pulse data. We kept fitting for components until the statistical tests pass or until a maximum of five components is reached. 

\subsubsection{Stage 3: Multi-component Fitting}
In cases where multiple components were found, we performed another stage of the fitting. Here,  we generated a combined model consisting of all the components and fit for all components by comparing our model with the original spectrogram. This combined model ($\mathcal{F_{\mathrm{all}}}$) was generated by summing together the individual component models ($\mathcal{F}_i$) for all $N$ components,
\begin{equation}
    \mathcal{F_{\mathrm{all}}} = \sum_i^N \mathcal{F}_i.
\end{equation}
Again, we used \texttt{scipy.curve\_fit} for fitting. Here, the fit results of the individual component fits from previous stages were used as the initial guess for the parameters in \texttt{curve\_fit}.

\subsubsection{Stage 4: MCMC}
While \texttt{scipy.curve\_fit} is sufficient for fitting in many scenarios, in our testing we found that the estimates and the errors reported by \texttt{scipy.curve\_fit} were not robust for our purposes. In many cases, the errors reported by \texttt{scipy.curve\_fit} were possibly underestimated, and fitted results were highly susceptible to the choice of input parameter bounds. This was especially true for low-significance bursts and multiple-component bursts, where the least-squares-minimization technique struggles to find a good solution. Therefore, we added another stage to our fitting procedure and used a Markov Chain Monte Carlo (MCMC) to obtain the final fitting results. We used the results of previous stages (that used \texttt{scipy.curve\_fit}) as initial estimates to determine the starting positions of the walkers for the MCMC. An advantage of the MCMC procedure is that it provides the full posterior distribution of all the fitted parameters, which we could then use to estimate the errors and further follow-up analysis of the burst sample. We used the Goodman and Weare affine invariant sampler \citep{Goodman2010} as implemented in \textsc{emcee} \citep{emcee}. We used uniform priors for all the parameters, with the ranges of the priors given in Table~\ref{tab:priors}. We used the log-likelihood function
\begin{equation}
    \ln{\mathcal{L}} = -0.5 \sum \left(\frac{\mathcal{S} - \mathcal{F}_{\mathrm{all}}}{\sigma}\right)^2,
\end{equation}
where $\mathcal{S}$ refers to the original spectrogram, $\mathcal{F}_{\mathrm{all}}$ refers to the model, and $\sigma$ is the off-pulse standard deviation of the measured spectrogram. The sum is over all the pixels in the two spectrograms. We used autocorrelation analysis to determine when the MCMC has converged\footnote{See \url{https://emcee.readthedocs.io/en/stable/tutorials/autocorr} for details.}. We then estimated the burn-in\footnote{To avoid the phenomenon known as ``burn-in'', where there is a high degree of correlation between neighboring samples in each MCMC chain, the initial values are typically discarded. This is especially important if the MCMC was initialized at a low probability region in the parameter space. Therefore, if the initial samples are not discarded, then those might bias the posterior distributions of MCMC samples. See section 7 of \citet{hogg2018} for more details.} using the autocorrelation time and used the remaining samples to determine the fitting results. To decide if scattering was present in a burst, we used the percentage of samples with \mbox{$\tau_\mathrm{sc}/\sigma_t < 6$.} If this percentage was greater than 50\%, we concluded that scattering was not present (or was very small) in that burst.
%MAM: This seems kind of random. Is there a reason for this criterion?
%KA: The reason behind this comes from computation of exponential. When tau/sigma < 6, the exponential in eq 3 starts to give inaccurate values as the ratio**2 gets very small. .     
We do not report scattering timescales for such bursts.      

We generated corner plots and fit-result plots to verify the quality of the fits, as shown in Figures~\ref{fig:fit_res} and \ref{fig:corner}. We provide all the results (output parameters, corner plots, fitting-result plots, etc.) from our analysis in a Github repository.\footnote{\url{https://github.com/thepetabyteproject/FRB121102}}

\begin{deluxetable}{ccc}
\tablewidth{0pc}
\label{tab:priors}
\tablecaption{Priors used in the MCMC fitting}
\tabletypesize{\footnotesize}
\tablehead{\colhead{Parameter} & \colhead{Minimum} & \colhead{Maximum}}
\startdata
\hline 
${S}$ & 0 & 500$\times$max(\texttt{time\_series})$\times\sigma_t^{\mathrm{fit}}$\\ 
$\mu_f$ & --2$\times$N$_f$ & 3$\times$N$_f$\\
$\sigma_f$ & 0 & 5$\times$N$_f$\\
$\mu_t$ & 0.8$\times\mu_t^{\mathrm{fit}}$ & $ 1.2\times\mu_t^{\mathrm{fit}}$\\
$\sigma_t$ & 0 & 1.2$\times(\sigma_t^{\mathrm{fit}} + \tau_\mathrm{sc}^{\mathrm{fit}})$\\
$\tau_\mathrm{sc}$ & 0 & 1.2$\times(\sigma_t^{\mathrm{fit}} + \tau_\mathrm{sc}^{\mathrm{fit}})$ \\
DM & 0.8$\times$DM$^\mathrm{fit}$ & 1.2$\times$DM$^\mathrm{fit}$ \\
\hline 
\enddata
\tablecomments{Superscript fit refers to the values obtained using fits done in previous stages. N$_f$ refers to the number of frequency channels. \texttt{time\_series} refers to the 1-D array obtained by summing the dedispersed cutout spectrogram along the frequency axis. Subscripts $t$ and $f$ are used for profile and spectra parameters.}
\end{deluxetable}

\subsubsection{Handling data saturation}
\label{sec:saturation}
\begin{figure}
    \centering
    \includegraphics{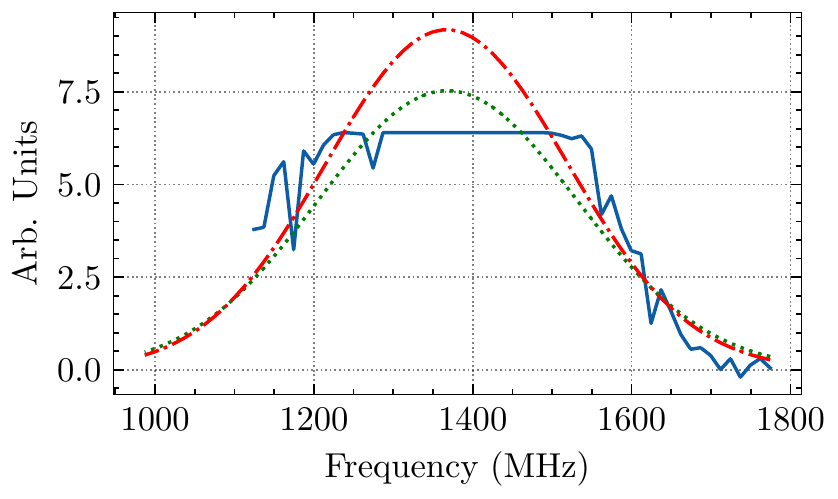}
    \caption{Plots of spectra and fits in case of data saturation. Blue solid line shows the spectrum of burst \textsc{B121}. The spectrum shows saturation between 1250~MHz and 1550~MHz, and any real structure in the spectrum is lost between those frequencies. The red dot-dashed line shows the model spectrum obtained when the fitting procedure incorporates the effect of saturation, while the dotted green line shows the spectrum obtained without considering saturation. The red curve better estimates the shape of the spectrum (assuming that the spectrum can be modeled using a Gaussian function), while the green curve underestimates the fluence and frequency width. See \S~\ref{sec:saturation} for more details.}
    \label{fig:saturation}
\end{figure}

The data we use in this analysis were recorded as 8-bit unsigned integers. Hence, the data range lies between 0--255, and any signal brighter than 255 is clipped at this value. We noticed data saturation for two bursts (B6 and B121), and hence this effect has been incorporated in our burst modeling. The spectrograms are subtracted by the off-pulse mean ($\mu_\mathrm{off}$) and divided by the off-pulse standard deviation ($\sigma_\mathrm{off}$). While making the spectro-temporal model, we clip the values greater than $(255 - \mu_\mathrm{off})/\sigma_\mathrm{off}$. This effect is visible in Figure~\ref{fig:saturation} for burst \textsc{B121} where the red dot-dashed curve and green dotted curve show the fit to the burst spectra with and without clipping, respectively. The fit performed without considering the saturation underestimates the burst's spectral width, leading to an underestimated burst energy. 
% A tacit assumption with this technique is that the saturated spectra are composed of a Gaussian function. %We cannot account for any additional components or complex functions as that information is lost due to saturation.

\begin{figure*}
    \centering
    \includegraphics[trim={0.5cm 10cm 0.0cm 6cm},clip, width=\textwidth]{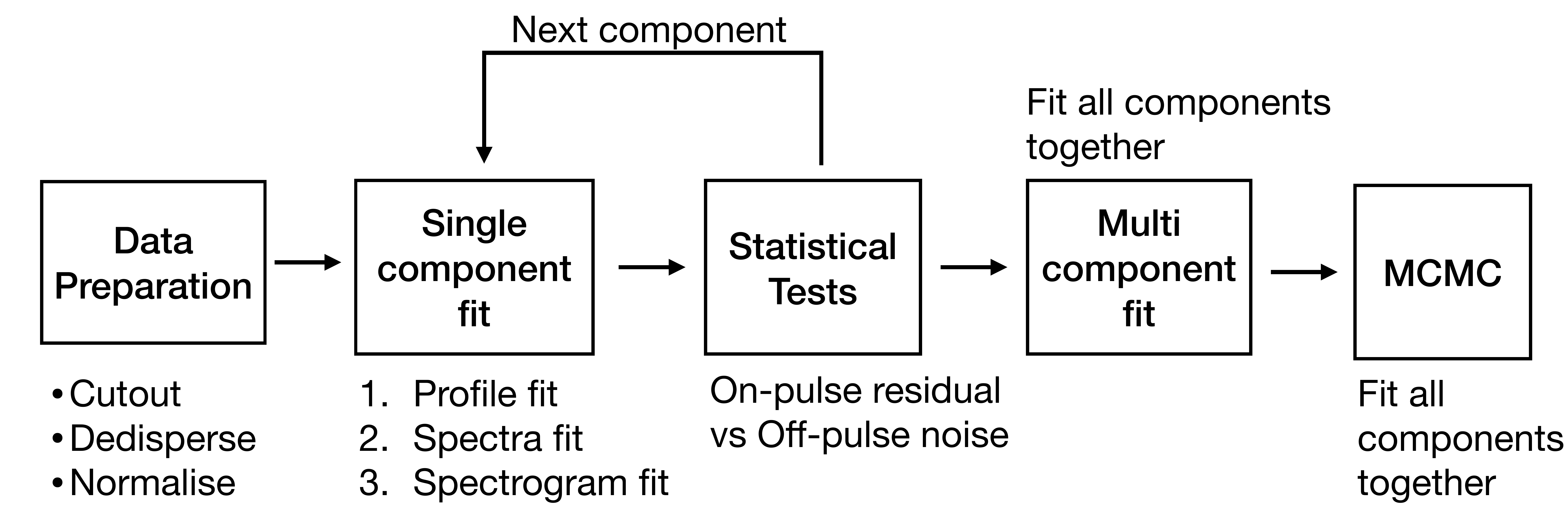}
    \caption{Flowchart showing the various stages of fitting in \textsc{burstfit}. See Section~\ref{sec:burstfit} for details.}
    \label{fig:bfflow}
\end{figure*}

\subsubsection{Caveats to our fitting analysis}
\label{sec:fitcaveats}
There are some caveats to our fitting procedure that are worth noting here. First, as with any model-dependent fitting, our analysis and results are dependent on the choice of the functions we use to model the data. We described those functions and our motivation for using them in Section~\ref{sec:stmodel}, but these are not the only proposed methods to model the spectrogram of an FRB. The spectrum of an FRB can also have power-law-like behavior \citep{chime1935}, and some other pulse broadening functions can also be used besides the exponential tail used to model the scattering effect \citep{Bhat2004}. 

Second, the emission of some bursts in our sample was present only in the top part of the band. In many such cases, the emission appeared similar to the tail of a Gaussian function, with the mean lying outside our observing band. Although we allowed for such a mean value to be estimated with the MCMC procedure, our fitting results for such bursts would inevitably be unconstrained. We therefore mark such bursts with a $\dagger$ in Table~\ref{tab:small_table} to highlight this. 

Third, in some cases, the MCMC procedure was unable to find a robust solution. This was due to the presence of RFI, which could not be removed as the FRB signal coincided with the channels heavily corrupted by RFI. It also occurred when there was significant baseline variation in the data close to a weak FRB pulse. This could dominate the MCMC likelihood estimate, and therefore the procedure could not converge on a solution. For such cases, we only used \texttt{scipy.curve\_fit} to perform the fits, and we modified the fitting bounds to obtain a visually good fit. We highlight these in Table~\ref{tab:small_table} with $*$, and note that the values could be unreliable.

\section{Results}
\label{sec:results}

\begin{deluxetable*}{llllllll}%cccccccc}
\tablecaption{Properties of the first 10 bursts. See Appendix for the full table.}
\label{tab:small_table}
\tablehead{\colhead{Burst}\tablenotemark{a} & \colhead{$\mu_f$} & \colhead{$\sigma_f$} & \colhead{$S$} & \colhead{$\mu_t$\tablenotemark{b}} & \colhead{$\sigma_t$ } & \colhead{$\tau$\tablenotemark{c}} & \colhead{DM}\\
\colhead{ID} & \colhead{(MHz)} & \colhead{(MHz)} & \colhead{(Jy ms)} & \colhead{(MJD)} & \colhead{(ms)} & \colhead{(ms)} & \colhead{(pc cm$^{-3}$)}}
\startdata
B1$*$ & $1560^{+30}_{-30}$ & $210^{+40}_{-40}$ & $0.09^{+0.02}_{-0.02}$ & 57644.408906976(1) & $0.0^{+0.02}_{-0.02}$ & $1.9^{+0.3}_{-0.3}$ & $565.3^{+0.4}_{-0.4}$\\
B2$*$ & $1200^{+10}_{-10}$ & $50^{+10}_{-10}$ & $0.043^{+0.007}_{-0.007}$ & 57644.40956768(1) & $1.35^{+0.05}_{-0.05}$ & - & $562.4^{+0.8}_{-0.8}$\\
B3.1$\dagger$ & $2900^{+300}_{-600}$ & $800^{+300}_{-200}$ & $0.6^{+0.7}_{-0.3}$ & 57644.409673699(3) & $0.4^{+0.2}_{-0.2}$ & $1.3^{+0.7}_{-0.7}$ & $566.8^{+0.8}_{-0.9}$\\
B3.2$\dagger$ & $1100^{+300}_{-1400}$ & $1000^{+2000}_{-1000}$ & $0.09^{+0.13}_{-0.04}$ & 57644.40967384(2) & $0.3^{+1.4}_{-0.2}$ & $0.3^{+0.7}_{-0.1}$ & $564.7^{+1.5}_{-0.4}$\\
B4$\dagger$ & $3100^{+200}_{-600}$ & $550^{+80}_{-110}$ & $2^{+4}_{-2}$ & 57644.410072889(4) & $1.1^{+0.2}_{-0.2}$ & $0.3^{+0.2}_{-0.1}$ & $564^{+1}_{-1}$\\
B5$\dagger$ & $2100^{+900}_{-1600}$ & $2700^{+900}_{-1200}$ & $0.19^{+0.05}_{-0.04}$ & 57644.410157834(4) & $0.7^{+0.3}_{-0.3}$ & $1.0^{+0.4}_{-0.3}$ & $562.1^{+0.7}_{-0.6}$\\
B6.1 & $1393^{+7}_{-7}$ & $183^{+7}_{-7}$ & $0.47^{+0.02}_{-0.03}$ & 57644.411071954(1) & $1.09^{+0.03}_{-0.04}$ & - & $562.3^{+0.2}_{-0.1}$\\
B6.2 & $1417^{+4}_{-5}$ & $102^{+5}_{-4}$ & $0.33^{+0.03}_{-0.02}$ & 57644.4110719755(9) & $0.57^{+0.03}_{-0.02}$ & - & $560.9^{+0.2}_{-0.1}$\\
B7.1$\dagger$ & $3100^{+200}_{-300}$ & $430^{+60}_{-80}$ & $10^{+20}_{-10}$ & 57644.412240214(5) & $0.7^{+0.3}_{-0.4}$ & $0.8^{+0.4}_{-0.4}$ & $569^{+3}_{-3}$\\
B7.2$\dagger$ & $1460^{+20}_{-20}$ & $90^{+30}_{-20}$ & $0.09^{+0.01}_{-0.01}$ & 57644.41224043(2) & $1.9^{+0.9}_{-1.0}$ & $1.2^{+0.6}_{-0.5}$ & $569^{+3}_{-2}$\\
B8$\dagger$ & $3000^{+200}_{-600}$ & $700^{+100}_{-100}$ & $1.1^{+1.5}_{-0.7}$ & 57644.414123628(4) & $1.0^{+0.3}_{-0.3}$ & $0.8^{+0.4}_{-0.3}$ & $567.5^{+0.9}_{-0.7}$\\
B9$*$ & $1430^{+10}_{-10}$ & $75^{+9}_{-9}$ & $0.076^{+0.003}_{-0.003}$ & 57644.41447161(2) & $2.0^{+0.6}_{-0.6}$ & $0.4^{+0.6}_{-0.6}$ & $564^{+2}_{-2}$\\
B10 & $1630^{+10}_{-10}$ & $82^{+8}_{-8}$ & $0.1^{+0.01}_{-0.01}$ & 57644.414475391(7) & $1.4^{+0.3}_{-0.3}$ & $0.5^{+0.3}_{-0.2}$ & $562^{+3}_{-3}$\\
\enddata
\tablecomments{1$\sigma$ errors on the fits are shown on superscript and subscript of each value in the table. For $\mu_t$, the error on the last significant digit is shown in parenthesis.}
\tablenotetext{a}{Burst IDs are chronological. Individual component number (N) for multi-component bursts are appended to the burst IDs. Bursts modeled only using \texttt{curve\_fit} are marked with $*$. Note that the errors on these bursts could be unreliable and may be either under or over-estimated}. Bursts that extend beyond the observable bandwidth can also have unreliable estimates of spectra parameters and fluence (see Section~\ref{sec:fitcaveats}). We mark those bursts with $\dagger$ to indicate that their fluence and spectra parameters could be unconstrained. 
\tablenotetext{b}{$\mu_t$ is the mean of the pulse profile in units of MJD. This can be considered as the arrival time of the pulse. It is referenced to the solar system barycenter, after correcting to infinite frequency using a DM of 560.5\dmunits.}
\tablenotetext{c}{$\tau$ is referred to 1~GHz.}
% \tablenotetext{c}{Bursts which did not require $\tau$ during modeling are }
\end{deluxetable*}

As mentioned previously, we tripled the number of published bursts from these two observations to 133, by detecting 93 new bursts. Dynamic spectra for some of the high-significance bursts are shown in Figure~\ref{fig:bursts}. 

\begin{figure*}
    \centering
    \includegraphics{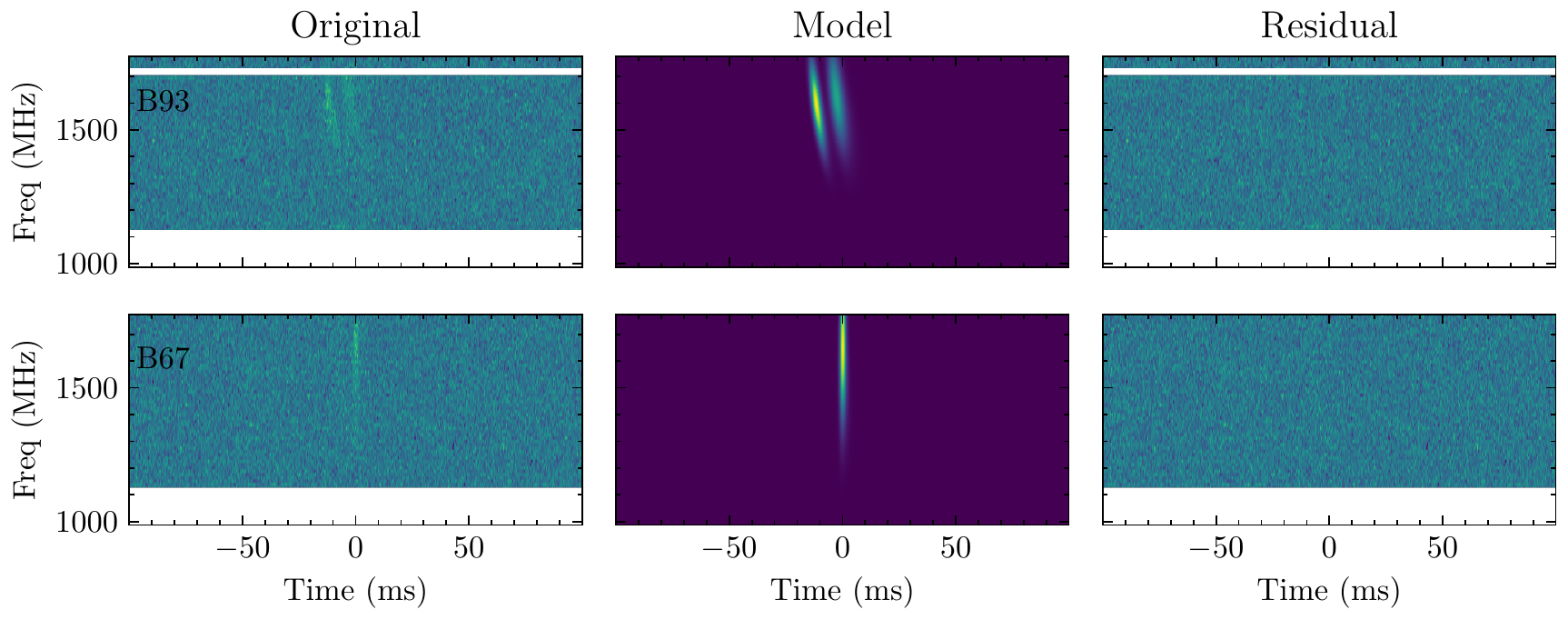}
    \caption{Results of spectro-temporal fits on two bursts (B93 and B67). The first column shows the original (normalized) dynamic spectra of the bursts. The burst can be seen in both cases towards the top of the band. B93 shows two components, separated by around 10~ms. The middle column shows the noise-free model spectrograms that best fit the original data. The last column shows the residual spectrogram obtained by subtracting the model from the original data. The residual spectrogram in both cases shows noise-like data with no remaining artifacts. See Section~\ref{sec:burstfit} for details of the fitting procedure.}
    \label{fig:fit_res}
\end{figure*}

\begin{figure*}
    \centering
    \includegraphics[width=1 \textwidth]{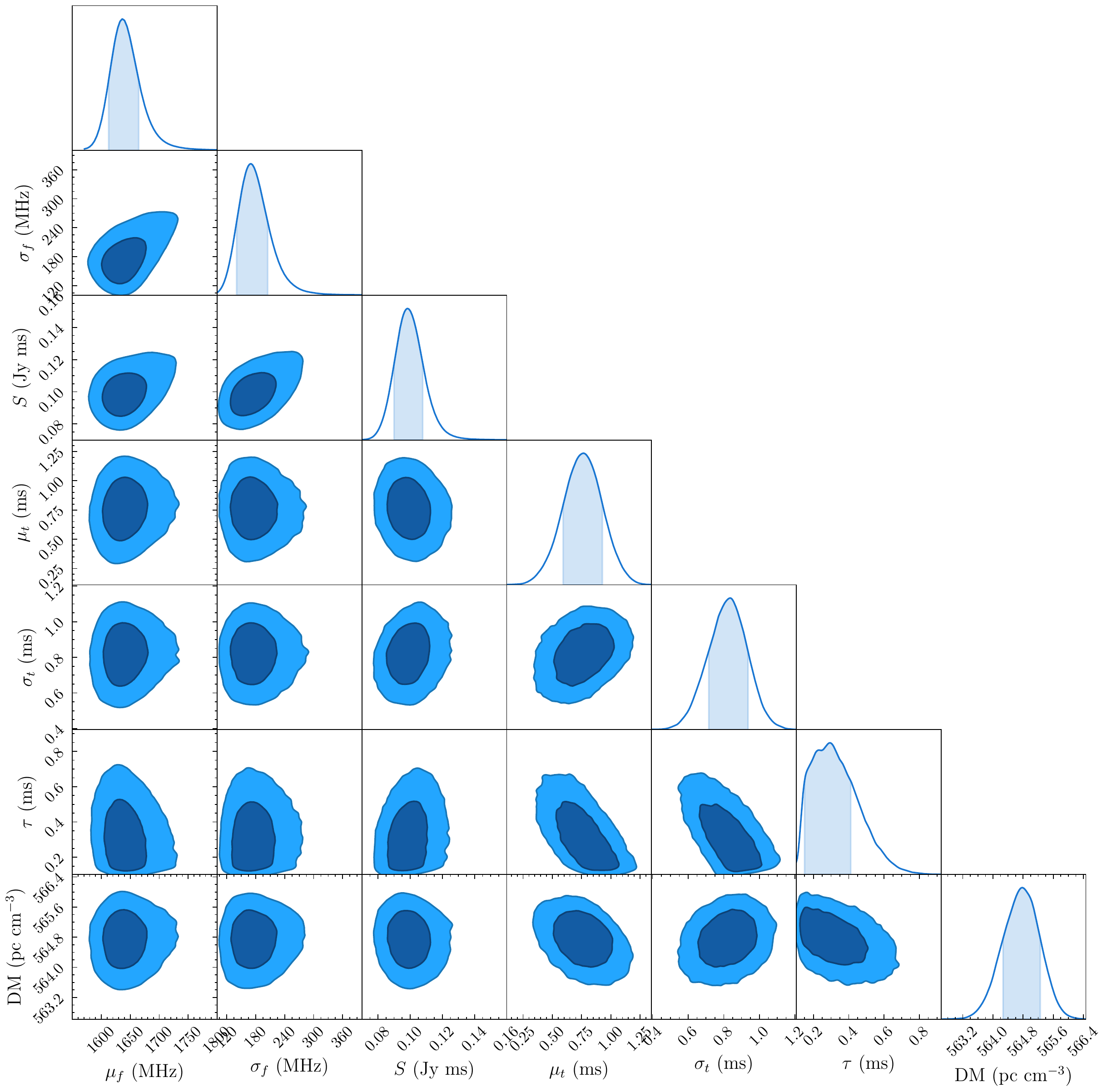}
    \caption{Corner plot generated using the MCMC samples obtained after fitting B67. The 1D histograms show the histogram of individual parameters, while the 2D plots show the correlations between parameters. The shaded and darkened region in 1D and 2D plots correspond to 1~$\sigma$. $\mu_t$ is in units of milliseconds and is with respect to MJD=57645.41732241795. The value of $\mu_t$ is referenced to the observatory site (Arecibo Observatory) and at the top of the frequency band i.e. 1774.53125~MHz. See Sections~\ref{sec:burstfit} and \ref{sec:results} for more details.}
    \label{fig:corner}
\end{figure*}

We used the burst modeling procedure described in the previous sections to estimate spectral and temporal properties for all the bursts. Figure~\ref{fig:fit_res} shows the fitting results for two bursts. Comparing the three columns, we can see that the modeled bursts look similar to the original burst signal and the residuals are noise-like, indicating that the models assumed for the burst spectrogram provide satisfactory fits. Figure~\ref{fig:corner} shows the posterior distribution of the burst properties obtained using MCMC for B67. The 1D plots show the parameter histogram, while the 2D plots show correlations between parameters. The properties of the bursts are given in Table~\ref{tab:small_table}.

\subsection{Burst sample properties}
\begin{figure*}
    \centering
    \includegraphics[width=1 \textwidth]{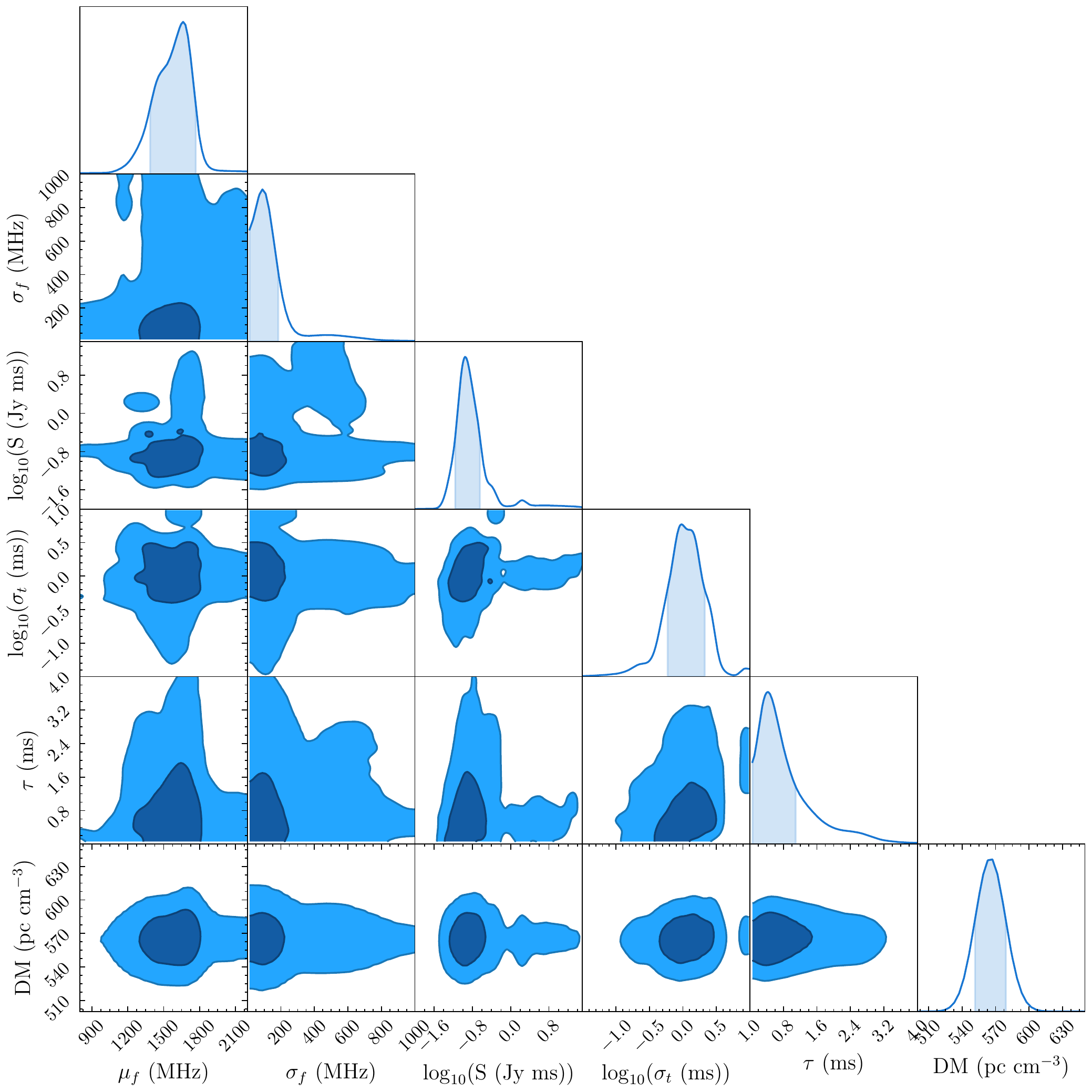}
    \caption{Corner plot generated using samples from the MCMC fit for all the bursts. The 1D histograms show the distribution of individual parameters for all the \src\ bursts, while the 2D plots show the correlations for different parameters. The shaded and darker regions in 1D and 2D plots correspond to $1~\sigma$. See Section~\ref{sec:results} for more details.}
    \label{fig:cumumative_corner}
\end{figure*}

We used the converged sample chains from the MCMC fitting for each burst to generate a cumulative corner plot with the whole burst sample properties. To do this, we randomly selected 1000 samples from the final 25\% of the MCMC chains and then concatenated such samples from all the bursts. We then generated a corner plot using these samples, as shown in Figure~\ref{fig:cumumative_corner}\footnote{Note that this corner plot is different from the one in Figure~\ref{fig:corner}. Figure~\ref{fig:corner} shows the samples from MCMC fit on only one burst, while Figure~\ref{fig:cumumative_corner} shows the cumulative samples obtained from MCMC fit on all the bursts.}. We can now use Figure~\ref{fig:cumumative_corner} to infer trends in various spectro-temporal properties of \src. Table~\ref{tab:burst_prop_cumulative} shows the summary statistics of all the bursts obtained from this analysis. We can see that the spectra of the bursts typically peak around 1650~MHz, and there is a dearth of burst emission below 1300~MHz. Most of the burst spectra peak within the top part of our observing band (i.e 1550--1780~MHz). This behavior possibly extends further to higher frequencies, as is evident from many burst spectra that increase towards the top part of the observing band with their spectral peak possibly lying outside our observing band. Interestingly, \citet{platts2021} have recently reported complex bifurcating structures in some \src\ bursts below 1250~MHz using higher-resolution data. It is, therefore, possible that the emission of \src\ shows a different behavior below these frequencies, which might also vary with time. 

\begin{figure*}
    \centering
    \includegraphics{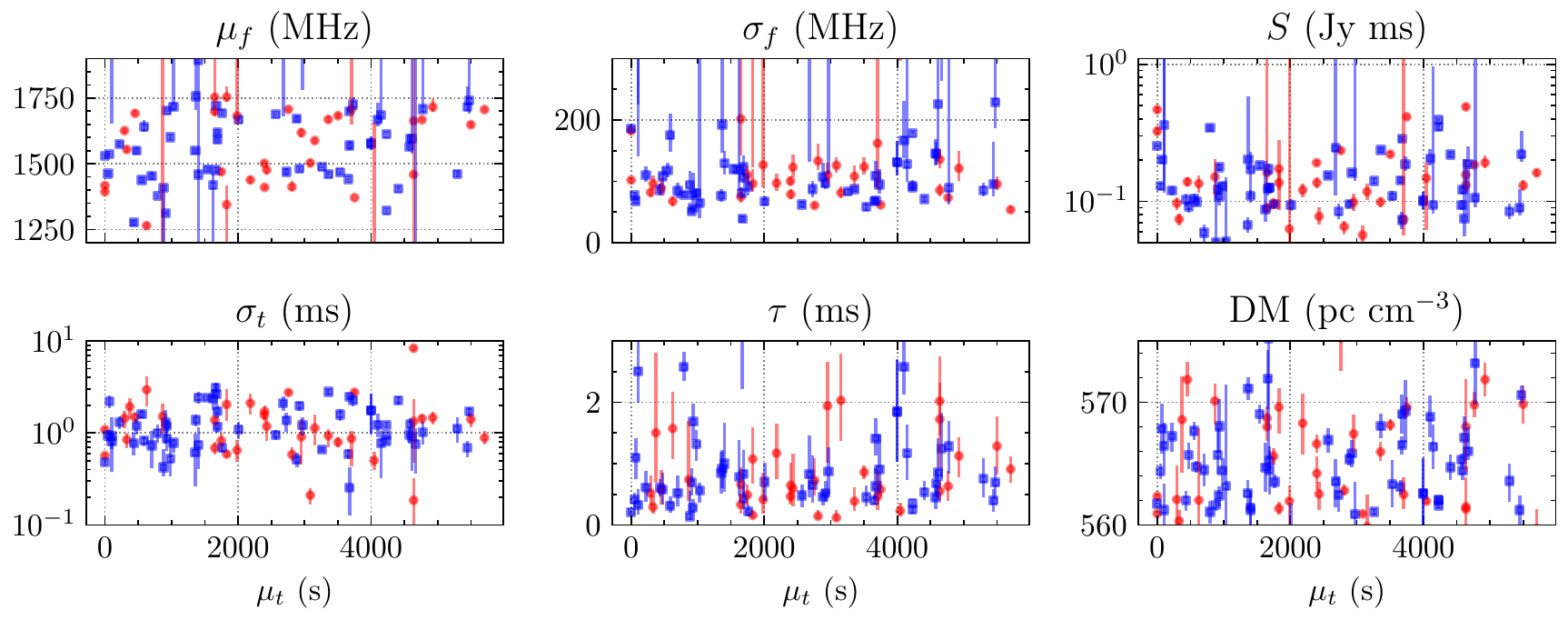}
    \caption{Scatter plots of fitted property of the bursts versus the fitted time of the burst. The data from the two observations are shown in different colors, with MJD~57644 in red and MJD~57645 in blue. $\mu_t$ is referred to the first detected burst in the observation. Only bursts fit using MCMC are shown here. The error bars represent $1~\sigma$ errors on the fit.
    %are shown as vertical and horizontal lines. 
    Errors on $\mu_t$ are very small, and hence are not visible. None of the six properties show any clear trend with respect to the burst time.}
    \label{fig:param_vs_time}
\end{figure*}

As already noted by \citet{Gourdji2019}, \src\ shows a variety of spectral widths. Using our modeling, we observe that most bursts are frequency-modulated and have a typical frequency width of $\sim230$~MHz. The bursts also show a wide range of intrinsic pulse widths, from $0.4$~ms to $20$~ms, and various scattering timescales, up to $3$~ms. The median dispersion measure of the bursts we observe was 564\dmunits, with a 1$\sigma$ variation of $\sim4$ \dmunits. This variation in DM is also apparent in Figures~\ref{fig:cumumative_corner} and \ref{fig:param_vs_time}. This value is consistent with the other published estimates \citep{li2021, platts2021, Cruces2020, Gourdji2019}. We also did not see any strong correlation between any two burst properties from Figure~\ref{fig:cumumative_corner}. Several of the bursts from this sample also show the characteristic sub-burst drift in frequency during the burst duration, sometimes referred to as the ``sad-trombone'' effect \citep{Hessels2019}. We did not detect any evidence of upward drifting as predicted by some FRB models \citep{cordes2017}, and reported by \citet{platts2021}.

Many bursts in our sample show multiple components, and we estimated the properties of these components using our fitting procedure. Nine bursts in our sample show two components, while there is one burst with three components (see Table~\ref{tab:small_table}). We also note that it is difficult to differentiate between multiple closely spaced bursts and different components from single bursts. This is further complicated by the detection of a very wide ($\sim$35~ms) burst reported by \citet{Cruces2020}. As there is no clear consensus on how to resolve this, we visually identified some bursts as components of a nearby burst and reported them as such in Table~\ref{tab:small_table}. We consider all the components as individual bursts for all the following analysis except the cumulative energy distribution analysis.

% \fixme{check the separation between bursts and separation between components, vs. the widest burst in our sample. Cruces et al. had a >35ms burst. }

% \fixme{Drift rates?}

Figure~\ref{fig:param_vs_time} shows the scatter plots of various burst properties with respect to the burst time. The bursts from two observations are shown in different colors, and the time is referenced to the first burst of the respective observation. The burst properties do not show any temporal evolution at the seconds-to-minutes time scale. We also did not observe any distinction between the distribution of properties of bursts detected on two consecutive days. 
% We also compared the distribution of burst properties for the two days using KS test \citep{kstest}, and found no difference in the distributions from these two days.   

\begin{deluxetable}{ccc}
% \tablewidth{0pc}
\label{tab:burst_prop_cumulative}
\tablecaption{Results from the burst sample analysis. The values represent the median values with $1\sigma$ errors.}
\tabletypesize{\footnotesize}
\tablehead{\colhead{Parameter} & \colhead{Units} & \colhead{Value}}
\startdata
\hline 
$\mu_f$ & MHz & $1608^{+100}_{-200}$\\
$\sigma_f$ & MHz & $102^{+130}_{-30}$\\
$S$ & Jy~ms & $0.13^{+0.13}_{-0.05}$\\
$\sigma_t$ & ms  & $1.1^{+0.9}_{-0.5}$ \\
$\tau$ & ms  & $0.7^{+0.9}_{-0.4}$\\
DM & pc~cm$^{-3}$ & $564^{+5}_{-3}$\\ 
% $\mu_f$ (MHz) & $1660^{+110}_{-280}$\\
% $\sigma_f$ (MHz) & $89^{+95}_{-76}$\\
% S (Jy ms) & $0.17^{+0.75}_{-0.00}$\\
% $\sigma_t$ (ms) & $0.84^{+0.92}_{-0.52}$ \\
% $\tau$ (ms) & $0.43^{+0.66}_{-0.37}$\\
% DM (pc cm$^{-3}$) & $567^{+12}_{-16}$\\ 
\hline 
\enddata
\end{deluxetable}

\section{Discussion}

\subsection{Cumulative energy distribution}
\label{sec:energydist}
Energy distributions can provide useful intuition into the emission mechanism of the source. Regular pulsar emission typically shows a log-normal distribution, whereas giant pulses show power-law cumulative distributions \citep{Burke-Spolaor2012}. Crab giant pulses have how evidence that the index depends on pulse width and energy, with flatter indices for weaker and shorter pulses \citep{Karuppuswamy2010, Popov2007,Bera2019}. High-energy magnetar emission has been described by power-law distributions with $\gamma$ ranging from roughly $-1.6$ to  $-1.8$ \citep{Cheng2020}. Previous studies of \frb\ energy distributions have used a single power-law fit ($\mathrm{N}(>\mathrm{E})\propto \mathrm{E}^{\gamma}$) to model the cumulative distribution and have obtained different values of $\gamma$ ranging from $-0.7$ in \citet{Law2017}, $-1.1$ in \citet{Cruces2020}, $-1.7$ in \citet{Oostrum2020} and $-1.8$ in \citet{Gourdji2019}. Another well-studied repeating source, FRB~180916, shows $\gamma=-1.3$ at 400~MHz, although recent observations have reported a flattening of the power-law at lower energies \citep{Chime/FRB2020,Marazuela2020}. 

We calculate the isotropic energy ($E$) of a burst as,
\begin{eqnarray}
    {E} = && 4\pi D_{L}^{2} \times {S}\,\mathrm{(Jy~s)}\times 2.355\,\sigma_f \mathrm{(Hz)}  \nonumber \\
    && \times 10^{-23}  \mathrm{(ergs}^{-1}\mathrm{cm}^{-2}\mathrm{Hz}^{-1}).
    \label{eq:energy}
\end{eqnarray}
Here, $D_{L}$ is the luminosity distance to \frb, 972 Mpc, as reported by \citet{Tendulkar2017}. ${S}$ and $2.355~\sigma_{f}$ are the fitted fluence and FWHM of the Gaussian spectra.

To make the cumulative energy distribution, we choose only bursts for which the $\pm 1\sigma$ bounds on the spectral peak fell within our observing band. This was done as our fluence estimates obtained from fitting are reliable for bursts within our band and because we are incomplete to the population of bursts that are partially outside our band \citep{banded_repeaters}. Therefore, from a total of 133 bursts, we obtained 60 bursts that satisfied this criteria. For each of the 60 such bursts, we used the posterior distribution of bandwidths and fluences from the MCMC based fitting analysis to calculate the distribution of energies (using Eq.~\ref{eq:energy}). %We boostrap the cumulative energy distribution by randomly drawing one energy from each burst's distribution. 
We then randomly sample one energy from the burst energy distributions of each of the 60 bursts and generate a cumulative energy distribution using those 60 energies. We repeated this process 1000 times and thereby generated 1000 cumulative energy distributions.

The previous studies of the cumulative energy distribution of \frb~have reported a break in power-law with a flattening towards low energies \citep{Gourdji2019, Cruces2020}. Also, it was visually evident in our data that the cumulative distribution flattened towards low energies. Therefore, a single power-law would not have been sufficient to accommodate the burst energy distribution. Therefore, we used \texttt{scipy.curve\_fit} to fit each of these 1000 energy distributions with a broken power-law of the form,

\begin{equation}
        N(\geq{E})= 
    \begin{cases}
         {E_{\rm scale}}\left(\frac{{E}}{{E_{\rm break}}}\right)^{\alpha},& \text{if } {E} < {E_{\rm break}} \\
        {E_{\rm scale}}\left(\frac{{E}}{{E_{\rm break}}}\right)^{\beta}, & \text{if } {E} > {E_{\rm break}}.
    \end{cases}
\end{equation}

Here, $\alpha$ and $\beta$ are the two power-law indices, $E_{\rm break}$ is the break energy, and $E_{\rm scale}$ is the energy scaling. Figure~\ref{fig:energydist} shows the cumulative energy distributions above an estimated completeness of $5.8\times10^{36}$~ergs, calculated from the aforementioned completeness limit on the fluence, 0.0216~Jy~ms (Section~\ref{sec:completeness}), and median bandwidth ($2.355\,\sigma_f$) of the bursts i.e. 240~MHz. It also shows (in red) the power-law fit to each cumulative energy distribution. 

The median of the distribution of fitted power-law indices and break energy (with 1$\sigma$ errors) are given by $\alpha=-0.4\pm0.1$, $\beta=-1.8\pm0.2$ and ${E_{\rm break}}=(2.28\pm0.19)\times 10^{37} \mathrm{ergs}$. This break at ${E_{\rm break}}$ could indicate the actual completeness energy limit of our observations, and we might therefore be incomplete to the bursts with energies below ${E_{\rm break}}$. This could be due to the incompleteness of our observations to the weak, band-limited bursts. The bursts above this energy are well fitted by the power-law of index $\beta=-1.8\pm0.2$.  %This argument was also used by \citet{Gourdji2019} to estimate the completeness limit. 
The break in energy distribution has also been reported for other repeating FRBs \citep{Marazuela2020}. It is worth noting that our higher-energy power-law index $\beta$ is also consistent with the power-law index estimated by \citet{Gourdji2019} above a completeness threshold of $2\times 10^{37}$~ergs. In the context of pulse-energy distributions, the similarity of the power-law indices of \frb\ with both those of Crab giant pulses and magnetar pulses might also imply a common origin \citep{Lyu2021}. 

\subsubsection{Testing for a high-energy break}
\begin{figure*}
    \centering
    \includegraphics{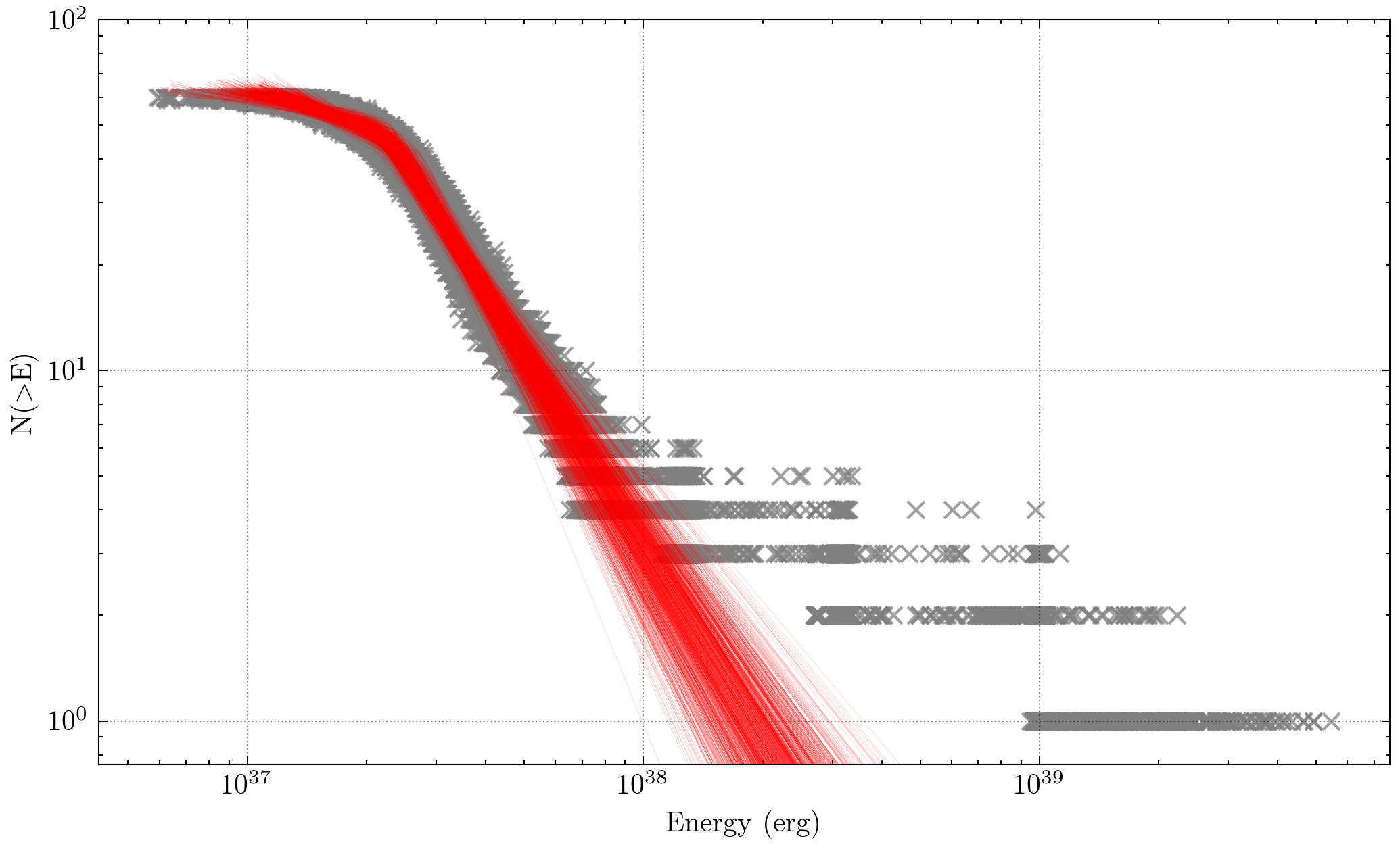}
    \caption{Cumulative  energy distribution of bursts above our completeness limit of $5.81\times10^{36}$~ergs. The grey points show the cumulative energy distributions created by bootstrapping the burst energies. The red lines show the broken power-law fit to each of these cumulative energy distributions. See Section~\ref{sec:energydist} for more details.}
    \label{fig:energydist}
\end{figure*}

Figure~\ref{fig:energydist} shows that two high-energy bursts deviate from the power-law fits. This has also been seen in Crab giant pulses, where this behavior was speculated to be due to supergiant pulses \citep{Mickaliger2012}. We therefore tested the presence of a high-energy break in the power-law (between $5-9\times 10^{37}$~ergs). We assumed that the break energy estimated above to be the completeness threshold, and only used the energies greater than that value. Then we repeated the bootstrapping method to fit the cumulative distribution of the remaining bursts using: a single power-law and a broken power-law. Note that in this test we were only fitting the bursts with energy greater than $2.3\times 10^{37} \mathrm{ergs}$.

% median of the distribution of fitted single power-law indices and break energy (with 1$\sigma$ errors) are given by $\alpha=-0.4\pm0.1$, $\beta=-1.8\pm0.2$ and ${E_{\rm break}}=(2.28\pm0.19)\times 10^{37} \mathrm{ergs}$. 

The fitted slope obtained for the single power-law fit was $-1.8^{+0.1}_{-0.2}$. The fitted slopes (below and above the break energy) for the double power-law fit were $-1.8^{+0.2}_{-0.2}$ and $-0.5^{+0.1}_{-0.3}$ with fitted break energy of $7.8^{+1.5}_{-1.9}\times 10^{37} \mathrm{ergs}$. The power-law slope obtained in the single power-law fit (and the lower energy slope in case of double power-law fit) was consistent with the higher energy slope reported earlier ($\beta=-1.8\pm0.2$). We also found that the reduced chi-square value for the single power-law case was $1.2^{+2.0}_{-0.6}$, while that for the double power-law fit was $0.08^{+0.08}_{-0.03}$. This indicates that the double power-law fit model over-fitted the data, and so a single power-law is sufficient. This test gives further confidence that above the energy of $2.3\times 10^{37} \mathrm{ergs}$, bursts from \src\ follow a single power-law with slope $\beta=-1.8\pm0.2$, and that there is no evidence for a higher energy break in the energy distribution.

%\fixme{two lines on implications}
% \maura{MAM: Not sure what else one can say about implications, really. Did you look for power-law index dependence on width?}
% KA: I divided the sample into two parts: narrow and wide widths. The power-laws in both cases got steeper than the complete sample power-law. But narrow pulses didn't give a flatter power-law, as is seen for Crab. 

\subsection{Wait-time distribution}
\label{subsec:waittime}
\begin{figure*}
    \centering    
    \includegraphics{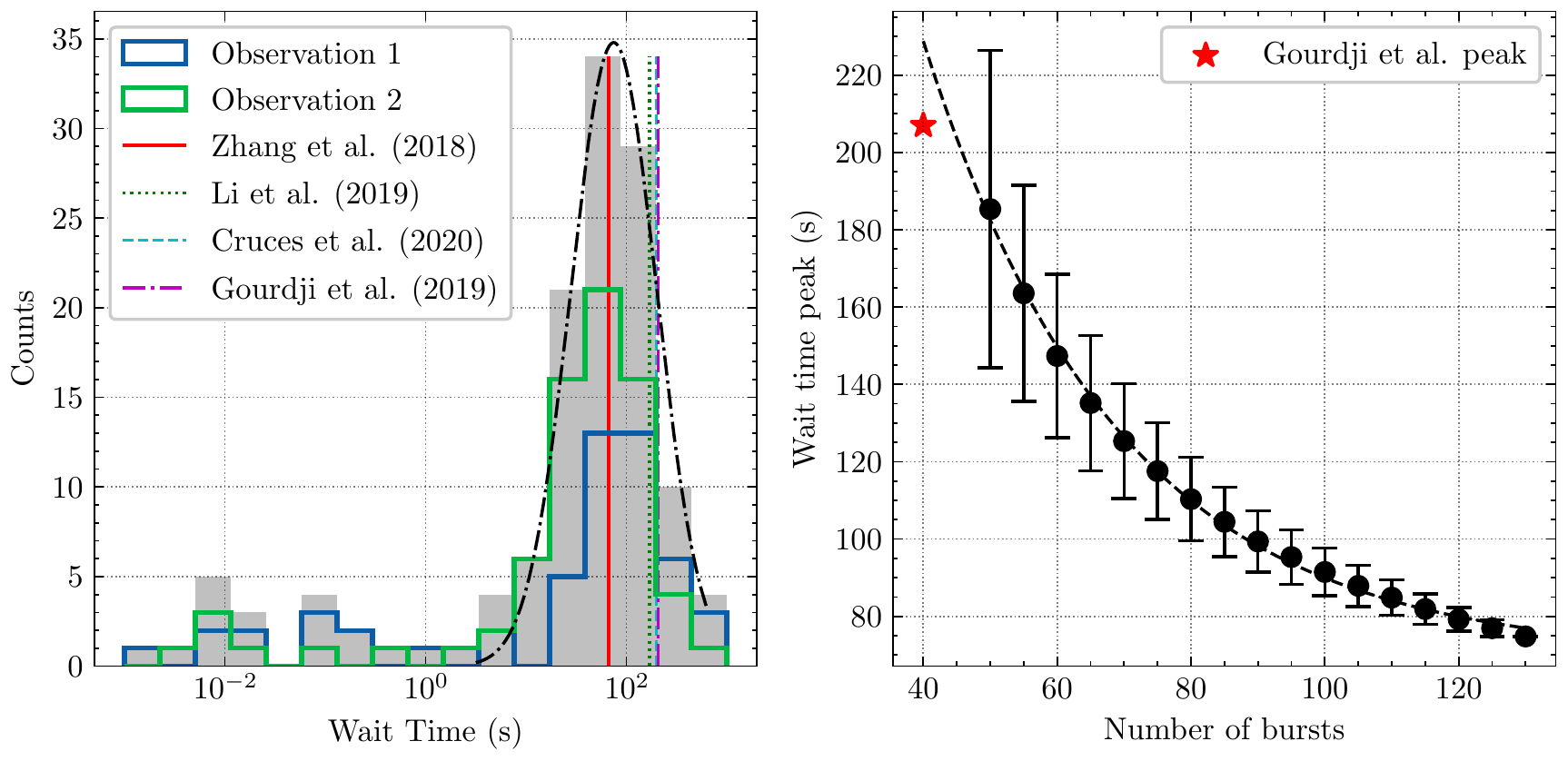} %{wt_final.png}
    \caption{Left: the wait time distribution for both observations is plotted in gray, with the fitted log-normal function peaking at $74.8 \pm 0.1$~s. The individual observations are plotted in green and blue, respectively. The fitted wait-time peaks from previous works are notated with vertical lines. Right: the average fitted wait-time peaks as a function of the number of bursts randomly selected from the full sample. The fitted exponential distribution is shown with a black dashed line, and the error bars represent one standard deviation of the fitted peaks. The fitted peak from \citet{Gourdji2019} is represented with a red star.}
    \label{fig:waittime}
\end{figure*}

The left panel in Figure~\ref{fig:waittime} shows the distribution of wait times between bursts, which follows a bi-modal distribution as also seen in previous studies \citep{Li2019, Gourdji2019}. On wait times greater than 1~s, we use \texttt{scipy.curve\_fit} to fit a log-normal function to the main distribution, finding a peak at $74.8 \pm 0.1$~s. The peak of our wait-time distribution is significantly lower than the $207 \pm  1$~s peak found in this data by \citet{Gourdji2019}, due to our increased sample of bursts filling in the wait-time gaps. As more bursts are discovered in an observation of constant length, the average time between bursts decreases, lowering the peak of the wait time distribution. Our findings most closely match the wait-time distribution of \citet{Zhang2018}, which peaks at $\sim 67$~s; while the original paper did not report an exact peak, we used their publicly available data\footnote{Table 2, accessible at \url{https://doi.org/10.3847/1538-4357/aadf31}} to perform our wait-time analysis. %and calculate a peak. 
These similarities suggest that careful single-pulse searches using machine-learning algorithms allow us to obtain a robust sample of bursts that accurately reflects the burst population.

As in previous studies \citep[e.g.,][]{Li2019, Katz2018, Gourdji2019}, we find a smaller population of bursts with sub-second separations. However, unlike the previously reported distributions, which cluster around tens of milliseconds, our sub-second burst separations span the range of tens of milliseconds up to nearly one second without as clear of a break between the two distributions. For this analysis, we assume that each closely spaced pair of bursts is composed of two separate bursts instead of components of a single broader burst, leading to the larger distribution of short wait times compared to other papers. This assumption will not drastically alter the fitted log-normal distribution since the sub-second wait-time population is small, and their removal will not significantly alter the shape of the main distribution. As in \citet{Gourdji2019} and \citet{Li2019}, we find that the wait time between bursts and their relative fluences are not correlated. 

To quantify the change in the wait-time peak as a function of the number of bursts in a sample, we perform the same wait-time analysis for a random selection of our bursts over a range of burst numbers, as seen in the right-hand panel of Figure~\ref{fig:waittime}. Each point represents a fit to 300 random selections of that number of bursts, with the error bars representing the standard deviation of all of the fitted peaks. Our findings show the expected effect that as more bursts are included in a sample from a constant-length observation, the average time between bursts will decrease along with the fitted wait-time distribution peak. We find that the distribution of fitted wait times peaks exponentially decays with added bursts with a timescale of $\sim29$ seconds. We observe the same effect in the \citet{Zhang2018} dataset, which shows a timescale of $\sim25.5$ seconds. The peak obtained by \citet{Gourdji2019} using this data is shown in the figure and matches with our fitted exponential curve. We perform a similar analysis by filtering out the lowest fluence bursts from our sample and find that the wait time peak increases as the minimum fluence limit is increased, and weaker bursts are excluded. This serves to explain the higher wait-time peaks calculated by previous papers with fewer bursts and higher fluence limits.

% \fixme{some notable points?}
% \begin{enumerate}
%     \item our number matches with zhang et al that was at a diff freq, both can be considered complete now?
%     \item what else?
% \end{enumerate}

\subsection{Short-period periodicity search}
In this section we discuss the results of various periodicity searches tested on this burst sample to search for a short-period periodicity (millisecond to hundreds of seconds). 

\subsubsection{Difference Search}
We first perform a periodicity search on the burst times ($\mu_t$ in Table~\ref{tab:small_table}) by calculating the differences between consecutive pulse times and searching over a range of trial periods to determine how many differences are evenly divisible, within some tolerance, by this period. We searched over trial periods starting with the minimum difference between pulses, in integer divisors down to the minimum difference divided by 256, after removing all differences less than 50~ms, in both the full set of pulses and 100 trials where only three-quarters of the bursts were randomly selected (to sample a more complete range of minimum differences). Differences less than 50 ms were removed, since potential single bursts with widths greater than 30 ms have been reported from \frb\ \citep[e.g.][]{Katz2018}, and the distinction between whether these bursts are single multi-component bursts or separate bursts is unclear. Furthermore, shorter trial periods are much more likely to return false positive results. To allow for a variety of possible emission mechanisms for \src, including a broad or multi-component pulse profile, we searched phase tolerances ranging from 1\% to 50\%. At any given phase tolerance, a trial period is considered to fit a difference between two pulses if the difference is an integer multiple of the trial period, within an error equal to the phase tolerance.

We also searched 1000 simulated time series of identical length, with the same number of pulses distributed randomly, using the same methodology in order to gauge the significance of any detected periodicities. By searching for periods in a set of bursts with no underlying period, we can evaluate whether our period search finds a real periodicity in the data, or if it is a coincidence.  Above a 50\% tolerance, we get many more pulse matches in all of the random timeseries than we do with the real data, likely due to the FRB pulse distribution not following a random distribution (see Figure~\ref{fig:waittime} for the distribution of pulse arrival time differences).

%The number of pulses fitting any of the detected periodicities in the real data searches did not exceed 95\% of the values in the  simulations for any tolerance value. 
The most significant period found was 658.838 milliseconds, which fit nine pulses at a tolerance of 3\%, with a false alarm probability (FAP) of 0.3\% for random trials at that tolerance. However, considering all 50 tolerance values searched over two observations, the effective FAP is 30\%, and we therefore conclude that no periodicity can be detected through this differencing method.

\subsubsection{Fast Folding Algorithm}
We ran a fast folding algorithm (FFA) on each observation using \texttt{riptide} \citep{morello2020}. Unlike the periodicity search in the previous section which uses the calculated pulse arrival times, \texttt{riptide} searches for periodic signals in the entire dedispersed time series. This allows us to efficiently search over a greater range of trial periods, and will not be affected by issues such as missing bursts, or considering closely spaced individual bursts as a single multi-component burst. However, while it is more sensitive to weak, time-averaged periodic emission, it is less sensitive to periodicities only found in the detected single pulses. The FFA folds each dedispersed time series over a range of trial periods to create an integrated pulse profile. For each observation, we searched time series with DMs ranging from 550~pc~cm$^{-3}$ to 580~pc~cm$^{-3}$ and at periods greater than 500~ms \citep[the approximate period at which folding algorithms are more sensitive than Fourier techniques; see][]{ParentFFA} and less than 20~s (to ensure a sufficient number of pulses across the observation for a pulsed detection).
%We searched for  periods ranging from 25--500~ms using 256 pulse phase bins, 
%We searched periods from 0.5--20~s using 4096 phase bins, and  periods from 20--500~s using 8192 phase bins. 
We used 1024 output bins, with boxcar filters providing sensitivity to pulses with widths ranging from 1 to 300~ms.
%\maura{MAM: I think you should also include results with a smaller number of bins. It doesn't make sense to me to just use one, since (similar to searching different tolerances) we do not know the width of the composite pulse.}
%\maura{MAM: Why are these two ranges called out separately if you are using the same number of pulse phase bins? Also I think you should also search to much larger numbers of bins. For a period of 20~s and 1-ms duration pulses, assuming they are in phase, 20,000 pulse phase bins are required to optimally detect the signal. Even if you assume that the composite profile width is 10 times greater than that of single pulses, you still require more bins. Even for a 1-s period, you require 1000 bins if you assume that the 1-ms pulses are in phase.} %%EL: Phase bins fixed.

First, candidates due to RFI, such as periods at exact integers and known RFI frequencies, were removed. Of the remaining candidates, a signal-to-noise cutoff of $10\sigma$ was applied for a total of 1,250 candidate periods between the two observations.
%\maura{MAM: I think it would be better to use some significance cutoff, like S/N $>$ 8 or similar.}
%\maura{MAM: A cutoff of 30$\sigma$ seems WAY too high. I think this is because you are not sifting out 0-DM candidates at this stage.}

We then folded the relevant dedispersed time series  for each candidate using the \texttt{prepfold} command from the \texttt{PRESTO} \citep{presto} package at the candidate period and DM identified by \texttt{riptide}. We allowed \texttt{prepfold} to search in DM in order to determine which candidates had signal-to-noise ratios which peaked for DMs inside of the searched range. We expanded our acceptable DM range from 520 to 610 pc cm$^{-3}$ to allow leeway in prepfold's search, leaving 91 periodic candidates in total.
%We visually examined the profiles for each candidate, folded with the \texttt{prepfold} command from the \texttt{PRESTO} \citep{presto} package at the candidate period and DM identified by \texttt{riptide}. \maura{MAM: How many bins were used? Same as in the FFA search? If you use prepfold's default bin number, it's very unlikely you will see the same signals.} 
% Of the XX original candidates, XX had signal-to-noise ratios which actually peaked in the 550~pc~cm$^{-3}$ to 580~pc~cm$^{-3}$ range. 
We visually inspected these profiles and found that all were consistent with RFI or noise, with no evidence of  %Of these, most  were the result of RFI, with no evidence of 
 emission at the same phase over time or frequency. %\maura{MAM: We know that the FRB emission won't be consistent with time, so I'd rephrase this.} For most of the candidate periods, the reduced chi-squared statistic of the detection peaked at a DM of zero rather than that of the FRB, further indicating that these candidates are the result of \maura{RFI}. 
 We compared the candidate periods found by the algorithm on both observations, and found that the only common periods 
 %above our cutoff of $30\sigma$ 
 were caused by RFI. 

\subsubsection{\texttt{frbpa}}
We used \texttt{frbpa} \citep{Aggarwal2020} to search for a short-period periodicity using the burst times (MJDs). We used two methods: search to find the period that minimizes the fractional width of folded profile \citep{Rajwade121102} and a Quadratic-Mutual-Information-based periodicity search technique \citep{Huijse2018}. In the first method, we phase-coherently folded  the burst times between trial periods and generated a set of profiles consisting of the source activity with respect to the trial period phase. We then measured the width of the source activity in each folded profile. Low width would indicate that source activity is concentrated in a small set of contiguous phase windows, indicating the presence of periodic activity \citep{Rajwade121102}. The second method uses quasi-mutual-information to estimate the period. It has been shown to be robust to noise and works well on sparsely sampled data as well \citep{Huijse2018}. We searched for periods between 1--1000~seconds on bursts from the two days individually and did not recover any significant period.
% Less significant detections were observed at periods of 7.53~s and its harmonics.
% \maura{MAM: In both observations? Is this significant? What does the folded profile look like?}

\subsubsection{Lomb-Scargle}
We also used \texttt{timeseries.LombScargle} from the \texttt{astropy} library \citep{astropy:2013, astropy:2018} to search for periods ranging from 100~ms to 1000~seconds on the bursts from the two observations separately. Note that sensitivity to 100-ms periods requires sampling frequencies higher than the traditional Nyquist limit. 
%In order to reach the magnitude of 100~ms seconds, we must adjust the frequency limit which is automatically set at the average Nyquist frequency by the Astropy software. 
This is possible because
the effective Nyquist frequency for  unevenly sampled data set is  much smaller than the traditional limit  \citep{VanderPlas_2018}.
%and therefore the effective Nyquist frequency for our unevenly sampled data set is actually much smaller than the average Nyquist frequency determined by Astropy.
The most significant periods are approximately 118
%144  
and 179
%328 
milliseconds. %\maura{MAM: Please quote errors on these!}
However,  false alarm probabilities  of around 3\% and 26\%, respectively, indicate that the detected periods are unlikely to be real. We therefore we conclude that we did not detect any significant periodicity in the bursts using the Lomb-Scargle method.

\subsection{Burst Rate}
Previous observations of \src\, have found significant evidence for pulse clustering on short time scales, where the burst separations deviate from a Poissonian distribution \citep{Oppermann2018, Oostrum2020, Cruces2020}. The Weibull distribution, as described in \citet{Oppermann2018}, is a modification of the Poisson distribution, with a shape parameter $k$ describing the degree of pulse clustering. Clustering is present for $k<1$  with lower values corresponding to more clustering, while $k =1$ reduces to the Poissonian case; a value of $k \gg1$  causes the distribution to peak more sharply at the event rate and indicates a periodic signal. A better understanding of the burst statistics may help us understand the progenitor of \src\ and help strategize the timing of future observations.

\begin{figure*}
    \centering
    \includegraphics{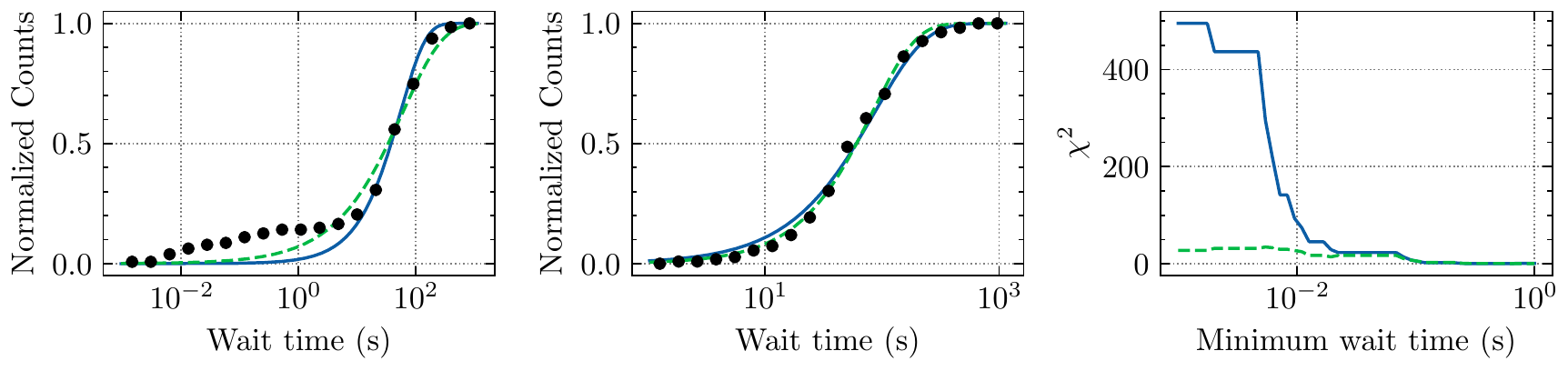} %{wait_times.png}
    \caption{The cumulative wait time distributions fitted by the Poisson (blue solid lines) and Weibull (green dashed lines) distributions for the entire sample of wait times (left) and for the population of wait times greater than one second (middle).
    The fitted values for the Poisson distribution for the left and middle plots are $r=65$~hr$^{-1}$ and $r=41$~hr$^{-1}$ respectively. The fitted values of the Weibull distribution are $k=0.634$, $r=42$~hr$^{-1}$ and $k=1.162$, $r=46$~hr$^{-1}$ respectively. Right: The reduced chi-squared statistic of the Poisson and Weibull fits, when fitted to wait times greater than the minimum wait time. While the middle and right panels show that both the Weibull and Poisson distributions fit the longer wait times well, the left panel shows that neither distribution accurately fits for the larger sample of sub-second wait times. The right panel shows that pulse clustering starts to impact the fits once wait times shorter than $\sim 50$~ms are considered.}
    \label{fig:rate}
\end{figure*}

Figure~\ref{fig:rate} shows the cumulative probability density of the wait times between consecutive bursts, fitted to the Weibull and Poisson cumulative density functions as defined by \citet{Oppermann2018}. The fitted values are given in Table~\ref{tab:burstrate}, as well as the reduced chi-squared statistic and the coefficient of determination, $r^2$, which ranges from zero to one with a value of one representing a perfect fit. In Figure~\ref{fig:rate}, we also plot the values of the reduced chi-squared statistic for the Poisson and Weibull distributions when fitted to only the wait times longer than a range of chosen minimum wait times. We observe that both the Poisson and Weibull distributions fit the main population of longer wait times much better than the entire set of bursts. The Weibull distribution's ability to account for clustering allows it to have a significantly better fit when including shorter wait times, as its reduced chi-squared statistic is a factor of 18 smaller than that of the corresponding Poisson distribution (see Table~\ref{tab:burstrate}). However, the Weibull distribution is only slightly favored over the Poisson distribution when fitting only longer wait times. We find that the reduced chi-squared statistic is equal to 0.368 for the Weibull fit to wait times greater than one second, indicating that the distribution is overfitted. These findings may indicate that the main distribution of bursts with longer wait times roughly follows a Poissonian distribution, while the entire burst rate distribution cannot be accurately described with solely a Poisson or Weibull distribution. This may result from our decision to consider each burst as a separate burst rather than a sub-component of a broad burst.

\begin{figure*}
    \centering
    \includegraphics{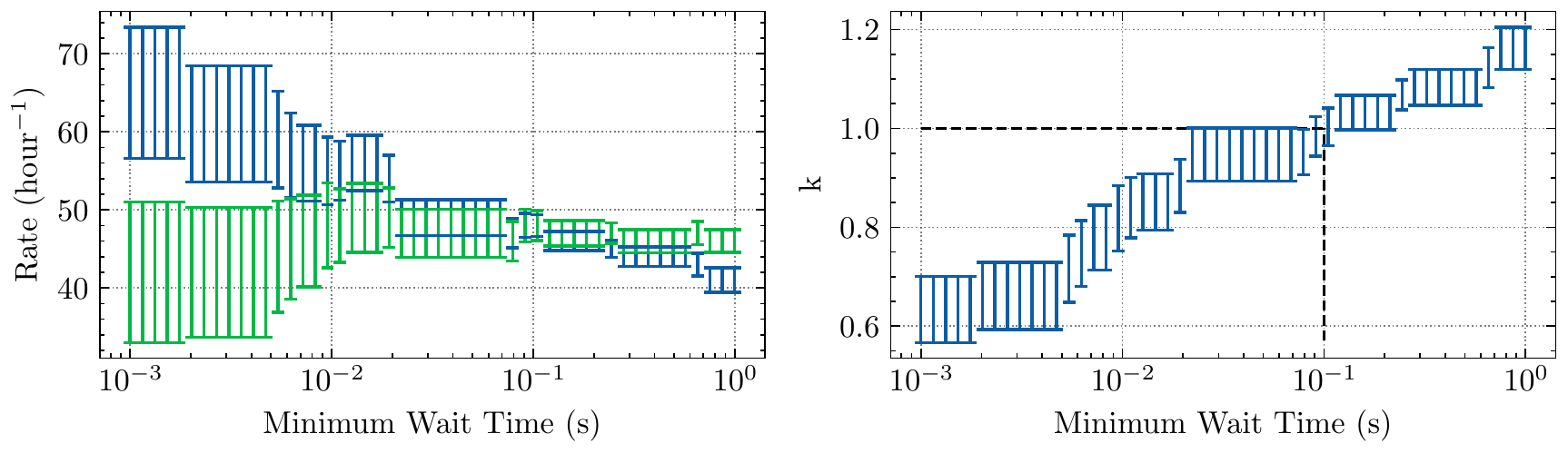} %{wait_times.png}
    \caption{Left: the fitted burst rate per hour for the Poisson (blue) and Weibull (green) distributions as a function of the minimum wait time used in the fitting. As the minimum wait time increases and we begin to sample only the main distribution of wait times, the burst rates converge to roughly 45 bursts per hour. Right: The Weibull shape parameter $k$ as a function of minimum wait time. The fitted value increases with the minimum wait time, and black dotted lines have been added to indicate that the Weibull parameter increases to a value of 1.0 at a minimum wait time of 0.1 seconds.}
    \label{fig:rate_k_vs_waittime}
\end{figure*}

In addition to their observations, \citet{Cruces2020} used the original dataset from \citet{Gourdji2019} to study the burst rate statistics and found that the addition or removal of the sub-second wait time population in each dataset significantly impacts the extrapolated burst rate behavior. In each case, removing these short wait times led to the fitted Weibull shape parameter $k$ increasing towards one, further indicating that the main distribution of pulses may follow a Poissonian distribution while the shorter distributions do not. However, the sample used in \citet{Cruces2020} only had two sub-second wait times; our more extensive sample of short wait times allows us to confirm this behavior with more statistical significance. In Figure~\ref{fig:rate_k_vs_waittime}, we plot the fitted burst rates for the Weibull and Poisson distributions and the Weibull shape parameter $k$ as a function of minimum wait time. As the minimum wait time increases, both burst rates converge to a rate of roughly 45 bursts per hour. We also find that the fitted value of $k$ increases with the minimum wait time, reaching a value of $k=1$ at a minimum wait time cutoff of roughly 0.1~s.
% As mentioned in Section~\ref{subsec:waittime}, some of these burst pairs may actually be sub-components of a single broad burst. 

\begin{deluxetable}{ccccc}[h]
%\tablewidth{0pc}
\label{tab:burstrate}
\tablecaption{Fitted burst rate distributions}
\tabletypesize{\footnotesize}
\tablehead{\colhead{} & \colhead{Rate (hour$^{-1}$)} & \colhead{$k$} & \colhead{$\chi^2$} & \colhead{$r^2$}}
\startdata
\hline 
Poisson (all) & $65 \pm 8.4$ & \nodata & 495 & 0.953 \\
Weibull (all) & $42 \pm 9$ & $0.63 \pm 0.07$ & 27.5 & 0.970 \\
Poisson ($\delta$t $ > 1$s) & $41 \pm 1.6$ & \nodata & 1.076 & 0.994 \\
Weibull ($\delta$t $ > 1$s) & $46 \pm 1.5$ & $1.16 \pm 0.04$ & 0.368 & 0.997 \\
\hline 
\enddata
\tablecomments{The posterior values for the Poisson and Weibull distributions as well as the reduced chi-squared statistic and $r^2$ value, fit both to the entire set of wait times as well as only wait times greater than one second. The errors represent $1~\sigma$ uncertainties.}
\end{deluxetable} 

\subsection{Implications for progenitor models}
Based on our results, any progenitor model proposed for \src\ would have to explain the following observations: (1) band-limited emission; (2) varying peak emission of the spectra and its lack below 1300MHz; (3) median scattering timescale of 0.7~ms, with a maximum value of around 2~ms; (4) rapid variability of these three properties at second timescales. Further, some of these observations have also been reported for other FRBs \citep{Kumar2021, Shannon2018, Marazuela2020}.

\subsection{Comparison to previous work}
\label{sec:comparison}
\begin{figure*}
    \centering
    \includegraphics{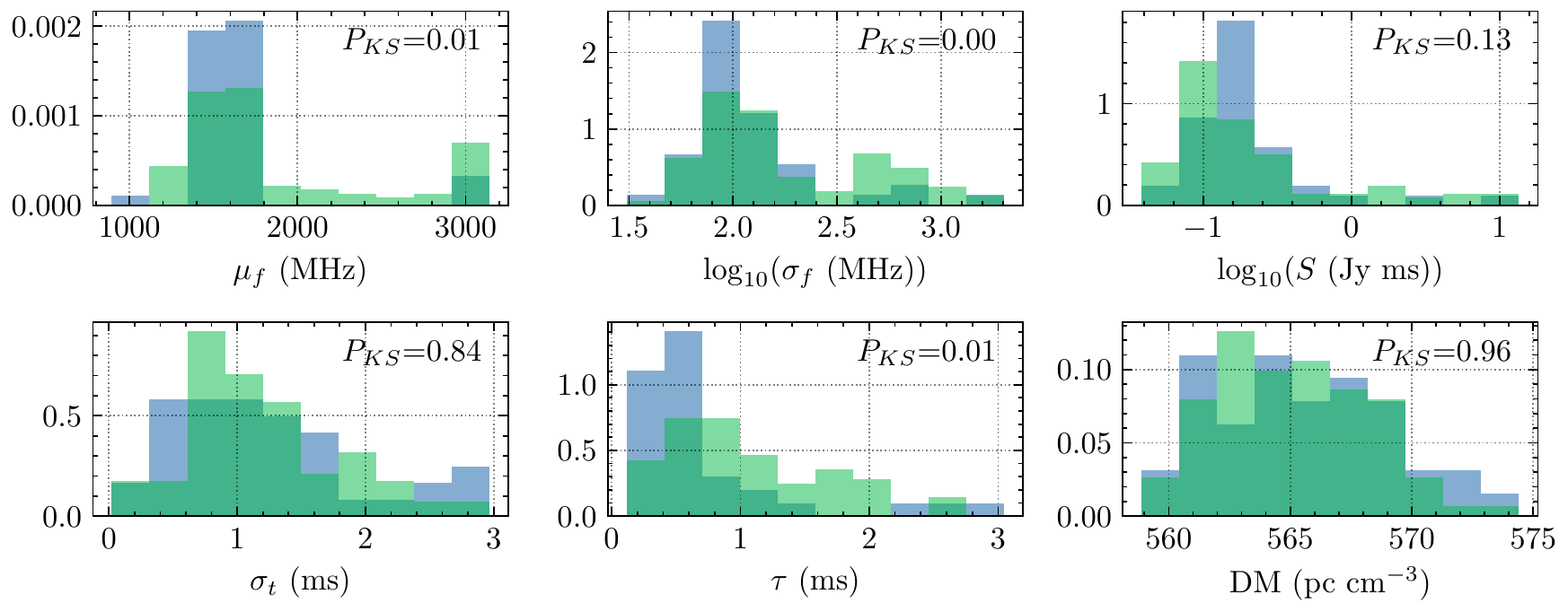} 
    \caption{Normalized histogram of burst properties. New bursts detected in this analysis are shown in green, while the ones published by \cite{Gourdji2019} are shown in blue. For all the bursts, the properties shown here were estimated using the fitting procedure described in Section~\ref{sec:burstfit}. We did a KS-test to compare the old and new burst distributions, for all the properties, with p-values obtained from these tests  given on the top right of each plot.}
    \label{fig:old_vs_new_params}
\end{figure*}

In this work, we have presented 93 additional bursts detected on reprocessing the data presented in \citet{Gourdji2019}. Figure~\ref{fig:old_vs_new_params} shows the distribution of properties of new bursts detected with our pipeline as compared to the ones already published by \citet{Gourdji2019}. We performed Kolmogorov-Smirnov (KS) tests to compare the distributions of fitted parameters of old versus the new bursts. The distribution of $S$, $\sigma_t$ and DM are similar, while those for $\mu_f$, $\sigma_f$ and $\tau$ are different. This indicates that the searches carried out by \citet{Gourdji2019} missed bursts that span the entire range of these parameter values, rather than just the weaker bursts.

Recently, \citet{li2021} reported a large sample of bursts from \src\ detected using the Five-hundred-meter Aperture Spherical radio Telescope. The mean of the wait-time distribution estimated from that larger sample of bursts ($70\pm12$s) is consistent with what we report in Section~\ref{subsec:waittime}. They also did not detect any short-term periodicity, similar to our findings. Further, they report a bimodal energy distribution for \src. \cite{banded_repeaters} highlighted that this bimodality disappears when burst bandwidth, instead of center frequency of the observing band, is used to calculate the energy. Moreover, the burst bandwidths reported in \citet{li2021} were estimated visually, and not using a fitting procedure. This can lead to observational biases that will make the interpretation of intrinsic energy distributions difficult \citep{banded_repeaters}. %With these caveats in mind, we used KS-test to compare the distributions of energies reported by \citet{li2021} and the ones reported in this work, and found that they are consistent with a p-value of XX.}

All the tools and software used in our pipeline are independent of the ones used in the original work \citep{Gourdji2019}. This brings the critical question of understanding why our pipeline detected more bursts or why the original work missed the bursts. While an exhaustive comparison of the two pipelines using a standard dataset consisting of simulated FRBs is beyond the scope of this work, here, we try to investigate the reasons for different results based on our understanding of the software used. We discuss three reasons that might contribute to very different recovery rates\footnote{By recovery rate, we mean the percentage of transients (above the completeness limit) correctly identified by the software. A perfect pipeline would have a recovery rate of 100\%, indicating that it can detect all transients present in the data.} of the two pipelines.

\subsubsection{Threshold signal-to-noise ratio}
Although \citet{Gourdji2019} used a S/N threshold of 6 for the search, they discarded candidate groups with a maximum S/N less than 8. We used a S/N threshold of 6 in our search. Therefore, they would have missed low S/N candidates. 

Assuming that the higher energy power-law slope of -1.8 (estimated in Section~\ref{sec:energydist}) is intrinsic to FRB, we can estimate the expected ratio of number of bursts with S/N greater than 6 to that above a S/N of 8. This is given by $(6/8)^{-1.8}=1.7$. The observed ratio of the number of bursts above S/N of 6 (N$=133$) to that above S/N of 8 (N$=70$) is: $133/70 = 1.9$. This implies that we detected more bursts between S/N of 6 and 8, than expected from the power-law distribution of energies. But, it is important to note that this simple estimate relies on the following assumptions:
\begin{enumerate}
    \item Our observations are complete to bursts with S/N $<8$. But the flatter energy distribution at low energies indicates that this might not be true.
    \item The fluence and energy distributions are similar. This is true only when there is no incompleteness due to banded nature of the burst spectra \citep{banded_repeaters}.  
    \item The burst energy distribution can be modeled by a single power-law, even at low energies.
\end{enumerate}

Further, we now estimate the number of purely noise candidates we expect above a S/N greater than 6 in our search. The chance probability of a purely noise candidate with S/N greater than 6 is P$=9.9\times10^{-10}$. The number of trials in our search can be calculated by:
\begin{equation}
    \mathrm{N}_\mathrm{trial} = \mathrm{N}_\mathrm{DM} \sum_{i=0}^{10} \frac{\mathrm{N}_\mathrm{time}}{2^i}
\end{equation}

Where, $\mathrm{N}_\mathrm{trial}$ is the total number of trials in our search, $\mathrm{N}_\mathrm{DM}$ is the number of trial DMs (65, between 450--650 \dmunits at a tolerance of 1\%) and $\mathrm{N}_\mathrm{time}$ is the total number of time samples in the data ($\sim3$ hrs at a time resolution of 81.92-$\mu$s). The sum is over the boxcar widths searched ($2^0$ to $2^{10}$, doubling at each step). Therefore, we expect P$\times\mathrm{N}_\mathrm{trial}\sim20$ purely noise events, with a S/N greater than 6, for these number of trials. But number of events we detected with S/N $>6$ in our single pulse search were 1,427 (with 133 FRB and 1,294 RFI candidates).

This shows that we detected much more candidates above a S/N of 6 than expected from pure Gaussian noise, implying that data is non-Gaussian. Most of the candidates we obtained were due to RFI, which is expected. It is possible that some weak events are still misidentified, however this will not influence the results of our analysis.

\subsubsection{Single-pulse search software}
\citet{Gourdji2019} used \texttt{single\_pulse\_search.py} \citep{presto} to search for the bursts, and manually verified the candidates. In contrast, we used \textsc{Heimdall} for the single-pulse search, and \textsc{Fetch} for classification. \citet{Keane2015} highlight several steps at which a single-pulse search pipeline might miss a transient. A few such steps are sizes of boxcar convolutional kernel, spacing between DM trials, the position of the boxcar convolution with respect to the phase of the pulse, clustering of redundant candidates, etc. %\fixme{\texttt{single\_pulse\_search.py} has been shown to miss transients that do not align with the boxcar phase}, While \textsc{Heimdall} is not sensitive to that effect (see figure 1 of \citet{Keane2015}). \textsc{Heimdall} based pipelines also demonstrate a high recovery rate of transients (see below), while similar results are not available for \texttt{single\_pulse\_search.py} based pipelines. 
Although recovery rates for both these search softwares have been shown to be $>$90\% in their respective pipelines \citep{patel2018, gb, gupta2021}, a thorough comparison of the two search strategies has so far not been done.

\subsubsection{RFI mitigation and classification}
\citet{Gourdji2019} used a different RFI mitigation strategy than our pipeline. They used the classifier (\textsc{sps}) presented in \citet{michilli2018} to filter the RFI candidates and then manually verified the remaining 125 candidates to search for real pulses. \textsc{sps} was designed specifically to search for Galactic single pulses in a LOFAR survey. Features in LOFAR data would be very different from the data used in this study. This is primarily due to different observing bands (1.4GHz for this study, compared to 100MHz for LOFAR), RFI environment, telescopes, and observing backends. Moreover, the dispersion in FRB pulses is typically much larger than that seen in Galactic transients. Due to these differences, it is not possible to translate the performance of \textsc{sps} on LOFAR data to the data used in this work. \citet{michilli2018} mention the use of specially designed filters for such datasets on which \textsc{sps} was not trained, but they did not report the performance of these filters on any such data. \citet{fetch2020} and \citet{Connor2018} also highlight that it is non-trivial to generalize a machine learning algorithm to unseen data without rigorous pre-processing and injection tests. It is therefore possible that \textsc{sps} missed to correctly identify real pulses that were correctly detected by \texttt{single\_pulse\_search.py}.

As mentioned previously, we used spectral kurtosis for RFI mitigation \citep{nita2010} and used \textsc{fetch} for classification. It is known that very strict RFI flagging can lead to a reduced recovery rate, as the RFI algorithm might flag real signal as RFI \citep{Rajwade2020}. Spectral kurtosis is robust to astrophysical signals and performs better than the simple median-based RFI thresholds \citep{nita2010}. \textsc{fetch} was developed to be robust to different telescopes, observing configurations and RFI environments \citep{fetch2020}. We accomplished this by a carefully designed pre-processing and training strategy. We also showed that the performance of \textsc{fetch} remains consistent on unseen data. This has been further established by the new FRBs discovered using \textsc{fetch} on different telescopes \citep{Kumar2019, law2020, Aggarwal2020, Rajwade121102, Kumar2021, Pleunis2021, Kirsten2021}. 

\subsubsection{General comments}
As mentioned in Section~\ref{sec:completeness}, a robust analysis of any single-pulse search software and pipeline requires exhaustive injection analysis. Such an analysis can highlight the percentage of transients recovered with respect to various physical parameters of interest, like DM, width, phase, etc. These metrics are essential in understanding possible inefficiencies and estimating the completeness of any single-pulse search. Such analysis done for specific pipelines that use \textsc{Heimdall} has reported a recovery rate of $>90\%$ \citep{gb, gupta2021}. While our pipeline is not identical, it is very similar to the one reported by \citet{gb}. On a less diverse dataset, the recovery rates for \texttt{single\_pulse\_search.py} were also reported to be $>90\%$ \citep{patel2018}, but a more rigorous analysis would better assess the robustness of this search software.  

% \fixme{On the contrary, we are not aware of the recovery rates for any such analysis done on \texttt{single\_pulse\_search.py} based pipelines. As also shown by \citet{Keane2015}, care should be taken while interpreting results from the software without understanding their recovery rate.   }

Additionally, if a machine learning classifier is deployed in a single-pulse search pipeline, then it is necessary to estimate the recovery rate for the classifier itself. %the advent of machine learning systems deployed in single pulse search pipelines to classify RFI vs. FRBs add another sensitivity estimation step. 
Such rigorous analyses have been performed for some classifiers \citep{fetch2020, gupta2021}, and provide insights into their recovery rate estimates. The robustness of a classifier to unseen data should also be carefully investigated (or verified by manual inspection), before reporting the completeness of any search. While human recovery rates for visual classification of thousands of candidates have not been estimated for FRBs, similar estimates are present in computer vision literature. Here, the neural networks (a top-5 error rate of 3.5\%) routinely outperform humans (a top-5 error rate of 5.1\%) in image classification tasks \citep{he2015, Russakovsky2014}. %It is still not trivial to compare the classification accuracy of a human to that of a classifier for a single pulse classification task without a comparative analysis using a standard dataset.  

Further, as most of the bursts detected in this sample are narrow-band (i.e., present only in a part of the frequency band), traditional searches might still miss bursts. Sub-banded searches would be more sensitive to detect such band-limited bursts, especially at wide-band systems  \citep[R. Anna-Thomas et al. 2021, in preparation; ][]{Kumar2021, Gourdji2019}.

\subsection{Caveats}
\label{sec:caveats}
Finally, it is appropriate to highlight and reiterate four main caveats to the analysis presented in this work. (1) The data used in this analysis was downsampled to 64 frequency channels. We did not have access to the native resolution data; therefore, all the search and analysis was performed on downsampled data. The sensitivity of single-pulse search would be higher on native resolution data; therefore, our pipeline may have missed some pulses. (2) We only performed a search on the data that averaged over the full bandwidth to create timeseries. Given the band-limited nature of many bursts, a sub-band search on this data might reveal weaker and narrower bursts. (3) As mentioned previously, our estimate of completeness limit is not robust, as such a robust estimate requires injection analysis that was not possible with the available data. (4) The reported properties of the bursts depend on the assumption that burst spectra and profile follow the assumed functional forms used for fitting.  

% \begin{enumerate}
    % \item we found XX more bursts
    % \item reasons those bursts were missed earlier?
    % \item importance of sensitivity analysis (keane and petroff, gupta 2021)
    % \item careful RFI mitigation (rajwade?)
    % \item Completeness and sensitivity of classifiers
    % \item subbanded searches (kumar et al , anna-thomas et al)
    % \item data is downsampled. it is possible that original data has more pulses. 
    % \item subband search might recover more pulses 
    % \item no injection analysis, so completeness isn't reliable 
    % \item fitting is only good as model 
% \end{enumerate}

% \subsection{Progenitor models}
% mandatory section with one parah about progenitor models 

\section{Conclusions}
This paper presents a dense sample of \src\ bursts detected at L-band using Arecibo Observatory, analyzed as a part of The Petabyte Project. More importantly, we report 93 new bursts detected with our single-pulse search pipeline, as compared to the published results \citep{Gourdji2019}, making a total of 133 burst detections in 3~hours of data. We have developed a robust burst fitting procedure to model the spectro-temporal properties of FRBs and provide it as a user-friendly python package \textsc{burstfit}\footnote{\url{https://github.com/thepetabyteproject/burstfit}}. We use the MCMC procedure implemented in \textsc{burstfit} to estimate the properties of all the bursts in our sample. We find that the burst spectra can be well modeled using a Gaussian function, with a median width of 230~MHz and a median peak at 1608~MHz. Most of the burst emission is present in the top of our band, and there is a lack of emission below 1300~MHz, consistent with other published results \citep{Gourdji2019, platts2021}. Many bursts also show a scattering tail, with a median timescale of 0.7~ms. Some bursts show complex structures like multiple components and frequency drift. The wait time distribution of the bursts shows two distributions, at millisecond and second timescales. The latter of the two follows a log-normal distribution, with the peak at 74.8~s, consistent with other published results \citep{Zhang2018}. We further note that the peak of the wait time changes significantly based on the number of bursts in an observation. We find that both Poisson and Weibull distributions fit the burst rate distributions at long wait times ($>1$ second) equally well, and neither accurately describes the whole burst rate distribution. We did not detect any short-period periodicity in the bursts. The cumulative burst energy distribution is well modeled by a double power-law with a break. We find the value of low and high-energy slopes to be $-0.4\pm0.1$ and $-1.8\pm0.2$ with a break at $(2.3\pm0.2)\times 10^{37}$~ergs. Our analysis reveals that only the bursts that are completely within the band should be used for energy distribution analysis. We discuss some possible differences between our single-pulse search pipeline and the one used by \citet{Gourdji2019}, to explain the different results obtained using the two approaches. All the analysis scripts and results presented in this paper are openly available in a Github repository\footnote{\url{https://github.com/thepetabyteproject/FRB121102}} for the community to use in their repeater burst analyses.   

% \begin{enumerate}
    % \item XX new bursts
    % \item careful and robust burst fitting, developed burstfit, publicly available 
    % \item properties of all bursts 
    % \item dearth of emission below 1300MHz
    % \item most bursts have gaussian spectra
    % \item DM consistent with lit 
    % \item many bursts show scattering, max was XX
    % \item wait time changes significantly using more bursts. future analysis should be careful. our numbers more consistent with other published ones 
    % \item bunch of bursts at millisecond separation, hard to say if multi component, or multi burst 
    % \item burst rate well modeled by both poisson and weibull. 
    % \item careful analysis of cumulative E distribution shows that only in band bursts shold be used. the slope was XX, consistent/inconsistnet with others? two power-laws estimate it better 
    % \item no short period periodicity was found 
    % \item Discuss why new bursts were found/missed
    % \item all the analysis scripts and results are present in Github repository.  
% \end{enumerate}

\section{Acknowledgements}
We would like to thank Kelly Gourdji and Jason Hessels for sharing the data. K.A, R.A.T, S.B.S acknowledge support from NSF grant AAG-1714897. S.B.S, K.A, and R.A.T are also supported by NSF grant \#2108673. S.B.S is a CIFAR Azrieli Global Scholar in the Gravity and the Extreme Universe program. EFL and MAM are supported by NSF AAG award \#2009425. This research was made possible by the NASA West Virginia Space Grant Consortium, Grant \#80NSSC20M0055. The Arecibo Observatory is a facility of the National Science Foundation operated under cooperative agreement by the University of Central Florida and in alliance with Universidad Ana G. Mendez, and Yang Enterprises, Inc. We are saddened by the extremely tragic collapse of the Arecibo Telescope in December 2020. 

\appendix
Table~\ref{tab:large_table} shows the fitted properties (with $1\sigma$ errors) of all the bursts reported in this analysis. 
\startlongtable
\begin{deluxetable*}{llllllll}
\tablecaption{Properties of all the bursts presented in this analysis.}
\label{tab:large_table}
\tablehead{\colhead{Burst}\tablenotemark{a} & \colhead{$\mu_f$} & \colhead{$\sigma_f$} & \colhead{$S$} & \colhead{$\mu_t$\tablenotemark{b}} & \colhead{$\sigma_t$ } & \colhead{$\tau$\tablenotemark{c}} & \colhead{DM}\\
\colhead{ID} & \colhead{(MHz)} & \colhead{(MHz)} & \colhead{(Jy ms)} & \colhead{(MJD)} & \colhead{(ms)} & \colhead{(ms)} & \colhead{(pc cm$^{-3}$)}}
\startdata
B1$*$ & $1560^{+30}_{-30}$ & $210^{+40}_{-40}$ & $0.09^{+0.02}_{-0.02}$ & 57644.408906976(1) & $0.0^{+0.02}_{-0.02}$ & $1.9^{+0.3}_{-0.3}$ & $565.3^{+0.4}_{-0.4}$\\
B2$*$ & $1200^{+10}_{-10}$ & $50^{+10}_{-10}$ & $0.043^{+0.007}_{-0.007}$ & 57644.40956768(1) & $1.35^{+0.05}_{-0.05}$ & - & $562.4^{+0.8}_{-0.8}$\\
B3.1$\dagger$ & $2900^{+300}_{-600}$ & $800^{+300}_{-200}$ & $0.6^{+0.7}_{-0.3}$ & 57644.409673699(3) & $0.4^{+0.2}_{-0.2}$ & $1.3^{+0.7}_{-0.7}$ & $566.8^{+0.8}_{-0.9}$\\
B3.2$\dagger$ & $1100^{+300}_{-1400}$ & $1000^{+2000}_{-1000}$ & $0.09^{+0.13}_{-0.04}$ & 57644.40967384(2) & $0.3^{+1.4}_{-0.2}$ & $0.3^{+0.7}_{-0.1}$ & $564.7^{+1.5}_{-0.4}$\\
B4$\dagger$ & $3100^{+200}_{-600}$ & $550^{+80}_{-110}$ & $2^{+4}_{-2}$ & 57644.410072889(4) & $1.1^{+0.2}_{-0.2}$ & $0.3^{+0.2}_{-0.1}$ & $564^{+1}_{-1}$\\
B5$\dagger$ & $2100^{+900}_{-1600}$ & $2700^{+900}_{-1200}$ & $0.19^{+0.05}_{-0.04}$ & 57644.410157834(4) & $0.7^{+0.3}_{-0.3}$ & $1.0^{+0.4}_{-0.3}$ & $562.1^{+0.7}_{-0.6}$\\
B6.1 & $1393^{+7}_{-7}$ & $183^{+7}_{-7}$ & $0.47^{+0.02}_{-0.03}$ & 57644.411071954(1) & $1.09^{+0.03}_{-0.04}$ & - & $562.3^{+0.2}_{-0.1}$\\
B6.2 & $1417^{+4}_{-5}$ & $102^{+5}_{-4}$ & $0.33^{+0.03}_{-0.02}$ & 57644.4110719755(9) & $0.57^{+0.03}_{-0.02}$ & - & $560.9^{+0.2}_{-0.1}$\\
B7.1$\dagger$ & $3100^{+200}_{-300}$ & $430^{+60}_{-80}$ & $10^{+20}_{-10}$ & 57644.412240214(5) & $0.7^{+0.3}_{-0.4}$ & $0.8^{+0.4}_{-0.4}$ & $569^{+3}_{-3}$\\
B7.2$\dagger$ & $1460^{+20}_{-20}$ & $90^{+30}_{-20}$ & $0.09^{+0.01}_{-0.01}$ & 57644.41224043(2) & $1.9^{+0.9}_{-1.0}$ & $1.2^{+0.6}_{-0.5}$ & $569^{+3}_{-2}$\\
B8$\dagger$ & $3000^{+200}_{-600}$ & $700^{+100}_{-100}$ & $1.1^{+1.5}_{-0.7}$ & 57644.414123628(4) & $1.0^{+0.3}_{-0.3}$ & $0.8^{+0.4}_{-0.3}$ & $567.5^{+0.9}_{-0.7}$\\
B9$*$ & $1430^{+10}_{-10}$ & $75^{+9}_{-9}$ & $0.076^{+0.003}_{-0.003}$ & 57644.41447161(2) & $2.0^{+0.6}_{-0.6}$ & $0.4^{+0.6}_{-0.6}$ & $564^{+2}_{-2}$\\
B10 & $1630^{+10}_{-10}$ & $82^{+8}_{-8}$ & $0.1^{+0.01}_{-0.01}$ & 57644.414475391(7) & $1.4^{+0.3}_{-0.3}$ & $0.5^{+0.3}_{-0.2}$ & $562^{+3}_{-3}$\\
B11 & $1550^{+10}_{-10}$ & $100^{+10}_{-10}$ & $0.074^{+0.008}_{-0.007}$ & 57644.414878803(4) & $0.8^{+0.2}_{-0.2}$ & $0.3^{+0.2}_{-0.1}$ & $560.3^{+0.9}_{-0.9}$\\
B12 & $3100^{+200}_{-300}$ & $450^{+60}_{-80}$ & $20^{+20}_{-10}$ & 57644.41537809(1) & $1.9^{+0.5}_{-0.5}$ & $1.5^{+1.3}_{-0.8}$ & $569^{+3}_{-4}$\\
B13 & $1692^{+13}_{-9}$ & $90^{+20}_{-10}$ & $0.14^{+0.01}_{-0.01}$ & 57644.416314736(5) & $1.5^{+0.2}_{-0.2}$ & $0.6^{+0.3}_{-0.2}$ & $572^{+1}_{-1}$\\
B14 & $1260^{+10}_{-10}$ & $67^{+9}_{-8}$ & $0.13^{+0.02}_{-0.02}$ & 57644.41830674(8) & $3^{+1}_{-1}$ & $1.6^{+0.6}_{-0.5}$ & $562^{+6}_{-5}$\\
B15.1$\dagger$ & $1550^{+10}_{-10}$ & $80^{+10}_{-10}$ & $0.07^{+0.01}_{-0.01}$ & 57644.418309206(7) & $0.9^{+0.3}_{-0.3}$ & $0.8^{+0.5}_{-0.4}$ & $565^{+2}_{-1}$\\
B15.2$\dagger$ & $2000^{+1000}_{-1000}$ & $3000^{+1000}_{-1000}$ & $0.12^{+0.03}_{-0.04}$ & 57644.418309273(2) & $0.2^{+0.2}_{-0.1}$ & $0.7^{+0.2}_{-0.2}$ & $569.4^{+0.4}_{-0.5}$\\
B16$\dagger$ & $3100^{+200}_{-600}$ & $510^{+60}_{-110}$ & $6^{+9}_{-5}$ & 57644.420508580(4) & $1.2^{+0.2}_{-0.2}$ & $0.5^{+0.3}_{-0.2}$ & $570^{+2}_{-2}$\\
B17 & $2000^{+1000}_{-2000}$ & $3000^{+1000}_{-1000}$ & $0.15^{+0.05}_{-0.04}$ & 57644.42110545(1) & $1.5^{+0.6}_{-0.5}$ & $0.7^{+1.0}_{-0.4}$ & $570^{+1}_{-2}$\\
B18$*$ & $1530^{+30}_{-30}$ & $140^{+30}_{-30}$ & $0.046^{+0.008}_{-0.008}$ & 57644.426303862(4) & $1.5^{+0.1}_{-0.1}$ & - & $564.8^{+0.8}_{-0.8}$\\
B19$*$ & $1330^{+30}_{-30}$ & $160^{+20}_{-20}$ & $0.11^{+0.02}_{-0.02}$ & 57644.42721003(2) & $1.9^{+0.8}_{-0.8}$ & $1.6^{+0.5}_{-0.5}$ & $564^{+1}_{-1}$\\
B20$\dagger$ & $2900^{+400}_{-1000}$ & $1000^{+2000}_{-400}$ & $0.23^{+0.47}_{-0.08}$ & 57644.427376859(4) & $0.7^{+0.3}_{-0.3}$ & $0.6^{+0.5}_{-0.3}$ & $565.6^{+0.8}_{-0.6}$\\
B21$\dagger$ & $3100^{+200}_{-400}$ & $600^{+100}_{-100}$ & $3^{+7}_{-2}$ & 57644.42794036(2) & $4.2^{+0.9}_{-1.0}$ & $1.8^{+1.7}_{-0.8}$ & $576^{+4}_{-5}$\\
B22$\dagger$ & $3000^{+300}_{-700}$ & $700^{+100}_{-200}$ & $1.3^{+1.7}_{-0.9}$ & 57644.428593784(8) & $1.9^{+0.4}_{-0.4}$ & $1.5^{+0.6}_{-0.6}$ & $568^{+2}_{-2}$\\
B23 & $1699^{+14}_{-9}$ & $70^{+20}_{-10}$ & $0.089^{+0.009}_{-0.008}$ & 57644.430170170(2) & $0.7^{+0.1}_{-0.1}$ & $0.3^{+0.2}_{-0.1}$ & $569.0^{+1.0}_{-1.0}$\\
B24$*$ & $1670^{+20}_{-20}$ & $80^{+20}_{-20}$ & $0.0384^{+0.0007}_{-0.0007}$ & 57644.430170303(1) & $0.02^{+0.05}_{-0.05}$ & $1.2^{+0.3}_{-0.3}$ & $563.0^{+0.5}_{-0.5}$\\
B25 & $1750^{+1380}_{-70}$ & $200^{+390}_{-60}$ & $0.16^{+3.67}_{-0.03}$ & 57644.430171419(4) & $1.4^{+0.2}_{-0.2}$ & $0.7^{+0.3}_{-0.3}$ & $568^{+1}_{-1}$\\
B26 & $1470^{+10}_{-10}$ & $110^{+20}_{-10}$ & $0.096^{+0.009}_{-0.009}$ & 57644.431295361(4) & $0.8^{+0.2}_{-0.2}$ & $0.5^{+0.2}_{-0.2}$ & $565.6^{+0.7}_{-0.7}$\\
B27 & $1340^{+70}_{-1400}$ & $1000^{+2000}_{-1000}$ & $0.17^{+0.11}_{-0.09}$ & 57644.43223490(2) & $2.0^{+1.0}_{-0.7}$ & $1.1^{+0.5}_{-0.4}$ & $570^{+2}_{-2}$\\
B28 & $1750^{+40}_{-20}$ & $100^{+20}_{-10}$ & $0.14^{+0.03}_{-0.02}$ & 57644.432242722(1) & $0.59^{+0.05}_{-0.05}$ & $0.16^{+0.09}_{-0.05}$ & $561.3^{+0.5}_{-0.5}$\\
B29 & $1680^{+1440}_{-30}$ & $130^{+480}_{-40}$ & $0.06^{+1.5}_{-0.01}$ & 57644.434045197(4) & $0.6^{+0.2}_{-0.2}$ & $0.4^{+0.4}_{-0.2}$ & $562^{+1}_{-3}$\\
B30 & $1440^{+10}_{-10}$ & $100^{+20}_{-10}$ & $0.12^{+0.01}_{-0.01}$ & 57644.43636575(2) & $2.1^{+0.6}_{-0.5}$ & $1.2^{+0.5}_{-0.4}$ & $568^{+2}_{-3}$\\
B31 & $1502^{+6}_{-6}$ & $100^{+8}_{-7}$ & $0.19^{+0.01}_{-0.01}$ & 57644.438794966(4) & $1.6^{+0.1}_{-0.2}$ & $0.5^{+0.1}_{-0.1}$ & $566.6^{+0.7}_{-0.7}$\\
B32 & $1410^{+8}_{-8}$ & $79^{+9}_{-8}$ & $0.14^{+0.01}_{-0.01}$ & 57644.43884518(1) & $1.7^{+0.3}_{-0.4}$ & $0.6^{+0.4}_{-0.3}$ & $564^{+1}_{-1}$\\
B33 & $1480^{+20}_{-20}$ & $120^{+20}_{-20}$ & $0.08^{+0.01}_{-0.01}$ & 57644.439212855(8) & $1.2^{+0.3}_{-0.4}$ & $0.6^{+0.6}_{-0.3}$ & $563.0^{+1.0}_{-1.0}$\\
B34$\dagger$ & $2600^{+600}_{-1400}$ & $2000^{+1000}_{-1000}$ & $0.27^{+0.1}_{-0.06}$ & 57644.440688613(5) & $1.3^{+0.4}_{-0.4}$ & $1.0^{+0.4}_{-0.3}$ & $567.0^{+0.8}_{-0.9}$\\
B35 & $1707^{+5}_{-5}$ & $61^{+6}_{-5}$ & $0.24^{+0.01}_{-0.01}$ & 57644.442997729(6) & $2.8^{+0.2}_{-0.2}$ & $0.7^{+0.4}_{-0.2}$ & $575^{+3}_{-3}$\\
B36$\dagger$ & $3100^{+200}_{-400}$ & $470^{+70}_{-90}$ & $10^{+20}_{-10}$ & 57644.44358918(2) & $3^{+2}_{-1}$ & $2^{+2}_{-1}$ & $574^{+11}_{-6}$\\
B37 & $1410^{+20}_{-20}$ & $130^{+30}_{-20}$ & $0.065^{+0.007}_{-0.008}$ & 57644.443590029(3) & $0.58^{+0.11}_{-0.09}$ & $0.14^{+0.1}_{-0.04}$ & $562.8^{+0.5}_{-0.4}$\\
B38 & $1620^{+20}_{-20}$ & $110^{+20}_{-20}$ & $0.1^{+0.01}_{-0.01}$ & 57644.445225058(7) & $0.9^{+0.5}_{-0.5}$ & $1.9^{+0.7}_{-0.9}$ & $567^{+2}_{-2}$\\
B39 & $1500^{+10}_{-20}$ & $130^{+10}_{-10}$ & $0.057^{+0.01}_{-0.006}$ & 57644.446788124(1) & $0.21^{+0.04}_{-0.04}$ & $0.12^{+0.13}_{-0.05}$ & $560.9^{+0.1}_{-0.2}$\\
B40 & $1590^{+10}_{-10}$ & $81^{+11}_{-9}$ & $0.12^{+0.02}_{-0.02}$ & 57644.447567822(8) & $1.1^{+0.5}_{-0.4}$ & $2.0^{+0.8}_{-0.7}$ & $560^{+3}_{-3}$\\
B41$*$ & $1379^{+1}_{-1}$ & $31^{+1}_{-1}$ & $0.4^{+0.02}_{-0.02}$ & 57644.44772750(2) & $0.8^{+0.2}_{-0.2}$ & $3.0^{+0.2}_{-0.2}$ & $559^{+2}_{-2}$\\
B42 & $1670^{+10}_{-10}$ & $110^{+20}_{-20}$ & $0.099^{+0.009}_{-0.008}$ & 57644.449915568(3) & $0.9^{+0.2}_{-0.2}$ & $0.4^{+0.2}_{-0.2}$ & $566.0^{+0.8}_{-0.7}$\\
B43 & $1683^{+12}_{-9}$ & $120^{+20}_{-10}$ & $0.22^{+0.01}_{-0.01}$ & 57644.451605444(1) & $0.79^{+0.1}_{-0.09}$ & $0.9^{+0.1}_{-0.1}$ & $568.2^{+0.5}_{-0.5}$\\
B44$\dagger$ & $3100^{+200}_{-400}$ & $520^{+50}_{-70}$ & $11^{+13}_{-7}$ & 57644.452389712(6) & $1.8^{+0.3}_{-0.3}$ & $1.1^{+0.5}_{-0.5}$ & $569^{+2}_{-4}$\\
B45 & $1710^{+1490}_{-60}$ & $160^{+470}_{-70}$ & $0.07^{+1.81}_{-0.02}$ & 57644.453937473(4) & $0.9^{+0.2}_{-0.4}$ & $0.5^{+0.4}_{-0.2}$ & $562^{+1}_{-1}$\\
B46 & $1371^{+2}_{-2}$ & $61^{+2}_{-2}$ & $0.41^{+0.01}_{-0.01}$ & 57644.454477404(8) & $2.8^{+0.2}_{-0.2}$ & $0.6^{+0.1}_{-0.1}$ & $569.6^{+0.8}_{-0.8}$\\
B47 & $900^{+800}_{-1100}$ & $2000^{+1000}_{-2000}$ & $0.15^{+0.04}_{-0.09}$ & 57644.457883227(3) & $0.5^{+0.1}_{-0.1}$ & $0.23^{+0.14}_{-0.09}$ & $561.9^{+0.3}_{-0.2}$\\
B48.1$\dagger$ & $2900^{+300}_{-1200}$ & $600^{+2500}_{-100}$ & $1^{+15}_{-1}$ & 57644.464507488(7) & $0.4^{+0.4}_{-0.3}$ & $0.7^{+0.6}_{-0.3}$ & $565^{+4}_{-2}$\\
B48.2$\dagger$ & $0^{+1300}_{-500}$ & $500^{+500}_{-400}$ & $0.7^{+26.2}_{-0.6}$ & 57644.4645075(2) & $0.9^{+3.2}_{-0.6}$ & $0.7^{+3.6}_{-0.4}$ & $600^{+100}_{-200}$\\
B49.1 & $1660^{+10}_{-10}$ & $140^{+20}_{-10}$ & $0.49^{+0.03}_{-0.03}$ & 57644.46475893(1) & $8.3^{+0.5}_{-0.5}$ & $2.0^{+0.7}_{-0.6}$ & $568^{+4}_{-4}$\\
B49.2 & $2400^{+700}_{-1400}$ & $2000^{+1000}_{-1000}$ & $0.16^{+0.05}_{-0.04}$ & 57644.464759903(1) & $0.2^{+0.1}_{-0.1}$ & $0.6^{+0.2}_{-0.1}$ & $561.3^{+0.3}_{-0.2}$\\
B50 & $1460^{+10}_{-10}$ & $90^{+10}_{-10}$ & $0.13^{+0.02}_{-0.01}$ & 57644.46476275(1) & $1.3^{+1.0}_{-0.9}$ & $1.7^{+0.6}_{-0.8}$ & $561^{+2}_{-2}$\\
B51$\dagger$ & $2500^{+600}_{-900}$ & $1000^{+1000}_{-700}$ & $0.3^{+0.2}_{-0.2}$ & 57644.465729923(7) & $1.4^{+0.6}_{-0.4}$ & $1.2^{+0.5}_{-0.4}$ & $569^{+1}_{-1}$\\
B52 & $1667^{+5}_{-4}$ & $73^{+6}_{-5}$ & $0.185^{+0.009}_{-0.009}$ & 57644.466222289(4) & $1.4^{+0.1}_{-0.2}$ & $0.6^{+0.2}_{-0.2}$ & $570^{+1}_{-1}$\\
B53 & $1720^{+30}_{-20}$ & $120^{+30}_{-20}$ & $0.19^{+0.03}_{-0.02}$ & 57644.468095365(4) & $1.5^{+0.2}_{-0.2}$ & $1.1^{+0.3}_{-0.3}$ & $572^{+1}_{-1}$\\
B54$*$ & $1600^{+600}_{-600}$ & $100^{+100}_{-100}$ & $0.0^{+0.1}_{-0.1}$ & 57644.47117767(7) & $0^{+10}_{-10}$ & $0.9^{+0.2}_{-0.2}$ & $560^{+40}_{-40}$\\
B55 & $1650^{+10}_{-10}$ & $100^{+10}_{-10}$ & $0.13^{+0.01}_{-0.01}$ & 57644.474717918(6) & $1.4^{+0.3}_{-0.2}$ & $1.3^{+0.5}_{-0.5}$ & $570^{+2}_{-2}$\\
B56 & $1706^{+4}_{-4}$ & $54^{+6}_{-5}$ & $0.16^{+0.01}_{-0.01}$ & 57644.477082041(3) & $0.9^{+0.2}_{-0.1}$ & $0.9^{+0.2}_{-0.2}$ & $560^{+2}_{-2}$\\
B57 & $1530^{+6}_{-6}$ & $186^{+6}_{-6}$ & $0.254^{+0.006}_{-0.005}$ & 57645.4110889611(4) & $0.48^{+0.03}_{-0.02}$ & $0.21^{+0.02}_{-0.02}$ & $561.75^{+0.06}_{-0.06}$\\
B58 & $1462^{+6}_{-6}$ & $78^{+6}_{-5}$ & $0.129^{+0.008}_{-0.008}$ & 57645.411651653(4) & $0.9^{+0.1}_{-0.1}$ & $0.4^{+0.1}_{-0.1}$ & $564.4^{+0.6}_{-0.7}$\\
B59 & $1537^{+5}_{-4}$ & $68^{+4}_{-4}$ & $0.2^{+0.01}_{-0.01}$ & 57645.411889086(9) & $2.2^{+0.4}_{-0.3}$ & $1.1^{+0.3}_{-0.3}$ & $568^{+2}_{-2}$\\
B60.1 & $3000^{+300}_{-1000}$ & $440^{+80}_{-220}$ & $4^{+6}_{-4}$ & 57645.412281872(4) & $0.8^{+0.2}_{-0.1}$ & $0.3^{+0.3}_{-0.2}$ & $566^{+2}_{-2}$\\
B60.2 & $2300^{+900}_{-600}$ & $500^{+300}_{-300}$ & $0.4^{+1.6}_{-0.2}$ & 57645.412281990(6) & $0.9^{+0.6}_{-0.5}$ & $2.5^{+0.6}_{-0.5}$ & $561^{+2}_{-2}$\\
B61$\dagger$ & $1000^{+1000}_{-1000}$ & $2800^{+900}_{-1100}$ & $0.25^{+0.08}_{-0.07}$ & 57645.41286869(2) & $2.0^{+2.0}_{-1.0}$ & $2^{+2}_{-1}$ & $580^{+3}_{-13}$\\
B62 & $1570^{+10}_{-10}$ & $110^{+10}_{-10}$ & $0.12^{+0.01}_{-0.009}$ & 57645.413644740(5) & $1.3^{+0.2}_{-0.2}$ & $0.6^{+0.3}_{-0.2}$ & $567^{+1}_{-1}$\\
B63 & $1280^{+10}_{-10}$ & $86^{+11}_{-9}$ & $0.1^{+0.01}_{-0.01}$ & 57645.41609534(1) & $0.8^{+0.2}_{-0.2}$ & $0.6^{+0.2}_{-0.1}$ & $562.0^{+0.8}_{-0.9}$\\
B64$\dagger$ & $2000^{+1000}_{-1000}$ & $3000^{+1000}_{-1000}$ & $0.32^{+0.09}_{-0.09}$ & 57645.41639521(1) & $1.4^{+0.5}_{-0.5}$ & $3^{+1}_{-1}$ & $564^{+3}_{-2}$\\
B65 & $1550^{+10}_{-10}$ & $110^{+20}_{-10}$ & $0.091^{+0.009}_{-0.009}$ & 57645.416564818(6) & $1.2^{+0.2}_{-0.2}$ & $0.6^{+0.3}_{-0.2}$ & $566.0^{+1.0}_{-1.0}$\\
B66 & $1440^{+10}_{-10}$ & $120^{+10}_{-10}$ & $0.105^{+0.009}_{-0.008}$ & 57645.417467306(6) & $1.6^{+0.2}_{-0.2}$ & - & $567.7^{+0.7}_{-0.7}$\\
B67 & $1640^{+30}_{-20}$ & $180^{+30}_{-30}$ & $0.099^{+0.009}_{-0.008}$ & 57645.417897463(2) & $0.8^{+0.1}_{-0.1}$ & $0.3^{+0.1}_{-0.1}$ & $564.8^{+0.5}_{-0.5}$\\
B68$\dagger$ & $2000^{+1000}_{-1000}$ & $2000^{+1000}_{-1000}$ & $0.2^{+0.05}_{-0.06}$ & 57645.419005902(8) & $1.0^{+0.6}_{-0.5}$ & $0.9^{+0.4}_{-0.4}$ & $567.9^{+0.9}_{-1.3}$\\
B69 & $1450^{+20}_{-10}$ & $80^{+10}_{-10}$ & $0.059^{+0.009}_{-0.009}$ & 57645.41920745(1) & $0.7^{+0.3}_{-0.3}$ & $0.5^{+0.3}_{-0.2}$ & $564^{+1}_{-1}$\\
B70$\dagger$ & $3100^{+200}_{-500}$ & $600^{+100}_{-100}$ & $0.9^{+1.7}_{-0.6}$ & 57645.419896226(4) & $0.8^{+0.3}_{-0.3}$ & $0.8^{+0.4}_{-0.3}$ & $564^{+1}_{-1}$\\
B71 & $1378^{+5}_{-4}$ & $78^{+5}_{-4}$ & $0.34^{+0.02}_{-0.02}$ & 57645.420265931(9) & $1.0^{+0.2}_{-0.2}$ & $2.6^{+0.2}_{-0.2}$ & $561.1^{+0.9}_{-1.0}$\\
B72$\dagger$ & $3100^{+200}_{-500}$ & $700^{+200}_{-100}$ & $1.2^{+3.4}_{-0.8}$ & 57645.420600439(9) & $1.2^{+0.7}_{-0.8}$ & $2^{+1}_{-1}$ & $559^{+2}_{-2}$\\
B73$\dagger$ & $3100^{+200}_{-500}$ & $600^{+100}_{-100}$ & $2^{+3}_{-1}$ & 57645.420679524(7) & $1.5^{+0.6}_{-0.8}$ & $1.7^{+0.8}_{-0.7}$ & $575^{+3}_{-3}$\\
B74$\dagger$ & $3100^{+200}_{-400}$ & $520^{+100}_{-90}$ & $5^{+12}_{-4}$ & 57645.420752363(6) & $1.2^{+0.8}_{-0.6}$ & $1.9^{+0.7}_{-0.9}$ & $551^{+11}_{-2}$\\
B75$\dagger$ & $2200^{+800}_{-1400}$ & $2700^{+900}_{-1200}$ & $0.39^{+0.07}_{-0.08}$ & 57645.421284056(4) & $0.9^{+0.3}_{-0.3}$ & $1.8^{+0.3}_{-0.3}$ & $567.6^{+0.9}_{-0.7}$\\
B76 & $1410^{+10}_{-350}$ & $90^{+20}_{-10}$ & $0.05^{+0.111}_{-0.006}$ & 57645.421290699(5) & $0.42^{+0.25}_{-0.09}$ & $0.14^{+0.42}_{-0.05}$ & $561.6^{+0.6}_{-0.5}$\\
B77 & $1312^{+5}_{-5}$ & $51^{+6}_{-6}$ & $0.12^{+0.01}_{-0.01}$ & 57645.42161300(2) & $1.3^{+0.5}_{-0.5}$ & $0.7^{+0.2}_{-0.2}$ & $566^{+2}_{-2}$\\
B78 & $1170^{+20}_{-40}$ & $80^{+40}_{-10}$ & $0.11^{+0.03}_{-0.02}$ & 57645.42184846(1) & $0.9^{+0.3}_{-0.2}$ & $0.28^{+0.14}_{-0.08}$ & $562.4^{+0.7}_{-0.9}$\\
B79 & $1702^{+4}_{-4}$ & $57^{+6}_{-5}$ & $0.18^{+0.01}_{-0.01}$ & 57645.421872443(3) & $1.2^{+0.2}_{-0.2}$ & $1.7^{+0.3}_{-0.3}$ & $568^{+2}_{-2}$\\
B80 & $1601^{+7}_{-8}$ & $81^{+8}_{-7}$ & $0.13^{+0.01}_{-0.01}$ & 57645.422454976(3) & $0.5^{+0.3}_{-0.2}$ & $1.3^{+0.2}_{-0.3}$ & $564.4^{+1.0}_{-0.8}$\\
B81 & $1720^{+390}_{-10}$ & $60^{+310}_{-20}$ & $0.05^{+0.1}_{-0.01}$ & 57645.423009490(6) & $0.78^{+0.08}_{-0.18}$ & $0.6^{+0.1}_{-0.2}$ & $563^{+8}_{-5}$\\
B82$\dagger$ & $3100^{+200}_{-400}$ & $580^{+60}_{-70}$ & $4^{+4}_{-2}$ & 57645.424145870(2) & $1.3^{+0.1}_{-0.1}$ & - & $569.2^{+0.7}_{-0.7}$\\
B83$\dagger$ & $2400^{+700}_{-1300}$ & $2000^{+2000}_{-1000}$ & $0.15^{+0.07}_{-0.04}$ & 57645.424617988(6) & $0.7^{+0.3}_{-0.3}$ & $0.4^{+0.2}_{-0.2}$ & $564.7^{+0.7}_{-0.5}$\\
B84$\dagger$ & $3000^{+300}_{-700}$ & $600^{+100}_{-200}$ & $1^{+3}_{-1}$ & 57645.425478263(8) & $1.0^{+0.8}_{-0.6}$ & $1.7^{+0.9}_{-0.9}$ & $563^{+2}_{-3}$\\
B85$*$ & $1310^{+20}_{-20}$ & $130^{+20}_{-20}$ & $0.11^{+0.02}_{-0.02}$ & 57645.426746546(2) & $0.01^{+0.06}_{-0.06}$ & $2.0^{+0.4}_{-0.4}$ & $566.1^{+0.2}_{-0.2}$\\
B86 & $1550^{+10}_{-10}$ & $80^{+10}_{-10}$ & $0.067^{+0.009}_{-0.008}$ & 57645.426792398(7) & $0.6^{+0.4}_{-0.3}$ & $0.9^{+0.3}_{-0.3}$ & $563^{+1}_{-1}$\\
B87 & $1760^{+440}_{-50}$ & $190^{+200}_{-50}$ & $0.2^{+0.38}_{-0.03}$ & 57645.426922532(4) & $1.4^{+0.2}_{-0.2}$ & $0.9^{+0.3}_{-0.3}$ & $571.1^{+0.9}_{-0.9}$\\
B88.1 & $2000^{+1000}_{-1000}$ & $2000^{+1000}_{-2000}$ & $0.17^{+0.06}_{-0.06}$ & 57645.427338932(7) & $0.7^{+0.4}_{-0.3}$ & $1.0^{+0.7}_{-0.5}$ & $561^{+2}_{-4}$\\
B88.2 & $1460^{+20}_{-10}$ & $130^{+20}_{-20}$ & $0.11^{+0.01}_{-0.01}$ & 57645.427339178(9) & $2.4^{+0.4}_{-0.3}$ & - & $561^{+1}_{-2}$\\
B89 & $1479^{+8}_{-8}$ & $120^{+8}_{-7}$ & $0.18^{+0.01}_{-0.01}$ & 57645.428904793(6) & $2.4^{+0.2}_{-0.2}$ & - & $569.0^{+0.7}_{-0.8}$\\
B90 & $1420^{+40}_{-500}$ & $120^{+2790}_{-30}$ & $0.09^{+0.11}_{-0.02}$ & 57645.42990044(2) & $2.4^{+0.6}_{-0.6}$ & $0.8^{+0.6}_{-0.3}$ & $565^{+2}_{-2}$\\
B91 & $1476^{+9}_{-9}$ & $81^{+10}_{-8}$ & $0.12^{+0.01}_{-0.01}$ & 57645.43034452(2) & $3.1^{+0.4}_{-0.4}$ & - & $572^{+2}_{-2}$\\
B92 & $1721^{+4}_{-4}$ & $39^{+6}_{-4}$ & $0.17^{+0.02}_{-0.02}$ & 57645.43044763(1) & $2.6^{+0.8}_{-0.7}$ & $3^{+1}_{-1}$ & $580^{+10}_{-10}$\\
B93.1 & $1591^{+9}_{-9}$ & $106^{+10}_{-9}$ & $0.126^{+0.008}_{-0.008}$ & 57645.430622487(3) & $1.2^{+0.1}_{-0.2}$ & $0.4^{+0.2}_{-0.2}$ & $565.3^{+0.8}_{-0.9}$\\
B93.2 & $1620^{+10}_{-10}$ & $120^{+20}_{-20}$ & $0.13^{+0.01}_{-0.01}$ & 57645.430622591(7) & $1.7^{+0.3}_{-0.3}$ & $0.8^{+0.4}_{-0.3}$ & $564^{+2}_{-2}$\\
B94$\dagger$ & $3100^{+200}_{-600}$ & $500^{+60}_{-120}$ & $7^{+10}_{-6}$ & 57645.431306274(4) & $1.1^{+0.2}_{-0.2}$ & $0.6^{+0.4}_{-0.3}$ & $565^{+2}_{-4}$\\
B95 & $1693^{+11}_{-8}$ & $90^{+20}_{-10}$ & $0.095^{+0.007}_{-0.007}$ & 57645.431478373(2) & $0.69^{+0.08}_{-0.08}$ & $0.22^{+0.12}_{-0.08}$ & $563.5^{+0.6}_{-0.5}$\\
B96$\dagger$ & $2000^{+1000}_{-2000}$ & $2800^{+800}_{-1100}$ & $0.2^{+0.06}_{-0.05}$ & 57645.43254531(2) & $2.1^{+0.8}_{-0.9}$ & $0.9^{+0.6}_{-0.3}$ & $565^{+2}_{-1}$\\
B97$\dagger$ & $2900^{+400}_{-800}$ & $800^{+500}_{-200}$ & $0.3^{+0.4}_{-0.1}$ & 57645.433086695(3) & $0.6^{+0.4}_{-0.4}$ & $1.0^{+0.4}_{-0.4}$ & $562.4^{+0.9}_{-0.5}$\\
B98 & $1667^{+7}_{-7}$ & $67^{+8}_{-7}$ & $0.094^{+0.01}_{-0.009}$ & 57645.434330453(5) & $1.1^{+0.2}_{-0.2}$ & $0.7^{+0.3}_{-0.3}$ & $559^{+2}_{-2}$\\
B99$\dagger$ & $2000^{+1000}_{-2000}$ & $2800^{+800}_{-1100}$ & $0.13^{+0.03}_{-0.03}$ & 57645.440066847(4) & $0.8^{+0.2}_{-0.2}$ & $0.3^{+0.2}_{-0.1}$ & $565.8^{+0.4}_{-0.4}$\\
B100 & $1688^{+4}_{-4}$ & $62^{+5}_{-5}$ & $0.154^{+0.008}_{-0.007}$ & 57645.440814641(2) & $1.0^{+0.1}_{-0.1}$ & $0.5^{+0.2}_{-0.2}$ & $567.0^{+1.0}_{-1.0}$\\
B101$*$ & $1510^{+20}_{-20}$ & $130^{+20}_{-20}$ & $0.042^{+0.006}_{-0.006}$ & 57645.441999925(3) & $0.87^{+0.02}_{-0.02}$ & - & $568.6^{+0.4}_{-0.4}$\\
B102 & $2000^{+1200}_{-300}$ & $300^{+300}_{-200}$ & $0.2^{+2.7}_{-0.1}$ & 57645.442100164(8) & $2.1^{+0.4}_{-0.4}$ & $0.8^{+0.6}_{-0.4}$ & $564^{+2}_{-2}$\\
B103 & $1470^{+10}_{-10}$ & $90^{+10}_{-10}$ & $0.08^{+0.01}_{-0.01}$ & 57645.44263325(1) & $1.4^{+0.3}_{-0.3}$ & $0.6^{+0.5}_{-0.3}$ & $562^{+1}_{-1}$\\
B104$*$ & $1460^{+60}_{-60}$ & $90^{+20}_{-20}$ & $0.07^{+0.09}_{-0.09}$ & 57645.4427413(1) & $0^{+2}_{-2}$ & $2.1^{+0.5}_{-0.5}$ & $569^{+8}_{-8}$\\
B105 & $1670^{+10}_{-10}$ & $110^{+20}_{-20}$ & $0.095^{+0.008}_{-0.007}$ & 57645.444480955(2) & $0.51^{+0.1}_{-0.09}$ & $0.5^{+0.1}_{-0.1}$ & $565.4^{+0.5}_{-0.5}$\\
B106 & $1480^{+7}_{-8}$ & $97^{+10}_{-9}$ & $0.16^{+0.01}_{-0.01}$ & 57645.444919472(7) & $2.0^{+0.2}_{-0.2}$ & $0.5^{+0.2}_{-0.1}$ & $566^{+1}_{-1}$\\
B107 & $2800^{+300}_{-1000}$ & $370^{+70}_{-260}$ & $10^{+10}_{-10}$ & 57645.445443070(5) & $1.2^{+0.3}_{-0.3}$ & $0.9^{+0.4}_{-0.4}$ & $561^{+3}_{-2}$\\
B108 & $1488^{+4}_{-4}$ & $83^{+4}_{-4}$ & $0.141^{+0.006}_{-0.006}$ & 57645.448802883(2) & $0.66^{+0.03}_{-0.03}$ & - & $561.1^{+0.3}_{-0.3}$\\
B109 & $1460^{+5}_{-5}$ & $87^{+5}_{-5}$ & $0.24^{+0.01}_{-0.01}$ & 57645.449987035(8) & $2.8^{+0.2}_{-0.2}$ & - & $568^{+1}_{-1}$\\
B110$*$ & $1620^{+10}_{-10}$ & $100^{+10}_{-10}$ & $0.1^{+0.01}_{-0.01}$ & 57645.451201066(8) & $1.7^{+0.4}_{-0.4}$ & $1.3^{+0.5}_{-0.5}$ & $561^{+2}_{-2}$\\
B111 & $1469^{+6}_{-6}$ & $58^{+7}_{-6}$ & $0.109^{+0.01}_{-0.009}$ & 57645.45198993(1) & $1.6^{+0.2}_{-0.2}$ & $0.5^{+0.2}_{-0.1}$ & $563^{+2}_{-2}$\\
B112 & $1442^{+4}_{-5}$ & $68^{+4}_{-4}$ & $0.142^{+0.008}_{-0.008}$ & 57645.453426198(4) & $0.59^{+0.08}_{-0.07}$ & $0.4^{+0.07}_{-0.06}$ & $563.1^{+0.6}_{-0.6}$\\
B113 & $1699^{+4}_{-4}$ & $68^{+6}_{-5}$ & $0.29^{+0.01}_{-0.01}$ & 57645.453639067(4) & $2.5^{+0.1}_{-0.2}$ & $0.6^{+0.3}_{-0.2}$ & $569^{+2}_{-2}$\\
B114 & $1570^{+20}_{-20}$ & $110^{+20}_{-20}$ & $0.07^{+0.01}_{-0.01}$ & 57645.453640216(3) & $0.3^{+0.2}_{-0.1}$ & $1.4^{+0.3}_{-0.3}$ & $566.5^{+0.7}_{-0.8}$\\
B115 & $1730^{+30}_{-10}$ & $90^{+20}_{-20}$ & $0.19^{+0.03}_{-0.02}$ & 57645.454258102(7) & $2.3^{+0.3}_{-0.3}$ & $0.9^{+0.6}_{-0.4}$ & $569^{+3}_{-3}$\\
B116$*$ & $1480^{+40}_{-40}$ & $260^{+60}_{-60}$ & $0.08^{+0.02}_{-0.02}$ & 57645.4544929400(9) & $0.05^{+0.07}_{-0.07}$ & $1.4^{+0.3}_{-0.3}$ & $565.7^{+0.2}_{-0.2}$\\
B117 & $1580^{+20}_{-20}$ & $130^{+30}_{-20}$ & $0.1^{+0.02}_{-0.01}$ & 57645.45736147(1) & $1.8^{+0.9}_{-0.8}$ & $1.8^{+0.8}_{-0.8}$ & $563^{+3}_{-3}$\\
B118 & $1580^{+30}_{-20}$ & $130^{+30}_{-20}$ & $0.1^{+0.02}_{-0.01}$ & 57645.45736147(1) & $1.7^{+0.9}_{-0.8}$ & $1.8^{+0.9}_{-0.8}$ & $563^{+3}_{-3}$\\
B119 & $1670^{+60}_{-20}$ & $170^{+60}_{-30}$ & $0.21^{+0.04}_{-0.02}$ & 57645.458536220(6) & $1.2^{+0.4}_{-0.3}$ & $2.6^{+0.5}_{-0.5}$ & $569^{+2}_{-2}$\\
B120 & $1690^{+1040}_{-30}$ & $130^{+380}_{-30}$ & $0.09^{+0.87}_{-0.01}$ & 57645.459103357(6) & $0.8^{+0.3}_{-0.5}$ & $1.2^{+0.5}_{-0.4}$ & $566^{+2}_{-2}$\\
B121.1 & $1612^{+2}_{-2}$ & $90^{+2}_{-2}$ & $0.394^{+0.007}_{-0.007}$ & 57645.460053270(1) & $0.82^{+0.04}_{-0.04}$ & $0.25^{+0.06}_{-0.07}$ & $562.1^{+0.2}_{-0.2}$\\
B121.2 & $1321^{+2}_{-3}$ & $178^{+2}_{-2}$ & $1.73^{+0.02}_{-0.02}$ & 57645.4600533283(4) & $0.95^{+0.02}_{-0.02}$ & $0.36^{+0.02}_{-0.01}$ & $561.98^{+0.03}_{-0.03}$\\
B121.3 & $1181^{+9}_{-11}$ & $93^{+9}_{-8}$ & $0.35^{+0.03}_{-0.02}$ & 57645.460053409(4) & $1.22^{+0.09}_{-0.07}$ & - & $561.6^{+0.2}_{-0.3}$\\
B122 & $1405^{+4}_{-4}$ & $71^{+5}_{-5}$ & $0.22^{+0.01}_{-0.01}$ & 57645.462106655(9) & $2.3^{+0.2}_{-0.2}$ & $0.5^{+0.2}_{-0.1}$ & $565^{+1}_{-1}$\\
B123$*$ & $1690^{+20}_{-20}$ & $90^{+20}_{-20}$ & $0.09^{+0.02}_{-0.02}$ & 57645.462556118(6) & $1.1^{+0.4}_{-0.4}$ & $2.0^{+0.6}_{-0.6}$ & $567^{+3}_{-3}$\\
B124 & $1560^{+10}_{-20}$ & $150^{+20}_{-20}$ & $0.094^{+0.008}_{-0.008}$ & 57645.464052738(3) & $1.0^{+0.2}_{-0.2}$ & $0.5^{+0.2}_{-0.2}$ & $565.3^{+0.5}_{-0.6}$\\
B125 & $1600^{+10}_{-10}$ & $140^{+20}_{-20}$ & $0.121^{+0.01}_{-0.009}$ & 57645.464188555(4) & $1.2^{+0.2}_{-0.2}$ & $0.7^{+0.2}_{-0.2}$ & $564.4^{+0.8}_{-0.8}$\\
B126 & $1600^{+1190}_{-60}$ & $230^{+2380}_{-90}$ & $0.07^{+0.13}_{-0.02}$ & 57645.464503855(6) & $0.9^{+0.4}_{-0.5}$ & $0.9^{+0.6}_{-0.4}$ & $567^{+2}_{-2}$\\
B127 & $2000^{+1000}_{-1000}$ & $2000^{+1000}_{-2000}$ & $0.19^{+0.06}_{-0.11}$ & 57645.465133243(7) & $0.8^{+0.4}_{-0.4}$ & $1.2^{+0.6}_{-0.5}$ & $566.0^{+1.9}_{-0.7}$\\
B128$\dagger$ & $1000^{+1000}_{-1000}$ & $2700^{+900}_{-1500}$ & $0.25^{+0.05}_{-0.09}$ & 57645.466100959(6) & $1.1^{+0.2}_{-0.2}$ & $0.9^{+0.3}_{-0.3}$ & $567.0^{+0.7}_{-0.9}$\\
B129 & $1710^{+1380}_{-20}$ & $90^{+450}_{-30}$ & $0.11^{+4.56}_{-0.02}$ & 57645.466379997(5) & $1.0^{+0.3}_{-0.3}$ & $1.3^{+0.4}_{-0.4}$ & $573^{+2}_{-3}$\\
B130$\dagger$ & $2000^{+1000}_{-1000}$ & $3000^{+1000}_{-1000}$ & $0.18^{+0.05}_{-0.05}$ & 57645.46833947(1) & $1.4^{+0.4}_{-0.4}$ & $0.6^{+0.4}_{-0.2}$ & $565.5^{+1.0}_{-0.8}$\\
B131 & $1460^{+10}_{-10}$ & $90^{+10}_{-10}$ & $0.08^{+0.01}_{-0.01}$ & 57645.47236212(1) & $1.1^{+0.4}_{-0.3}$ & $0.7^{+0.3}_{-0.3}$ & $564^{+1}_{-2}$\\
B132 & $1720^{+80}_{-20}$ & $100^{+70}_{-20}$ & $0.09^{+0.03}_{-0.01}$ & 57645.474138540(3) & $0.7^{+0.1}_{-0.1}$ & $0.4^{+0.2}_{-0.2}$ & $561^{+1}_{-1}$\\
B133 & $1740^{+180}_{-50}$ & $230^{+110}_{-40}$ & $0.22^{+0.1}_{-0.03}$ & 57645.474448229(4) & $1.7^{+0.2}_{-0.2}$ & $0.7^{+0.3}_{-0.3}$ & $570.6^{+0.7}_{-0.7}$\\
\enddata
\tablecomments{1$\sigma$ errors on the fits are shown on superscript and subscript of each value in the table. For $\mu_t$, the error on the last significant digit is shown in parenthesis.}
\tablenotetext{a}{Burst IDs are chronological. Individual component number (N) for multi-component bursts are appended to the burst IDs. Bursts modeled only using \texttt{curve\_fit} are marked with $*$. Note that the errors on these bursts could be unreliable and may be either under or over-estimated}. Bursts that extend beyond the observable bandwidth can also have unreliable estimates of spectra parameters and fluence (see Section~\ref{sec:fitcaveats}). We mark those bursts with $\dagger$ to indicate that their fluence and spectra parameters could be unconstrained. 
\tablenotetext{b}{$\mu_t$ is the mean of the pulse profile in units of MJD. This can be considered as the arrival time of the pulse. It is referenced to the solar system barycenter, after correcting to infinite frequency using a DM of 560.5\dmunits.}
\tablenotetext{c}{$\tau$ is referred to 1~GHz.}
\end{deluxetable*}

\facilities{Arecibo}

\software{Astropy \citep{astropy:2013, astropy:2018}, Numpy \citep{harris2020}, Matplotlib \citep{Hunter:2007}, Pandas \citep{pandas2010, reback2020pandas}, \textsc{your} \citep{your}, \textsc{fetch} \citep{fetch2020}, Emcee \citep{emcee}, ChainConsumer \citep{Hinton2016}, SciencePlots \citep{SciencePlots}}

\bibliography{frb}{}

\begin{thebibliography}{}
\expandafter\ifx\csname natexlab\endcsname\relax\def\natexlab#1{#1}\fi
\providecommand{\url}[1]{\href{#1}{#1}}
\providecommand{\dodoi}[1]{doi:~\href{http://doi.org/#1}{\nolinkurl{#1}}}
\providecommand{\doeprint}[1]{\href{http://ascl.net/#1}{\nolinkurl{http://ascl.net/#1}}}
\providecommand{\doarXiv}[1]{\href{https://arxiv.org/abs/#1}{\nolinkurl{https://arxiv.org/abs/#1}}}

\bibitem[{{Agarwal} {et~al.}(2020{\natexlab{a}}){Agarwal}, {Aggarwal},
  {Burke-Spolaor}, {Lorimer}, \& {Garver-Daniels}}]{fetch2020}
{Agarwal}, D., {Aggarwal}, K., {Burke-Spolaor}, S., {Lorimer}, D.~R., \&
  {Garver-Daniels}, N. 2020{\natexlab{a}}, Monthly Notices of the Royal
  Astronomical Society, 497, 1661, \dodoi{10.1093/mnras/staa1856}

\bibitem[{{Agarwal} {et~al.}(2020{\natexlab{b}}){Agarwal}, {Lorimer}, {Surnis},
  {Pei}, {Karastergiou}, {Golpayegani}, {Werthimer}, {Cobb}, {McLaughlin},
  {White}, {Armour}, {MacMahon}, {Siemion}, \& {Foster}}]{gb}
{Agarwal}, D., {Lorimer}, D.~R., {Surnis}, M.~P., {et~al.} 2020{\natexlab{b}},
  \mnras, 497, 352, \dodoi{10.1093/mnras/staa1927}

\bibitem[{{Aggarwal}(2021)}]{banded_repeaters}
{Aggarwal}, K. 2021, arXiv e-prints, arXiv:2108.04474.
\newblock \doarXiv{2108.04474}

\bibitem[{{Aggarwal} {et~al.}(2020){Aggarwal}, {Law}, {Burke-Spolaor}, {Bower},
  {Butler}, {Demorest}, {Linford}, \& {Lazio}}]{Aggarwal2020}
{Aggarwal}, K., {Law}, C.~J., {Burke-Spolaor}, S., {et~al.} 2020, Research
  Notes of the American Astronomical Society, 4, 94,
  \dodoi{10.3847/2515-5172/ab9f33}

\bibitem[{Aggarwal {et~al.}(2020)Aggarwal, Agarwal, Kania, Fiore, Thomas,
  Ransom, Demorest, Wharton, Burke-Spolaor, Lorimer, Mclaughlin, \&
  Garver-Daniels}]{your}
Aggarwal, K., Agarwal, D., Kania, J.~W., {et~al.} 2020, Journal of Open Source
  Software, 5, 2750, \dodoi{10.21105/joss.02750}

\bibitem[{{Aggarwal} {et~al.}(2021){Aggarwal}, {Burke-Spolaor}, {Law}, {Bower},
  {Butler}, {Demorest}, {Lazio}, {Linford}, {Sydnor}, \&
  {Anna-Thomas}}]{rfclustering}
{Aggarwal}, K., {Burke-Spolaor}, S., {Law}, C.~J., {et~al.} 2021, arXiv
  e-prints, arXiv:2104.07046.
\newblock \doarXiv{2104.07046}

\bibitem[{{Astropy Collaboration} {et~al.}(2013){Astropy Collaboration},
  {Robitaille}, {Tollerud}, {Greenfield}, {Droettboom}, {Bray}, {Aldcroft},
  {Davis}, {Ginsburg}, {Price-Whelan}, {Kerzendorf}, {Conley}, {Crighton},
  {Barbary}, {Muna}, {Ferguson}, {Grollier}, {Parikh}, {Nair}, {Unther},
  {Deil}, {Woillez}, {Conseil}, {Kramer}, {Turner}, {Singer}, {Fox}, {Weaver},
  {Zabalza}, {Edwards}, {Azalee Bostroem}, {Burke}, {Casey}, {Crawford},
  {Dencheva}, {Ely}, {Jenness}, {Labrie}, {Lim}, {Pierfederici}, {Pontzen},
  {Ptak}, {Refsdal}, {Servillat}, \& {Streicher}}]{astropy:2013}
{Astropy Collaboration}, {Robitaille}, T.~P., {Tollerud}, E.~J., {et~al.} 2013,
  Astronomy \& Astrophysics, 558, A33, \dodoi{10.1051/0004-6361/201322068}

\bibitem[{{Barsdell}(2012)}]{barsdell2012heimdall}
{Barsdell}, B.~R. 2012, PhD thesis, Swinburne University of Technology

\bibitem[{{Bera} \& {Chengalur}(2019)}]{Bera2019}
{Bera}, A., \& {Chengalur}, J.~N. 2019, \mnras, 490, L12,
  \dodoi{10.1093/mnrasl/slz140}

\bibitem[{{Bhat} {et~al.}(2004){Bhat}, {Cordes}, {Camilo}, {Nice}, \&
  {Lorimer}}]{Bhat2004}
{Bhat}, N.~D.~R., {Cordes}, J.~M., {Camilo}, F., {Nice}, D.~J., \& {Lorimer},
  D.~R. 2004, \apj, 605, 759, \dodoi{10.1086/382680}

\bibitem[{{Burke-Spolaor} {et~al.}(2012){Burke-Spolaor}, {Johnston}, {Bailes},
  {Bates}, {Bhat}, {Burgay}, {Champion}, {D'Amico}, {Keith}, {Kramer}, {Levin},
  {Milia}, {Possenti}, {Stappers}, \& {van Straten}}]{Burke-Spolaor2012}
{Burke-Spolaor}, S., {Johnston}, S., {Bailes}, M., {et~al.} 2012, \mnras, 423,
  1351, \dodoi{10.1111/j.1365-2966.2012.20998.x}

\bibitem[{{Cheng} {et~al.}(2020){Cheng}, {Zhang}, \& {Wang}}]{Cheng2020}
{Cheng}, Y., {Zhang}, G.~Q., \& {Wang}, F.~Y. 2020, \mnras, 491, 1498,
  \dodoi{10.1093/mnras/stz3085}

\bibitem[{{CHIME/FRB Collaboration} {et~al.}(2019){CHIME/FRB Collaboration},
  {Amiri}, {Bandura}, {Bhardwaj}, {Boubel}, {Boyce}, {Boyle}, {. Brar},
  {Burhanpurkar}, {Cassanelli}, {Chawla}, {Cliche}, {Cubranic}, {Deng},
  {Denman}, {Dobbs}, {Fandino}, {Fonseca}, {Gaensler}, {Gilbert}, {Gill},
  {Giri}, {Good}, {Halpern}, {Hanna}, {Hill}, {Hinshaw}, {H{\"o}fer},
  {Josephy}, {Kaspi}, {Landecker}, {Lang}, {Lin}, {Masui}, {Mckinven},
  {Mena-Parra}, {Merryfield}, {Michilli}, {Milutinovic}, {Moatti}, {Naidu},
  {Newburgh}, {Ng}, {Patel}, {Pen}, {Pinsonneault-Marotte}, {Pleunis},
  {Rafiei-Ravandi}, {Rahman}, {Ransom}, {Renard}, {Scholz}, {Shaw}, {Siegel},
  {Smith}, {Stairs}, {Tendulkar}, {Tretyakov}, {Vanderlinde}, \& {Yadav}}]{r2}
{CHIME/FRB Collaboration}, {Amiri}, M., {Bandura}, K., {et~al.} 2019, \nat,
  566, 235, \dodoi{10.1038/s41586-018-0864-x}

\bibitem[{{CHIME/FRB Collaboration} {et~al.}(2020){CHIME/FRB Collaboration},
  {Andersen}, {Bandura}, {Bhardwaj}, {Bij}, {Boyce}, {Boyle}, {Brar},
  {Cassanelli}, {Chawla}, {Chen}, {Cliche}, {Cook}, {Cubranic}, {Curtin},
  {Denman}, {Dobbs}, {Dong}, {Fandino}, {Fonseca}, {Gaensler}, {Giri}, {Good},
  {Halpern}, {Hill}, {Hinshaw}, {H{\"o}fer}, {Josephy}, {Kania}, {Kaspi},
  {Landecker}, {Leung}, {Li}, {Lin}, {Masui}, {McKinven}, {Mena-Parra},
  {Merryfield}, {Meyers}, {Michilli}, {Milutinovic}, {Mirhosseini},
  {M{\"u}nchmeyer}, {Naidu}, {Newburgh}, {Ng}, {Patel}, {Pen},
  {Pinsonneault-Marotte}, {Pleunis}, {Quine}, {Rafiei-Ravandi}, {Rahman},
  {Ransom}, {Renard}, {Sanghavi}, {Scholz}, {Shaw}, {Shin}, {Siegel}, {Singh},
  {Smegal}, {Smith}, {Stairs}, {Tan}, {Tendulkar}, {Tretyakov}, {Vanderlinde},
  {Wang}, {Wulf}, \& {Zwaniga}}]{chime1935}
{CHIME/FRB Collaboration}, {Andersen}, B.~C., {Bandura}, K.~M., {et~al.} 2020,
  \nat, 587, 54, \dodoi{10.1038/s41586-020-2863-y}

\bibitem[{{Chime/Frb Collaboration} {et~al.}(2020){Chime/Frb Collaboration},
  {Amiri}, {Andersen}, {Bandura}, {Bhardwaj}, {Boyle}, {Brar}, {Chawla},
  {Chen}, {Cliche}, {Cubranic}, {Deng}, {Denman}, {Dobbs}, {Dong}, {Fandino},
  {Fonseca}, {Gaensler}, {Giri}, {Good}, {Halpern}, {Hessels}, {Hill},
  {H{\"o}fer}, {Josephy}, {Kania}, {Karuppusamy}, {Kaspi}, {Keimpema},
  {Kirsten}, {Landecker}, {Lang}, {Leung}, {Li}, {Lin}, {Marcote}, {Masui},
  {McKinven}, {Mena-Parra}, {Merryfield}, {Michilli}, {Milutinovic},
  {Mirhosseini}, {Naidu}, {Newburgh}, {Ng}, {Nimmo}, {Paragi}, {Patel}, {Pen},
  {Pinsonneault-Marotte}, {Pleunis}, {Rafiei-Ravandi}, {Rahman}, {Ransom},
  {Renard}, {Sanghavi}, {Scholz}, {Shaw}, {Shin}, {Siegel}, {Singh}, {Smegal},
  {Smith}, {Stairs}, {Tendulkar}, {Tretyakov}, {Vanderlinde}, {Wang}, {Wang},
  {Wulf}, {Yadav}, \& {Zwaniga}}]{Chime/FRB2020}
{Chime/Frb Collaboration}, {Amiri}, M., {Andersen}, B.~C., {et~al.} 2020, \nat,
  582, 351, \dodoi{10.1038/s41586-020-2398-2}

\bibitem[{{Connor} \& {van Leeuwen}(2018)}]{Connor2018}
{Connor}, L., \& {van Leeuwen}, J. 2018, \aj, 156, 256,
  \dodoi{10.3847/1538-3881/aae649}

\bibitem[{{Cordes} {et~al.}(2017){Cordes}, {Wasserman}, {Hessels}, {Lazio},
  {Chatterjee}, \& {Wharton}}]{cordes2017}
{Cordes}, J.~M., {Wasserman}, I., {Hessels}, J.~W.~T., {et~al.} 2017, \apj,
  842, 35, \dodoi{10.3847/1538-4357/aa74da}

\bibitem[{Cruces {et~al.}(2020)Cruces, Spitler, Scholz, Lynch, Seymour,
  Hessels, Gouiffés, Hilmarsson, Kramer, \& Munjal}]{Cruces2020}
Cruces, M., Spitler, L.~G., Scholz, P., {et~al.} 2020, Monthly Notices of the
  Royal Astronomical Society, 500, 448, \dodoi{10.1093/mnras/staa3223}

\bibitem[{{Farah} {et~al.}(2019){Farah}, {Flynn}, {Bailes}, {Jameson},
  {Bateman}, {Campbell-Wilson}, {Day}, {Deller}, {Green}, {Gupta}, {Hunstead},
  {Lower}, {Os{\l}owski}, {Parthasarathy}, {Price}, {Ravi}, {Shannon},
  {Sutherland}, {Temby}, {Krishnan}, {Caleb}, {Chang}, {Cruces}, {Roy},
  {Morello}, {Onken}, {Stappers}, {Webb}, \& {Wolf}}]{farah2019}
{Farah}, W., {Flynn}, C., {Bailes}, M., {et~al.} 2019, \mnras, 488, 2989,
  \dodoi{10.1093/mnras/stz1748}

\bibitem[{{Fonseca} {et~al.}(2020){Fonseca}, {Andersen}, {Bhardwaj}, {Chawla},
  {Good}, {Josephy}, {Kaspi}, {Masui}, {Mckinven}, {Michilli}, {Pleunis},
  {Shin}, {Tendulkar}, {Bandura}, {Boyle}, {Brar}, {Cassanelli}, {Cubranic},
  {Dobbs}, {Dong}, {Gaensler}, {Hinshaw}, {Landecker}, {Leung}, {Li}, {Lin},
  {Mena-Parra}, {Merryfield}, {Naidu}, {Ng}, {Patel}, {Pen}, {Rafiei-Ravandi},
  {Rahman}, {Ransom}, {Scholz}, {Smith}, {Stairs}, {Vanderlinde}, {Yadav}, \&
  {Zwaniga}}]{fonseca2020}
{Fonseca}, E., {Andersen}, B.~C., {Bhardwaj}, M., {et~al.} 2020, \apjl, 891,
  L6, \dodoi{10.3847/2041-8213/ab7208}

\bibitem[{{Foreman-Mackey} {et~al.}(2013){Foreman-Mackey}, {Hogg}, {Lang}, \&
  {Goodman}}]{emcee}
{Foreman-Mackey}, D., {Hogg}, D.~W., {Lang}, D., \& {Goodman}, J. 2013, \pasp,
  125, 306, \dodoi{10.1086/670067}

\bibitem[{{Gajjar} {et~al.}(2018){Gajjar}, {Siemion}, {Price}, {Law},
  {Michilli}, {Hessels}, {Chatterjee}, {Archibald}, {Bower}, {Brinkman},
  {Burke-Spolaor}, {Cordes}, {Croft}, {Enriquez}, {Foster}, {Gizani},
  {Hellbourg}, {Isaacson}, {Kaspi}, {Lazio}, {Lebofsky}, {Lynch}, {MacMahon},
  {McLaughlin}, {Ransom}, {Scholz}, {Seymour}, {Spitler}, {Tendulkar},
  {Werthimer}, \& {Zhang}}]{Gajjar2018}
{Gajjar}, V., {Siemion}, A.~P.~V., {Price}, D.~C., {et~al.} 2018, \apj, 863, 2,
  \dodoi{10.3847/1538-4357/aad005}

\bibitem[{Garrett \& Peng(2021)}]{SciencePlots}
Garrett, J.~D., \& Peng, H.-H. 2021, \dodoi{10.5281/zenodo.4106649}

\bibitem[{{Goodman} \& {Weare}(2010)}]{Goodman2010}
{Goodman}, J., \& {Weare}, J. 2010, Communications in Applied Mathematics and
  Computational Science, 5, 65, \dodoi{10.2140/camcos.2010.5.65}

\bibitem[{Gourdji {et~al.}(2019)Gourdji, Michilli, Spitler, Hessels, Seymour,
  Cordes, \& Chatterjee}]{Gourdji2019}
Gourdji, K., Michilli, D., Spitler, L.~G., {et~al.} 2019, The Astrophysical
  Journal, 877, L19, \dodoi{10.3847/2041-8213/ab1f8a}

\bibitem[{{Gupta} {et~al.}(2021){Gupta}, {Flynn}, {Farah}, {Jameson},
  {Venkatraman Krishnan}, {Bailes}, {Bateman}, {Deller}, {Mandlik}, \&
  {Sutherland}}]{gupta2021}
{Gupta}, V., {Flynn}, C., {Farah}, W., {et~al.} 2021, \mnras, 501, 2316,
  \dodoi{10.1093/mnras/staa3683}

\bibitem[{Harris {et~al.}(2020)Harris, Millman, van~der Walt, Gommers,
  Virtanen, Cournapeau, Wieser, Taylor, Berg, Smith, Kern, Picus, Hoyer, van
  Kerkwijk, Brett, Haldane, del R{\'{i}}o, Wiebe, Peterson,
  G{\'{e}}rard-Marchant, Sheppard, Reddy, Weckesser, Abbasi, Gohlke, \&
  Oliphant}]{harris2020}
Harris, C.~R., Millman, K.~J., van~der Walt, S.~J., {et~al.} 2020, Nature, 585,
  357, \dodoi{10.1038/s41586-020-2649-2}

\bibitem[{{He} {et~al.}(2015){He}, {Zhang}, {Ren}, \& {Sun}}]{he2015}
{He}, K., {Zhang}, X., {Ren}, S., \& {Sun}, J. 2015, arXiv e-prints,
  arXiv:1512.03385.
\newblock \doarXiv{1512.03385}

\bibitem[{{Heintz} {et~al.}(2020){Heintz}, {Prochaska}, {Simha}, {Platts},
  {Fong}, {Tejos}, {Ryder}, {Aggerwal}, {Bhandari}, {Day}, {Deller},
  {Kilpatrick}, {Law}, {Macquart}, {Mannings}, {Marnoch}, {Sadler}, \&
  {Shannon}}]{Heintz2020}
{Heintz}, K.~E., {Prochaska}, J.~X., {Simha}, S., {et~al.} 2020, \apj, 903,
  152, \dodoi{10.3847/1538-4357/abb6fb}

\bibitem[{{Hessels} {et~al.}(2019){Hessels}, {Spitler}, {Seymour}, {Cordes},
  {Michilli}, {Lynch}, {Gourdji}, {Archibald}, {Bassa}, {Bower}, {Chatterjee},
  {Connor}, {Crawford}, {Deneva}, {Gajjar}, {Kaspi}, {Keimpema}, {Law},
  {Marcote}, {McLaughlin}, {Paragi}, {Petroff}, {Ransom}, {Scholz}, {Stappers},
  \& {Tendulkar}}]{Hessels2019}
{Hessels}, J.~W.~T., {Spitler}, L.~G., {Seymour}, A.~D., {et~al.} 2019, \apjl,
  876, L23, \dodoi{10.3847/2041-8213/ab13ae}

\bibitem[{{Hinton}(2016)}]{Hinton2016}
{Hinton}, S.~R. 2016, The Journal of Open Source Software, 1, 00045,
  \dodoi{10.21105/joss.00045}

\bibitem[{{Hogg} \& {Foreman-Mackey}(2018)}]{hogg2018}
{Hogg}, D.~W., \& {Foreman-Mackey}, D. 2018, \apjs, 236, 11,
  \dodoi{10.3847/1538-4365/aab76e}

\bibitem[{{Houben} {et~al.}(2019){Houben}, {Spitler}, {ter Veen}, {Rachen},
  {Falcke}, \& {Kramer}}]{Houben2019}
{Houben}, L.~J.~M., {Spitler}, L.~G., {ter Veen}, S., {et~al.} 2019, \aap, 623,
  A42, \dodoi{10.1051/0004-6361/201833875}

\bibitem[{{Huijse} {et~al.}(2018){Huijse}, {Est{\'e}vez}, {F{\"o}rster},
  {Daniel}, {Connolly}, {Protopapas}, {Carrasco}, \&
  {Pr{\'\i}ncipe}}]{Huijse2018}
{Huijse}, P., {Est{\'e}vez}, P.~A., {F{\"o}rster}, F., {et~al.} 2018, \apjs,
  236, 12, \dodoi{10.3847/1538-4365/aab77c}

\bibitem[{Hunter(2007)}]{Hunter:2007}
Hunter, J.~D. 2007, Computing in Science \& Engineering, 9, 90,
  \dodoi{10.1109/MCSE.2007.55}

\bibitem[{{Karuppusamy} {et~al.}(2010){Karuppusamy}, {Stappers}, \& {van
  Straten}}]{Karuppuswamy2010}
{Karuppusamy}, R., {Stappers}, B.~W., \& {van Straten}, W. 2010, \aap, 515,
  A36, \dodoi{10.1051/0004-6361/200913729}

\bibitem[{Katz(2018)}]{Katz2018}
Katz, J.~I. 2018, Monthly Notices of the Royal Astronomical Society, 476, 1849,
  \dodoi{10.1093/mnras/sty366}

\bibitem[{{Keane} \& {Petroff}(2015)}]{Keane2015}
{Keane}, E.~F., \& {Petroff}, E. 2015, \mnras, 447, 2852,
  \dodoi{10.1093/mnras/stu2650}

\bibitem[{{Kirsten} {et~al.}(2021){Kirsten}, {Snelders}, {Jenkins}, {Nimmo},
  {van den Eijnden}, {Hessels}, {Gawro{\'n}ski}, \& {Yang}}]{Kirsten2021}
{Kirsten}, F., {Snelders}, M.~P., {Jenkins}, M., {et~al.} 2021, Nature
  Astronomy, 5, 414, \dodoi{10.1038/s41550-020-01246-3}

\bibitem[{{Kramer} {et~al.}(1994){Kramer}, {Wielebinski}, {Jessner}, {Gil}, \&
  {Seiradakis}}]{kramer1994}
{Kramer}, M., {Wielebinski}, R., {Jessner}, A., {Gil}, J.~A., \& {Seiradakis},
  J.~H. 1994, \aaps, 107, 515

\bibitem[{{Kumar} {et~al.}(2019){Kumar}, {Shannon}, {Os{\l}owski}, {Qiu},
  {Bhandari}, {Farah}, {Flynn}, {Kerr}, {Lorimer}, {Macquart}, {Ng},
  {Phillips}, {Price}, \& {Spiewak}}]{Kumar2019}
{Kumar}, P., {Shannon}, R.~M., {Os{\l}owski}, S., {et~al.} 2019, \apjl, 887,
  L30, \dodoi{10.3847/2041-8213/ab5b08}

\bibitem[{{Kumar} {et~al.}(2021){Kumar}, {Shannon}, {Flynn}, {Os{\l}owski},
  {Bhandari}, {Day}, {Deller}, {Farah}, {Kaczmarek}, {Kerr}, {Phillips},
  {Price}, {Qiu}, \& {Thyagarajan}}]{Kumar2021}
{Kumar}, P., {Shannon}, R.~M., {Flynn}, C., {et~al.} 2021, \mnras, 500, 2525,
  \dodoi{10.1093/mnras/staa3436}

\bibitem[{{Law} {et~al.}(2017){Law}, {Abruzzo}, {Bassa}, {Bower},
  {Burke-Spolaor}, {Butler}, {Cantwell}, {Carey}, {Chatterjee}, {Cordes},
  {Demorest}, {Dowell}, {Fender}, {Gourdji}, {Grainge}, {Hessels}, {Hickish},
  {Kaspi}, {Lazio}, {McLaughlin}, {Michilli}, {Mooley}, {Perrott}, {Ransom},
  {Razavi-Ghods}, {Rupen}, {Scaife}, {Scott}, {Scholz}, {Seymour}, {Spitler},
  {Stovall}, {Tendulkar}, {Titterington}, {Wharton}, \& {Williams}}]{Law2017}
{Law}, C.~J., {Abruzzo}, M.~W., {Bassa}, C.~G., {et~al.} 2017, \apj, 850, 76,
  \dodoi{10.3847/1538-4357/aa9700}

\bibitem[{{Law} {et~al.}(2020){Law}, {Butler}, {Prochaska}, {Zackay},
  {Burke-Spolaor}, {Mannings}, {Tejos}, {Josephy}, {Andersen}, {Chawla},
  {Heintz}, {Aggarwal}, {Bower}, {Demorest}, {Kilpatrick}, {Lazio}, {Linford},
  {Mckinven}, {Tendulkar}, \& {Simha}}]{law2020}
{Law}, C.~J., {Butler}, B.~J., {Prochaska}, J.~X., {et~al.} 2020, \apj, 899,
  161, \dodoi{10.3847/1538-4357/aba4ac}

\bibitem[{{Levin}(2012)}]{levin2012}
{Levin}, L. 2012, PhD thesis, Swinburne University of Technology

\bibitem[{Li {et~al.}(2019)Li, Li, Zhang, Geng, Song, Huang, \& Yang}]{Li2019}
Li, B., Li, L.-B., Zhang, Z.-B., {et~al.} 2019, International Journal of
  Cosmology, Astronomy and Astrophysics, 1, 22, \dodoi{10.18689/ijcaa-1000108}

\bibitem[{{Li} {et~al.}(2021){Li}, {Wang}, {Zhu}, {Zhang}, {Zhang}, {Duan},
  {Zhang}, {Feng}, {Tang}, {Chatterjee}, {Cordes}, {Cruces}, {Dai}, {Gajjar},
  {Hobbs}, {Jin}, {Kramer}, {Lorimer}, {Miao}, {Niu}, {Niu}, {Pan}, {Qian},
  {Spitler}, {Werthimer}, {Zhang}, {Wang}, {Xie}, {Yue}, {Zhang}, {Zhi}, \&
  {Zhu}}]{li2021}
{Li}, D., {Wang}, P., {Zhu}, W.~W., {et~al.} 2021, arXiv e-prints,
  arXiv:2107.08205.
\newblock \doarXiv{2107.08205}

\bibitem[{{Luo} {et~al.}(2020){Luo}, {Wang}, {Men}, {Zhang}, {Jiang}, {Xu},
  {Wang}, {Lee}, {Han}, {Zhang}, {Caballero}, {Chen}, {Chen}, {Gan}, {Guo},
  {Hao}, {Huang}, {Jiang}, {Li}, {Li}, {Li}, {Luo}, {Pan}, {Pei}, {Qian},
  {Sun}, {Wang}, {Wang}, {Wen}, {Xu}, {Xu}, {Yan}, {Yan}, {Yu}, {Yuan},
  {Zhang}, \& {Zhu}}]{luo2020}
{Luo}, R., {Wang}, B.~J., {Men}, Y.~P., {et~al.} 2020, \nat, 586, 693,
  \dodoi{10.1038/s41586-020-2827-2}

\bibitem[{{Lyu} {et~al.}(2021){Lyu}, {Meng}, {Tang}, {Li}, {Wei}, {Geng},
  {Lin}, {Deng}, \& {Wu}}]{Lyu2021}
{Lyu}, F., {Meng}, Y.-Z., {Tang}, Z.-F., {et~al.} 2021, Frontiers of Physics,
  16, 24503, \dodoi{10.1007/s11467-020-1039-4}

\bibitem[{{McKinnon}(2014)}]{McKinnon2014}
{McKinnon}, M.~M. 2014, \pasp, 126, 476, \dodoi{10.1086/676975}

\bibitem[{{Michilli} {et~al.}(2018){Michilli}, {Hessels}, {Lyon}, {Tan},
  {Bassa}, {Cooper}, {Kondratiev}, {Sanidas}, {Stappers}, \& {van
  Leeuwen}}]{michilli2018}
{Michilli}, D., {Hessels}, J.~W.~T., {Lyon}, R.~J., {et~al.} 2018, \mnras, 480,
  3457, \dodoi{10.1093/mnras/sty2072}

\bibitem[{{Mickaliger} {et~al.}(2012){Mickaliger}, {McLaughlin}, {Lorimer},
  {Langston}, {Bilous}, {Kondratiev}, {Lyutikov}, {Ransom}, \&
  {Palliyaguru}}]{Mickaliger2012}
{Mickaliger}, M.~B., {McLaughlin}, M.~A., {Lorimer}, D.~R., {et~al.} 2012,
  \apj, 760, 64, \dodoi{10.1088/0004-637X/760/1/64}

\bibitem[{{Morello} {et~al.}(2020){Morello}, {Barr}, {Stappers}, {Keane}, \&
  {Lyne}}]{morello2020}
{Morello}, V., {Barr}, E.~D., {Stappers}, B.~W., {Keane}, E.~F., \& {Lyne},
  A.~G. 2020, \mnras, 497, 4654, \dodoi{10.1093/mnras/staa2291}

\bibitem[{{Nita} \& {Gary}(2010)}]{nita2010}
{Nita}, G.~M., \& {Gary}, D.~E. 2010, Monthly Notices of the Royal Astronomical
  Society, 406, L60, \dodoi{10.1111/j.1745-3933.2010.00882.x}

\bibitem[{Oostrum {et~al.}(2020)Oostrum, Maan, van Leeuwen, Connor, Petroff,
  Attema, Bast, Gardenier, Hargreaves, \& Kooistra}]{Oostrum2020}
Oostrum, L.~C., Maan, Y., van Leeuwen, J., {et~al.} 2020, Astronomy \&
  Astrophysics, 635, A61, \dodoi{10.1051/0004-6361/201937422}

\bibitem[{Oppermann {et~al.}(2018)Oppermann, Yu, \& Pen}]{Oppermann2018}
Oppermann, N., Yu, H.-R., \& Pen, U.-L. 2018, Monthly Notices of the Royal
  Astronomical Society, 475, 5109, \dodoi{10.1093/mnras/sty004}

\bibitem[{{Parent} {et~al.}(2018){Parent}, {Kaspi}, {Ransom}, {Krasteva},
  {Patel}, {Scholz}, {Brazier}, {McLaughlin}, {Boyce}, {Zhu}, {Pleunis},
  {Allen}, {Bogdanov}, {Caballero}, {Camilo}, {Camuccio}, {Chatterjee},
  {Cordes}, {Crawford}, {Deneva}, {Ferdman}, {Freire}, {Hessels}, {Jenet},
  {Knispel}, {Lazarus}, {van Leeuwen}, {Lyne}, {Lynch}, {Seymour}, {Siemens},
  {Stairs}, {Stovall}, \& {Swiggum}}]{ParentFFA}
{Parent}, E., {Kaspi}, V.~M., {Ransom}, S.~M., {et~al.} 2018, \apj, 861, 44,
  \dodoi{10.3847/1538-4357/aac5f0}

\bibitem[{{Pastor-Marazuela} {et~al.}(2020){Pastor-Marazuela}, {Connor}, {van
  Leeuwen}, {Maan}, {ter Veen}, {Bilous}, {Oostrum}, {Petroff}, {Straal},
  {Vohl}, {Attema}, {Boersma}, {Kooistra}, {van der Schuur}, {Sclocco},
  {Smits}, {Adams}, {Adebahr}, {de Blok}, {Coolen}, {Damstra}, {D{\'e}nes},
  {Hess}, {van der Hulst}, {Hut}, {Ivashina}, {Kutkin}, {Marcel Loose},
  {Lucero}, {Mika}, {Moss}, {Mulder}, {Norden}, {Oosterloo}, {Orr{\'u}},
  {Ruiter}, \& {Wijnholds}}]{Marazuela2020}
{Pastor-Marazuela}, I., {Connor}, L., {van Leeuwen}, J., {et~al.} 2020, arXiv
  e-prints, arXiv:2012.08348.
\newblock \doarXiv{2012.08348}

\bibitem[{{Patel} {et~al.}(2018){Patel}, {Agarwal}, {Bhardwaj}, {Boyce},
  {Brazier}, {Chatterjee}, {Chawla}, {Kaspi}, {Lorimer}, {McLaughlin},
  {Parent}, {Pleunis}, {Ransom}, {Scholz}, {Wharton}, {Zhu}, {Alam}, {Caballero
  Valdez}, {Camilo}, {Cordes}, {Crawford}, {Deneva}, {Ferdman}, {Freire},
  {Hessels}, {Nguyen}, {Stairs}, {Stovall}, \& {van Leeuwen}}]{patel2018}
{Patel}, C., {Agarwal}, D., {Bhardwaj}, M., {et~al.} 2018, \apj, 869, 181,
  \dodoi{10.3847/1538-4357/aaee65}

\bibitem[{Petroff \& Chatterjee(2021)}]{frb_letter}
Petroff, E., \& Chatterjee, S. 2021, Cornell University Library,
  \dodoi{10.7298/5N8E-XX87}

\bibitem[{{Platts} {et~al.}(2021){Platts}, {Caleb}, {Stappers}, {Main},
  {Weltman}, {Shock}, {Kramer}, {Bezuidenhout}, {Jankowski}, {Morello},
  {Possenti}, {Rajwade}, {Rhodes}, \& {Wu}}]{platts2021}
{Platts}, E., {Caleb}, M., {Stappers}, B.~W., {et~al.} 2021, arXiv e-prints,
  arXiv:2105.11822.
\newblock \doarXiv{2105.11822}

\bibitem[{{Pleunis} {et~al.}(2021){Pleunis}, {Michilli}, {Bassa}, {Hessels},
  {Naidu}, {Andersen}, {Chawla}, {Fonseca}, {Gopinath}, {Kaspi}, {Kondratiev},
  {Li}, {Bhardwaj}, {Boyle}, {Brar}, {Cassanelli}, {Gupta}, {Josephy},
  {Karuppusamy}, {Keimpema}, {Kirsten}, {Leung}, {Marcote}, {Masui},
  {Mckinven}, {Meyers}, {Ng}, {Nimmo}, {Paragi}, {Rahman}, {Scholz}, {Shin},
  {Smith}, {Stairs}, \& {Tendulkar}}]{Pleunis2021}
{Pleunis}, Z., {Michilli}, D., {Bassa}, C.~G., {et~al.} 2021, \apjl, 911, L3,
  \dodoi{10.3847/2041-8213/abec72}

\bibitem[{{Popov} \& {Stappers}(2007)}]{Popov2007}
{Popov}, M.~V., \& {Stappers}, B. 2007, \aap, 470, 1003,
  \dodoi{10.1051/0004-6361:20066589}

\bibitem[{{Price-Whelan} {et~al.}(2018){Price-Whelan}, {Sip{\H{o}}cz},
  {G{\"u}nther}, {Lim}, {Crawford}, {Conseil}, {Shupe}, {Craig}, {Dencheva},
  {Ginsburg}, {VanderPlas}, {Bradley}, {P{\'e}rez-Su{\'a}rez}, {de Val-Borro},
  {Paper Contributors}, {Aldcroft}, {Cruz}, {Robitaille}, {Tollerud},
  {Coordination Committee}, {Ardelean}, {Babej}, {Bach}, {Bachetti}, {Bakanov},
  {Bamford}, {Barentsen}, {Barmby}, {Baumbach}, {Berry}, {Biscani}, {Boquien},
  {Bostroem}, {Bouma}, {Brammer}, {Bray}, {Breytenbach}, {Buddelmeijer},
  {Burke}, {Calderone}, {Cano Rodr{\'\i}guez}, {Cara}, {Cardoso}, {Cheedella},
  {Copin}, {Corrales}, {Crichton}, {D{\textquoteright}Avella}, {Deil},
  {Depagne}, {Dietrich}, {Donath}, {Droettboom}, {Earl}, {Erben}, {Fabbro},
  {Ferreira}, {Finethy}, {Fox}, {Garrison}, {Gibbons}, {Goldstein}, {Gommers},
  {Greco}, {Greenfield}, {Groener}, {Grollier}, {Hagen}, {Hirst}, {Homeier},
  {Horton}, {Hosseinzadeh}, {Hu}, {Hunkeler}, {Ivezi{\'c}}, {Jain}, {Jenness},
  {Kanarek}, {Kendrew}, {Kern}, {Kerzendorf}, {Khvalko}, {King}, {Kirkby},
  {Kulkarni}, {Kumar}, {Lee}, {Lenz}, {Littlefair}, {Ma}, {Macleod},
  {Mastropietro}, {McCully}, {Montagnac}, {Morris}, {Mueller}, {Mumford},
  {Muna}, {Murphy}, {Nelson}, {Nguyen}, {Ninan}, {N{\"o}the}, {Ogaz}, {Oh},
  {Parejko}, {Parley}, {Pascual}, {Patil}, {Patil}, {Plunkett}, {Prochaska},
  {Rastogi}, {Reddy Janga}, {Sabater}, {Sakurikar}, {Seifert}, {Sherbert},
  {Sherwood-Taylor}, {Shih}, {Sick}, {Silbiger}, {Singanamalla}, {Singer},
  {Sladen}, {Sooley}, {Sornarajah}, {Streicher}, {Teuben}, {Thomas},
  {Tremblay}, {Turner}, {Terr{\'o}n}, {van Kerkwijk}, {de la Vega}, {Watkins},
  {Weaver}, {Whitmore}, {Woillez}, {Zabalza}, \& {Contributors}}]{astropy:2018}
{Price-Whelan}, A.~M., {Sip{\H{o}}cz}, B.~M., {G{\"u}nther}, H.~M., {et~al.}
  2018, The Astronomical Journal, 156, 123, \dodoi{10.3847/1538-3881/aabc4f}

\bibitem[{{Rajwade} {et~al.}(2020{\natexlab{a}}){Rajwade}, {Mickaliger},
  {Stappers}, {Bassa}, {Breton}, {Karastergiou}, \& {Keane}}]{Rajwade2020}
{Rajwade}, K.~M., {Mickaliger}, M.~B., {Stappers}, B.~W., {et~al.}
  2020{\natexlab{a}}, \mnras, 493, 4418, \dodoi{10.1093/mnras/staa616}

\bibitem[{{Rajwade} {et~al.}(2020{\natexlab{b}}){Rajwade}, {Mickaliger},
  {Stappers}, {Morello}, {Agarwal}, {Bassa}, {Breton}, {Caleb}, {Karastergiou},
  {Keane}, \& {Lorimer}}]{Rajwade121102}
---. 2020{\natexlab{b}}, \mnras, 495, 3551, \dodoi{10.1093/mnras/staa1237}

\bibitem[{{Ransom}(2011)}]{presto}
{Ransom}, S. 2011, {PRESTO: PulsaR Exploration and Search TOolkit}.
\newblock \doeprint{1107.017}

\bibitem[{Reback {et~al.}(2021)Reback, McKinney, {Jbrockmendel}, Bossche,
  Augspurger, Cloud, {Gfyoung}, Hawkins, {Sinhrks}, Roeschke, Klein, {Terji
  Petersen}, Tratner, She, Ayd, Naveh, Garcia, Schendel, Hayden, Saxton, {,
  Patrick}, Jancauskas, McMaster, Battiston, {Skipper Seabold}, Gorelli, {Kaiqi
  Dong}, {Chris-B1}, {H-Vetinari}, \& Hoyer}]{reback2020pandas}
Reback, J., McKinney, W., {Jbrockmendel}, {et~al.} 2021, pandas-dev/pandas:
  Pandas 1.2.1,  Zenodo, \dodoi{10.5281/ZENODO.3509134}

\bibitem[{{Russakovsky} {et~al.}(2014){Russakovsky}, {Deng}, {Su}, {Krause},
  {Satheesh}, {Ma}, {Huang}, {Karpathy}, {Khosla}, {Bernstein}, {Berg}, \&
  {Fei-Fei}}]{Russakovsky2014}
{Russakovsky}, O., {Deng}, J., {Su}, H., {et~al.} 2014, arXiv e-prints,
  arXiv:1409.0575.
\newblock \doarXiv{1409.0575}

\bibitem[{{Shannon} {et~al.}(2018){Shannon}, {Macquart}, {Bannister}, {Ekers},
  {James}, {Os{\l}owski}, {Qiu}, {Sammons}, {Hotan}, {Voronkov}, {Beresford},
  {Brothers}, {Brown}, {Bunton}, {Chippendale}, {Haskins}, {Leach},
  {Marquarding}, {McConnell}, {Pilawa}, {Sadler}, {Troup}, {Tuthill},
  {Whiting}, {Allison}, {Anderson}, {Bell}, {Collier}, {G{\"u}rkan}, {Heald},
  \& {Riseley}}]{Shannon2018}
{Shannon}, R.~M., {Macquart}, J.~P., {Bannister}, K.~W., {et~al.} 2018, \nat,
  562, 386, \dodoi{10.1038/s41586-018-0588-y}

\bibitem[{Spitler {et~al.}(2014)Spitler, Cordes, Hessels, Lorimer, McLaughlin,
  Chatterjee, Crawford, Deneva, Kaspi, Wharton, Allen, Bogdanov, Brazier,
  Camilo, Freire, Jenet, Karako-Argaman, Knispel, Lazarus, Lee, van Leeuwen,
  Lynch, Ransom, Scholz, Siemens, Stairs, Stovall, Swiggum, Venkataraman, Zhu,
  Aulbert, \& Fehrmann}]{Spitler2014}
Spitler, L.~G., Cordes, J.~M., Hessels, J. W.~T., {et~al.} 2014, The
  Astrophysical Journal, 790, 101, \dodoi{10.1088/0004-637x/790/2/101}

\bibitem[{Spitler {et~al.}(2016)Spitler, Scholz, Hessels, Bogdanov, Brazier,
  Camilo, Chatterjee, Cordes, Crawford, Deneva, Ferdman, Freire, Kaspi,
  Lazarus, Lynch, Madsen, McLaughlin, Patel, Ransom, Seymour, Stairs, Stappers,
  van Leeuwen, \& Zhu}]{Spitler2016}
Spitler, L.~G., Scholz, P., Hessels, J. W.~T., {et~al.} 2016, Nature, 531, 202,
  \dodoi{10.1038/nature17168}

\bibitem[{Tendulkar {et~al.}(2017)Tendulkar, Bassa, Cordes, Bower, Law,
  Chatterjee, Adams, Bogdanov, Burke-Spolaor, Butler, Demorest, Hessels, Kaspi,
  Lazio, Maddox, Marcote, McLaughlin, Paragi, Ransom, Scholz, Seymour, Spitler,
  van Langevelde, \& Wharton}]{Tendulkar2017}
Tendulkar, S.~P., Bassa, C.~G., Cordes, J.~M., {et~al.} 2017, The Astrophysical
  Journal, 834, L7, \dodoi{10.3847/2041-8213/834/2/l7}

\bibitem[{{The CHIME/FRB Collaboration} {et~al.}(2021){The CHIME/FRB
  Collaboration}, {:}, {Amiri}, {Andersen}, {Bandura}, {Berger}, {Bhardwaj},
  {Boyce}, {Boyle}, {Brar}, {Breitman}, {Cassanelli}, {Chawla}, {Chen},
  {Cliche}, {Cook}, {Cubranic}, {Curtin}, {Deng}, {Dobbs}, {Fengqiu}, {Dong},
  {Eadie}, {Fandino}, {Fonseca}, {Gaensler}, {Giri}, {Good}, {Halpern}, {Hill},
  {Hinshaw}, {Josephy}, {Kaczmarek}, {Kader}, {Kania}, {Kaspi}, {Landecker},
  {Lang}, {Leung}, {Li}, {Lin}, {Masui}, {Mckinven}, {Mena-Parra},
  {Merryfield}, {Meyers}, {Michilli}, {Milutinovic}, {Mirhosseini},
  {M{\"u}nchmeyer}, {Naidu}, {Newburgh}, {Ng}, {Patel}, {Pen}, {Petroff},
  {Pinsonneault-Marotte}, {Pleunis}, {Rafiei-Ravandi}, {Rahman}, {Ransom},
  {Renard}, {Sanghavi}, {Scholz}, {Shaw}, {Shin}, {Siegel}, {Sikora}, {Singh},
  {Smith}, {Stairs}, {Tan}, {Tendulkar}, {Vanderlinde}, {Wang}, {Wulf}, \&
  {Zwaniga}}]{chime_cat}
{The CHIME/FRB Collaboration}, {:}, {Amiri}, M., {et~al.} 2021, arXiv e-prints,
  arXiv:2106.04352.
\newblock \doarXiv{2106.04352}

\bibitem[{VanderPlas(2018)}]{VanderPlas_2018}
VanderPlas, J.~T. 2018, The Astrophysical Journal Supplement Series, 236, 16,
  \dodoi{10.3847/1538-4365/aab766}

\bibitem[{{W}es {M}c{K}inney(2010)}]{pandas2010}
{W}es {M}c{K}inney. 2010, in {P}roceedings of the 9th {P}ython in {S}cience
  {C}onference, ed. {S}t\'efan van~der {W}alt \& {J}arrod {M}illman, 56 -- 61,
  \dodoi{10.25080/Majora-92bf1922-00a}

\bibitem[{Zhang {et~al.}(2018)Zhang, Gajjar, Foster, Siemion, Cordes, Law, \&
  Wang}]{Zhang2018}
Zhang, Y.~G., Gajjar, V., Foster, G., {et~al.} 2018, The Astrophysical Journal,
  866, 149, \dodoi{10.3847/1538-4357/aadf31}

\end{thebibliography}
\bibliographystyle{aasjournal}

\end{document}